%% file: ms.tex
\documentclass[english, 12pt, a4paper, 
numbers=noenddot
]{scrartcl}
\usepackage[T1]{fontenc}
\usepackage[utf8]{inputenc}  
\usepackage[english]{babel} 
\usepackage{lmodern} 

\usepackage[onehalfspacing]{setspace} 
\usepackage{scrlayer-scrpage}
\usepackage{amsmath, amssymb, amsthm, amstext} 
\usepackage{thmtools}
\usepackage{bm}

\usepackage[table]{xcolor}
\usepackage{float} 
\usepackage[section]{placeins} 
\usepackage{multirow}
\usepackage{booktabs} 
\usepackage{longtable} 
\usepackage{tabularx} 
\newcolumntype{Y}{>{\centering\arraybackslash}X}

\usepackage{dcolumn}
\newcolumntype{d}[1]{D{.}{.}{#1}}

\usepackage{siunitx}
\usepackage{graphicx}  
\usepackage{epstopdf} 

\usepackage[inline]{enumitem} 

\usepackage[babel, english = british]{csquotes} 

\usepackage[backend=biber,
			style=apa, 
			sorting = nyt,
			maxnames = 2,
            maxbibnames=8,
            minbibnames=4,
			doi=false,
			isbn=false,
			url=false,
            dashed=false,
			eprint=true,
            uniquename=false
            ]{biblatex}
\DeclareDelimFormat{nameyeardelim}{\addcomma\space} 
\usepackage{varioref} 
\definecolor{darkblue}{rgb}{0.0, 0.0, 0.55} 
\usepackage[pdfencoding=auto, psdextra, 
   hyperfootnotes=false, 
   colorlinks=true,
   linkcolor=darkblue,   
   citecolor=darkblue,  
   urlcolor=darkblue 
   ]{hyperref} 
\usepackage{nameref}
\usepackage[capitalize, nameinlink]{cleveref}

\newtheoremstyle{Definition}
  {0.2cm}                 
  {0.2cm}                 
  {\normalfont}           
  {}                      
  {\bfseries}  						
  {.}                     
  { }              				
  {}
                          
\newtheoremstyle{Theorem}
  {0.2cm}                 
  {0.2cm}                 
  {\itshape}           	  
  {}                      
  {\bfseries}  						
  {.}                     
  { }              				
  {}
\theoremstyle{Theorem}
    	\Crefname{cor}{Corollary}{Corollaries}
		\Crefname{prop}{Proposition}{Propositions}
		\Crefname{lem}{Lemma}{Lemmas}
    	\Crefname{thm}{Theorem}{Theorems}
		\Crefname{assum}{Assumption}{Assumptions}
	
		\Crefname{conjec}{Conjecture}{Conjectures}
		\Crefname{defn}{Definition}{Definitions}

\theoremstyle{Definition}
    	\Crefname{example}{Example}{Examples}	
    	\Crefname{rem}{Remark}{Remarks}      
		\Crefname{algo}{Algorithm}{Algorithms}

\renewcommand*{\theenumi}{\arabic{enumi}} 
\renewcommand*{\theenumii}{(\alph{enumii})}
\renewcommand*{\theenumiii}{(\roman{enumiii})}

\makeatletter
\renewcommand*{\p@enumii}{\theenumi\,} 
\renewcommand*{\p@enumiii}{\p@enumii.\theenumii} 
\renewcommand*{\p@enumiv}{\p@enumiii.\theenumiii} 
\makeatother

\renewcommand\floatpagefraction{0.85}

\allowdisplaybreaks
\graphicspath{{./figures/}} 
\hypersetup{pdfauthor = {Martin C. Arnold and Thilo Reinschluessel},
            pdftitle = {Adaptive Unit Root Inference in Autoregressions using the Lasso Solution Path}
}

\input{ee.sty} 

\makeatletter
\renewenvironment{abstract}{%
\if@abstrt
    \small
    \begin{center}
      {\normalfont\sectfont\nobreak\abstractname 
        \vspace{-.5em}\vspace{\z@}}%
    \end{center}
\fi
    \singlespacing\quotation
}{%
\endquotation
} 
\makeatother

\newsavebox\extrainfobox

\newcommand{\DOI}[1]{%
  \gdef\@DOI{DOI: #1}%
}

\addtokomafont{disposition}{\rmfamily}
\addtokomafont{title}{}
\addtokomafont{author}{\large}
\addtokomafont{date}{\large}
\addtokomafont{date}{\vspace*{1em}}
\title{Adaptive Unit Root Inference in Autoregressions using the Lasso Solution Path\protect\footnotemark\footnotetext{We thank Christoph Hanck for valuable comments that significantly improved the paper. \input{funding_TRR391.tex} }}
\author{Martin Arnold\thanks{Faculty of Business Administration and Economics, University of Duisburg-Essen, Universit\"atsstra{\ss}e 12, 45117 Essen, Germany}\\{\vspace{-1ex}\small \href{mailto:martin.arnold@vwl.uni-due.de}{martin.arnold@vwl.uni-due.de}}  \and Thilo Reinschlüssel\footnotemark[2]~\thanks{RGS Econ - Ruhr Graduate School in Economics, Research Academy Ruhr, Universitätsstra{\ss}e 150, 44801 Bochum, Germany}\\{\small \href{mailto:thilo.reinschluessel@vwl.uni-due.de}{thilo.reinschluessel@vwl.uni-due.de}}}
\date{October 12, 2026} 

\begin{document}
\maketitle
\vspace*{-1em}

\begin{abstract}
\noindent\textit{\textbf{Abstract}}\\[.5ex]
\input{A_abstract.tex}
\\[2ex]
 \end{abstract}
\renewcommand*{\thefootnote}{\arabic{footnote}}
\setcounter{footnote}{0}
\newpage %
\input{B_corpus.tex}
%
\clearpage
\appendix
\addcontentsline{toc}{section}{Appendix}
\renewcommand{\thesubsection}{\Alph{subsection}} 
\renewcommand{\theequation}{\thesubsection\arabic{equation}} 
\setcounter{table}{0}
\setcounter{figure}{0}
\renewcommand{\thetable}{\thesubsection\arabic{table}}
\renewcommand{\thefigure}{\thesubsection\arabic{figure}}
%
\input{C_appendix.tex}
%
\newpage
\printbibliography
\end{document}

%% file: funding_TRR391.tex
The authors acknowledge partial financial support from TRR 391 Spatio-temporal Statistics for the Transition of Energy and Transport (520388526) by the Deutsche Forschungsgemeinschaft (DFG, German Research Foundation).

%% file: A_abstract.tex
We show that the activation knot of a potentially non-stationary regressor on the adaptive Lasso solution path in autoregressions can be leveraged for selection-free inference about a unit root. The resulting test has asymptotic power against local alternatives in $1/T$ neighbourhoods, unlike post-selection inference methods based on consistent model selection. Exploiting the information enrichment principle devised by Reinschlüssel and Arnold [\href{https://doi.org/10.48550/arXiv.2402.16580}{arXiv:2402.16580}, \citeyear{Reinschlussel2024}] to improve the Lasso-based selection of ADF models, we propose a composite statistic and analyse its asymptotic distribution and local power function. Monte Carlo evidence shows that the combined test dominates the comparable post-selection inference methods of Tibshirani et al. [JASA, 2016, 514, 600--620] and may surpass the power of established unit root tests against local alternatives. We apply the new tests to groundwater level time series for Germany and find evidence rejecting stochastic trends to explain observed long-term declines in mean water levels.

%% file: B_corpus.tex
\section{Introduction}

Penalised estimators and their econometric applications have become increasingly important in recent decades. Many theoretical contributions are extensions of Lasso-type estimators \parencite{FuKnight2000}, which go back to the seminal contribution of \textcite{Tibshirani1996}. Despite their significant empirical relevance for economic disciplines such as macroeconomics and finance, methodological contributions to the penalised estimation of time series models are less frequent than for cross-section models. A majority of this research considers stationary time series. Examples which apply Lasso estimators to autoregressions are \textcite{Wangetal2007}, \textcite{Hsuetal2008}, \textcite{RenZhang2010}, and \textcite{MedeirosMendes2012}. In high-dimensional settings, \textcite{Hecqetal2023} address testing for Granger causality in vector autoregressions (VARs) using the concept of double-selection introduced by \textcite{Bellonietal2014}. \textcite{Adameketal2023} use the desparsified Lasso for inference in time series models with general error processes where the number of regressors may exceed the time dimension $T$. 

Lasso-type estimators for time series models involving non-stationary data gain attention as well. One challenge here is to account for the different convergence rates of estimators when there are stationary and non-stationary regressors. For example, \textcite{LiaoPhillips2015} use Lasso-type regularisation for simultaneous selection of the cointegration rank and the lag order in vector error correction models. \textcite{Kooetal2020} show the consistency of the Lasso in a high-dimensional predictive regression setup. \textcite{Leeetal2022} propose using the adaptive Lasso in a double post-selection procedure for estimating predictive regressions with mixed-root regressors and multiple cointegration groups. 

\textcite{CanerKnight2013} show that Bridge estimators can be tuned for conservative model selection of augmented Dickey-Fuller (ADF) regressions. \textcite{kock2016consistent} proves that the adaptive Lasso \parencite{zou2006adaptive} has the oracle property in ADF models and thus correctly selects the lag pattern and between stationary and non-stationary data $y_t$ with probability approaching one as $T\to\infty$. These methods require tuning penalty functions to infer the order of integration from the estimated coefficient of the lagged level $y_{t-1}$. 

We, in turn, suggest a tuning-free test for the adaptive Lasso to distinguish stationary and non-stationary autoregressions. The method exploits that the adaptive penalty weights drive the stochastic properties of knots on the Lasso solution path such that nuisance-parameter-free testing of the non-stationarity hypothesis is feasible. The only comparable approach we know of is post-selection inference with the spacing test by \textcite{Tibshiranietal2016} for sequential regression procedures that can be computed by a least angle regression (LAR) algorithm. Their idea is to test for zero partial correlation of a predictor with the LAR residual at its activation knot, adjusting for other variables currently in the active set. The null hypothesis of the spacing test is conditional on the inherently stochastic variable selection at a LAR knot. Our test differs from this approach since the null is unconditional on the remainder of the model, as we elaborate below.

Transferring the information enrichment principle of \textcite{Reinschlussel2024}, we further suggest a combined test, enhancing the activation knot test with an auxiliary unit root test statistic via the penalty weight of $y_{t-1}$. We show that our tuning-free activation threshold tests have (asymptotic) power against local alternatives, contrary to the adaptive Lasso tuned for consistent model selection; see Theorem 5 in \textcite{kock2016consistent}. Monte Carlo studies establish that the new tests are superior to the LAR post-selection methods of \textcite{Leeetal2016,Tibshiranietal2016,Tibshiranietal2018}. We further verify the usefulness of the principle by simulations, showing that the information-enriched test, in particular, closely tracks and may even outperform established unit root tests such as the nearly efficient ADF test of \textcite{Elliottetal1996} under local alternatives.

The remainder of this article is as follows. \Cref{sec:alurt_aleoA} discusses the setup and recaps the estimation of autoregressions using the adaptive Lasso. In \Cref{sec:alurt_tfnutlsp}, we establish theoretical results on the solution path of the adaptive Lasso in ADF models underlying the new tests and discuss their implementation. Monte Carlo results are presented in \Cref{sec:alurt_mce}. \Cref{sec:alurt_atggl} illustrates the tests in an empirical application to German groundwater level time series. \Cref{sec:alurt_cao} concludes.

We will use the following conventions throughout the manuscript. $\xrightarrow{p}$ and $\xrightarrow{d}$ denote convergence in probability and distribution. $\lambda=\Theta(T^\kappa)$ denotes $0<\liminf_{T\rightarrow\infty}\lambda/T^\kappa\leq\limsup_{T\rightarrow\infty}\lambda/T^\kappa<\infty$, $\kappa\in\mathbb{R}$. $\lVert\vx\rVert_q$ denotes the $\ell_q$ norm of a vector $\vx$. Coefficients of the data-generating model have a superscript $\star$. We refer to an estimator $\widehat{\mathcal{M}}$ as model selection consistent if $\lim_{T\to\infty}\Prob{\widehat{\mathcal{M}} = \mathcal{M}} = 1$ and conservative if $\lim_{T\to\infty}\Prob{\widehat{\mathcal{M}}\supseteq\mathcal{M}} = 1$ with $\lim_{T\to\infty}\Prob{\widehat{\mathcal{M}}\supset\mathcal{M}} > 0$, where $\mathcal{M}:=\left\{i:\beta_i^\star\neq0\right\}$ is the set of relevant variables. Several (estimated) quantities, often denoted by Greek symbols, depend on the $\ell_1$ penalty and the sample size, which we express by sub-indexing with $\lambda$ and $T$ when helpful for the exposition. We define by $\lambda_\mathcal{O}$ the continuum of sequences within a specific range of stochastic orders providing for the oracle property as $T\to\infty$. For ease of comprehension, we denote the set of sequences asymptotically leading to consistent ($\lambda = \Theta(\lambda_\mathcal{O})$) or conservative ($\lambda = o_p(\lambda_\mathcal{O})$) model selection by $\lambda\in[0,\,\lambda_\mathcal{O}]$.

\section{Adaptive Lasso estimation of autoregressions}
\label{sec:alurt_aleoA}

This paper considers time series $y_t$ with an autoregressive (AR) data-generating process (DGP) of the form
    \begin{align}
        \label{eq:ALURT_thedgp}
        y_t = \vz_t'\vtheta^\star + x_t, \qquad x_t = \varrho^\star x_{t-1} + u_t, \qquad t=1,\dots,T.
    \end{align}
	The process \eqref{eq:ALURT_thedgp} incorporates a deterministic component $\vz_t = (1,t,\dots,t^q)'$ with coefficient vector $\vtheta^\star$, and a stochastic component $x_t$ driven by the error process $u_t$. The stochastic component is a first-order AR process with parameter $\varrho^\star$. We consider $\varrho^\star\in(-1, 1]$, i.e., $x_t$ is stationary when $\lvert\varrho^\star\rvert<1$ and non-stationary when $\varrho^\star=1$.
 The error process $u_t$ satisfies the following assumptions.
    \begin{restatable}[Linear process errors]{assum}{}
        \label{assum:lperrorsChangPark}
        $u_t = \phi(L)\varepsilon_t$, where the lag polynomial $\phi(L)$ satisfies $\phi(z)\neq0$ for all $\lvert z\rvert\leq1$ and $\sum_{j=1}^\infty j^s\lvert\phi_j\rvert<\infty$ for some $s\geq1$. The innovations $\varepsilon_t$ are a martingale difference sequence (MDS) such that $\E(\varepsilon_t^2)=\sigma^2$, $T^{-1}\sum_t \varepsilon_t^2 \xrightarrow{p} \sigma^2$ and $\E\lvert\varepsilon_t\rvert^r<K_r$ with $r\geq4$ and some $K_r<\infty$.
     \end{restatable}
	 \Cref{assum:lperrorsChangPark} is adopted from \textcite{ChangPark2002} and covers a general class of weakly stationary linear error processes, including ARMA processes with non-\-geo\-metrically decaying coefficients. For the MDS innovations $\varepsilon_t$ considered, $u_t$ has \enquote{long-run} variance (LRV) $\omega^2 := \lim_{T\rightarrow\infty} \E\left[T^{-1}(\sum_{t=1}^T u_t)^2\right]<\infty$. \Cref{assum:lperrorsChangPark} further ensures that $u_t$ admits an AR($\infty$) representation. Hence, the DGP~\eqref{eq:ALURT_thedgp} can be written as an ADF regression, i.e.,
	 \begin{align}
        	\Delta y_t =& \, d_t + \rho^\star y_{t-1} + \sum_{j=1}^{\infty} \delta_j^\star \Delta y_{t-j} + \varepsilon_{t},
        	\label{eq:adfmod}
	 \end{align} 
	 where $d_t$ is a deterministic component, $\rho^\star\in(-2,0]$, and $\sum_{j=1}^\infty \delta_j^\star < 1$. 
	 
	 Following \textcite{ChangPark2002}, we consider sparse approximations of \eqref{eq:adfmod},
	  \begin{align}
        	\Delta y_t =& \, d_t + \rho_p y_{t-1} + \sum_{j=1}^p \delta_{p,\,j} \Delta y_{t-j} + \varepsilon_{p,\,t},
        	\label{eq:ALURT_adfreg}
	    \end{align} 
     with lag order $p$ suitably increasing as $T\rightarrow\infty$.
     Note that the coefficients $\rho_p$ and $\delta_{p,\,1}, \dots, \delta_{p,\,p}$ depend on the truncation lag $p$ and are thus subject to approximation errors $\rho_p - \rho^\star$ and $\delta_{p,\,j} - \delta_j^\star$. The truncation additionally flaws the innovations with a truncation error $\varepsilon_{p,\,t} - \varepsilon_t$, cf. Remark 2.1 in \textcite{ChangPark2002}. All approximation errors and truncation errors vanish as $p\to\infty$.   
	 \begin{restatable}[AR order]{assum}{}
	 		\label{assum:aroder}
            $p$ satisfies $p=o(T^{1/3})$ and $p\to\infty$ as $T\to\infty$.
	 \end{restatable}
        \textcite{ChangPark2002} prove that under \Cref{assum:aroder}, ordinary least squares (OLS) estimation of \eqref{eq:ALURT_adfreg} is consistent for all coefficients by eliminating the errors from approximation, truncation and estimation concurrently. They further show these conditions to suffice for the ADF t-ratio and the normalised coefficient statistic (\cite{SaidDickey1984}) to attain their respective limit distributions under the null hypothesis $H_0$: $\rho^\star = 0$, as first studied in \textcite{DickeyFuller1979,DickeyFuller1981}.

The adaptive Lasso (AL) estimator (\cite{zou2006adaptive}) in model \eqref{eq:ALURT_adfreg} obtains as
	\begin{align}
		\widehat{\vbeta}_\lambda := (\widehat{\rho}_\lambda, \widehat\vdelta_\lambda')' = \arg\min_{\dot\rho,\, \dot\vdelta} \Psi_T(\dot\rho,\dot\vdelta\vert\lambda), \label{eq:ALURT_ALestimator}
	\end{align}
	where $\Psi_T$ is the adaptively penalised loss function
	\begin{align}
		\Psi_T(\dot\rho,\dot\vdelta\vert\lambda) := \sum_t \left(\Delta y_t - \dot\rho y_{t-1} - \sum_{j=1}^p \dot\delta_j \Delta y_{t-j} \right)^2 +&\, 2\lambda\left( w_1^{\gamma_1} \lvert\dot\rho\rvert + \sum_{j=1}^p w_{2,\,j}^{\gamma_2} \left\lvert\dot\delta_j\right\rvert\right),
    \label{eq:ALURT_adaptive_lassooptim}
  \end{align}
with $\lambda\geq0$ governing the $\ell_1$ penalty, adaptive weights $w_1 := 1/\lvert\widehat{\rho}\rvert$, $w_{2,\,j} := 1/\left\lvert\widehat{\delta}_j\right\rvert$ determined by the initial OLS estimates $\widehat{\rho}$ and $\widehat{\delta}_j$, and adjustment parameters $\gamma_1,\gamma_2\in\mathbb{R}^+$. This loss function is equivalent to the one given in \textcite{kock2016consistent}, except that we add a factor of 2 in front of the $\ell_1$ penalty terms for mathematical convenience.    

\textcite{kock2016consistent} shows that the AL solution \eqref{eq:ALURT_ALestimator} has the oracle property under suitable rate conditions for the $\ell_1$ penalty parameter $\lambda$. Consequently, AL can select the relevant variables in \eqref{eq:ALURT_adfreg} w.p.~1 as $T\to\infty$ and consistently estimates their coefficients. This includes the decision whether $y_{t-1}$ should be included in the model and, therefore, whether we have a stationary model with $\rho^\star\in(-2,0)$ or the unit root case $\rho^\star=0$, so that $y_{t-1}$ is excluded.

\section{Testing for non-stationarity using the Lasso path}
\label{sec:alurt_tfnutlsp}

One central issue with inference in Lasso-type estimation is that conventional tests rely on marginal distributions, assuming fixed hypotheses. This inference is invalid for sequential regression estimators where solution paths are obtained from the LARS algorithm (\cite{Efronetal2004}): they do not account for the uncertainty about the model selected at a specific knot, rendering null hypotheses (typically stating the irrelevance of coefficients) in sub-models characterising the solution path random. This issue is crucial in testing for non-stationarity of autoregressions based on a solution to \eqref{eq:ALURT_adaptive_lassooptim} where inference on the relevance of $y_{t-1}$ depends on the selected lags.

A recent branch of the literature on step-wise estimators considers post-selection methods that address uncertainty about the selection in sequential regression, e.g., \textcite{Leeetal2016,Tibshiranietal2016,Tibshiranietal2018}. Their core result enables inference \emph{conditional} on the stochastic variable selection using truncated Gaussian null distributions. Special cases of such tests are significance tests for a variable in the model selected at its LAR activation knot, i.e., just as the variable enters the set of active regressors.

We next show that a significance test for $\rho^\star$ in ADF regressions estimated by the adaptive Lasso can be derived immediately from the activation knot of $y_{t-1}$ on the Lasso regularisation path. The new testing principle exploits two features of the AL solution. 
First, the (appropriately scaled) activation knot of \textit{any irrelevant} regressor in \eqref{eq:ALURT_adfreg} asymptotically follows a distribution that is independent of the set of regressors selected up to this knot, under consistent estimation of the coefficients. This allows identifying a nuisance-parameter-free limiting null distribution for $\rho^\star=0$. Second, the activation knots of the relevant regressors in \eqref{eq:ALURT_adfreg} are of higher stochastic orders than those of the irrelevant regressors, which gives rise to the oracle property. This asymptotic sorting of activation knots allows the construction of consistent tests against the alternative $\rho^\star\in(-2,0)$.

\subsection{An adaptive Lasso activation knot unit root test}

Consider the LARS-computed sequence of penalty parameter values $\lambda_1$> $\lambda_2>\dots>\lambda_K$ determining the regularisation path of the adaptive Lasso estimator \eqref{eq:ALURT_ALestimator}. The knots $\lambda_l$ are obtained from the corresponding Karush-Kuhn-Tucker conditions for a minimum of \eqref{eq:ALURT_adaptive_lassooptim}. Since variables are activated one at a time, the solution path features at least $K=p+1$ knots. The Lasso modification of the LARS algorithm may entail that the solution path also features knots where variables are \emph{deactivated}. These variables are re-activated further down the solution path, resulting in $K\geq p+3$ in such cases.\footnote{If a coefficient undergoes a sign change between activation knots, LAR in Lasso mode drops the variable and recomputes the OLS direction. See Section 3.1 in \textcite{Efronetal2004} for details.} Hereafter, we will be interested in the stochastic properties of a set of \textit{activation knots} as specified by \Cref{def:ALURT_lambda0_definition}.
	\begin{restatable}[Activation knots]{defn}{ALURT_lambda0_definition}
    \label{def:ALURT_lambda0_definition}
    Let $x_i\in(y_{t-1}, \Delta y_{t-1}, \dots, \Delta y_{t-p})'$, $\dot\vbeta := (\dot\rho,\,\dot\vdelta')'$ and consider the coefficient functions $\dot{\beta}_i(\lambda) \in \mathbb{R}$, $i=1,\dots,p+1$ along the adaptive Lasso solution path to \eqref{eq:ALURT_adaptive_lassooptim}. Define $\lambda_{0,\,\beta_i^\star}$ as the earliest activation knot of $x_i$, i.e.,
	\begin{align*}
        \lambda_{0,\, \beta_i^\star} :=& \max \Lambda_{0,\, \beta_i^\star}, \\
	\Lambda_{0,\, \beta_i^\star} :=& \left\{\lambda : \ \ \partial_{+} \dot\beta_i(\lambda) = 0 \quad \land \quad \partial_{-} \dot\beta_i(\lambda) \neq 0 \quad \land \quad \dot\beta_i(\lambda) = 0\right\},
	\end{align*}
    where $\partial_{+}\dot\beta_i(\lambda)$ and $\partial_{-}\dot\beta_i(\lambda)$ denote derivatives of $\dot\beta_i(\lambda)$ with respect to $\lambda$ from above and from below, respectively. 
\end{restatable}

The virtue of the adaptive Lasso is that the adaptive penalty weights $w_1$, $w_{2,\,j}$, $j=1,\dots,p$ in \eqref{eq:ALURT_adaptive_lassooptim} lead to distinct stochastic orders of the activation knots of a relevant ($\lambda_{0,\,\beta_i^\star\neq0}$) and an irrelevant variable $x_i$ ($\lambda_{0,\,\beta_i^\star=0}$). These distinct orders give rise to tuning parameter sequences $\lambda_\mathcal{O}$ satisfying the oracle conditions
    \begin{equation}
        \frac{\lambda_\mathcal{O}}{\sqrt{T}}\rightarrow0, \quad \frac{\lambda_\mathcal{O}}{T^{1-\gamma_1}}\rightarrow\infty, \quad \frac{\lambda_\mathcal{O}}{T^{\frac{1-\gamma_2}{2}}}\rightarrow\infty,\label{eq:lambdaOracle}
    \end{equation}
    requiring that $\gamma_1>1/2$ and $\gamma_2>0$ \parencite{kock2016consistent}. In particular, $\lambda_\mathcal{O}$ ensures consistent model selection. Since our aim is to distinguish stationary from non-stationary time series, we focus on the knots of $y_{t-1}$. We consider its first activation knot $\lambda_{0,\,\rho^\star}$ as stated in \Cref{def:ALURT_lambda0_definition} since it bears the most information regarding the shrinkage on the coefficient of $y_{t-1}$. We have
		\begin{align}
			\label{eq:foc_lambda0_anatomy}
			\lambda_{0,\,\rho^\star} 
            :=&\, w_1^{-\gamma_1} \left\lvert \sum_t y_{t-1} \left(\Delta y_t - \sum_{j=1}^{p}  \widehat{\delta}_{\lambda,\,j} \Delta y_{t-j} \right)\right\rvert\\
            =&\, w_1^{-\gamma_1} \left\lvert \sum_t y_{t-1} \left(\rho^\star y_{t-1} + \sum_{j=1}^{p} \left(\delta_j^\star - \widehat{\delta}_{\lambda,\,j}\right) \Delta y_{t-j} + \varepsilon_{p,\, t} \right)\right\rvert,\label{eq:foc_lambda0_anatomy2}
		\end{align}
		where, by definition, $\widehat\rho_{\lambda} = 0$ for $\lambda = \lambda_{0,\,\rho^\star}$. Based on $\lambda_{0,\,\rho^\star}$, we propose a new route to classify stationary and non-stationary time series using the adaptive Lasso. The idea is to conduct inference on $\rho^\star$ using that the properly scaled activation knot $\lambda_{0,\,\rho^\star}$ is stochastically bounded under non-stationarity but diverges under stationarity as $T\to\infty$. In particular, $$\lambda_{0,\,\rho^\star = 0} = O_p\left(T^{1-\gamma_1}\right),\ \ \textup{and} \ \ \lambda_{0,\,\rho^\star\in(-2,0)}=\Theta\left(T\right),$$
        by Proposition 2 in \textcite{Reinschlussel2024}. Consider $H_0: \rho^\star = 0$ and $H_1: \rho^\star\in(-2,0)$ and define $\tau_{\gamma_1}:=T^{\gamma_1-1}\lambda_{0,\,\rho^\star}$. Then,
        \begin{align*}
          \tau_{\gamma_1}\vert H_0 = O_p(1), \ \ \textup{and} \ \  \tau_{\gamma_1}\vert H_1 = \Theta(T^{\gamma_1}),
        \end{align*}
        which allows for right-sided testing of $H_0$ against $H_1$ using $\tau_{\gamma_1}$.

We next examine the anatomy of the statistic $\tau_{\gamma_1}$ for non-stationary data and sketch its asymptotic null distribution $F_{\tau_{\gamma_1}}$. Under $H_0: \rho^\star = 0$, we use \cref{eq:foc_lambda0_anatomy2} to find
  \begin{align}
    \tau_{\gamma_1}\vert H_0 =&\, T^{\gamma_1 - 1}\bigg\lvert\underbrace{\vphantom{\sum_{j=1}^p} \widehat{\rho}}_{=:A_T}\bigg\rvert^{\gamma_1}\biggl\lvert\underbrace{\vphantom{\sum_{j=1}^p}\sum_t y_{t-1}\varepsilon_{p,\,t}}_{=:B_{T,\,\lambda}} + \underbrace{\sum_{j=1}^{p} \left(\delta_j^\star - \widehat{\delta}_{\lambda,\,j}\right)\sum_t y_{t-1} \Delta y_{t-j}}_{=:C_{T,\,\lambda}} \biggr\rvert\\ 
    =&\, \left\lvert T A_T\right\rvert^{\gamma_1} \left\lvert T^{-1} B_{T,\,\lambda} + T^{-1} C_{T,\,\lambda} \right\rvert. \label{eq:tauH0anatomy}
  \end{align}

Note that $T\cdot A_T = O_p(1)$ under \Cref{assum:lperrorsChangPark,assum:aroder} as it is a continuous mapping of the unscaled Dickey-Fuller coefficient statistic $T\widehat{\rho}$, cf. \textcite{ChangPark2002}. For identification of the limits of $T^{-1}B_{T,\,\lambda}$ and $T^{-1}C_{T,\,\lambda}$, we require consistent coefficient estimates $\widehat{\vbeta}_\lambda$ for $\lambda = \lambda_{0,\,\rho^\star=0}$: Consistency ensures $\varepsilon_{p,\,t}\xrightarrow[]{p}\varepsilon_t$ so that $T^{-1}B_{T,\,\lambda}=O_p(1)$ converges to a functional of a Brownian motion. Further, consistency of $\widehat{\vbeta}_\lambda$ implies the nuisance term $T^{-1}C_{T,\,\lambda}$ to vanish asymptotically as $\delta_j - \widehat{\delta}_{\lambda,\,j}\xrightarrow[]{p}0$. 

Because an oracle excludes any irrelevant variable, the activation knot of a unit root regressor $y_{t-1}$ (with coefficient $\rho^\star = 0$) must satisfy
    \begin{align}
      \liminf_{T\to\infty}  \lambda_{0,\,\rho^\star=0}=0 \ \ \textup{and} \ \ \limsup_{T\to\infty}  \lambda_{0,\,\rho^\star=0} \in \left[ 0,\lambda_{\mathcal{O}}\right).
    \end{align}
Therefore, $\widehat{\vbeta}_{\lambda}\xrightarrow[]{p}\vbeta^\star$ for all $\lambda = \lambda_{0,\,\rho^\star = 0}$ implies the existence of a distribution $F_{\lambda_{0,\,\rho^\star=0}}(\lambda)$ with all probability mass on the interval $[0, \lambda_\mathcal{O})$. Scaling $\lambda$ with $T^{\gamma_1 -1}$ delivers a stable and identifiable asymptotic null distribution $F_{\tau_{\gamma_1}}$,
    \begin{align}
        \left\lvert TA_T \right\rvert^{\gamma_1} \left\lvert T^{-1}B_{T,\,\lambda} + o_p(1)\right\rvert \, \overset{d}{\approx} \, F_{\tau_{\gamma_1}}(T^{\gamma_1-1}\lambda), \quad \lambda\in[0,\lambda_\mathcal{O}). \label{eq:taufsd}
    \end{align}
  \Cref{eq:taufsd} states that $F_{\tau_{\gamma_1}}$ is characterised by terms that solely hinge on $w_1$, $\gamma_1$ and the true innovations $\varepsilon_t$, and thus is independent of the unknown model $\left\{\Delta y_{t-i}:\, i\in\mathcal{M}\right\}$. 

The gist of the new test is that selection of $y_{t-1}$ rather than selection of the entire model avoids the need to tune $\lambda$: we may compute the solution path and select $y_{t-1}$ if $\tau_{\gamma_1}\geq T^{\gamma_1 -1} \lambda_{\alpha} := F_{\tau_{\gamma_1}}^{-1}(1-\alpha)$, i.e., \emph{without} a data-driven choice of $\lambda$. In this regard, using $\tau_{\gamma_1}$ for inference on whether $\rho^\star = 0$  is tuning-free but comes at the price of conservative selection since $\tau_{\gamma_1}$ selects an irrelevant $y_{t-1}$ with fixed probability $\alpha$ as $T\to\infty$. This is, however, inherent to classification using a significance test and may be attractive if the researcher wants to control the type-I-error rate. Another perspective is that $\tau_{\gamma_1}$ provides inference on $\rho^\star$ that could be reported alongside a subsequently selected model.   

\begin{restatable}[]{rem}{ConsSelatLambda0}
    Since $\lambda_\alpha = o(\lambda_\mathcal{O})$ for any $\alpha\in(0,1]$, setting $\lambda=\lambda_{\alpha}$ in \eqref{eq:ALURT_ALestimator} yields tuning-free but conservative selection of \emph{the whole model} \eqref{eq:ALURT_adfreg}. $\widehat{\vbeta}_{\lambda = \lambda_{\alpha}}$ could hence be an alternative to tuning $\lambda$ for conservative selection, e.g., using the AIC. Both approaches ensure that $\widehat{\rho}_\lambda$ has asymptotic power against local alternatives $\rho^\star=O(T^{-1})$. This contrasts consistent tuning where the adaptive Lasso has no asymptotic power against local alternatives in $1/T$ neighbourhoods; see also \Cref{rem:localpower}. A comparison of the AIC-tuned $\widehat{\vbeta}_\lambda$ with $\widehat{\vbeta}_{\lambda = \lambda_{\alpha}}$ under local alternatives would thus be interesting but is outside the scope of this paper.
\end{restatable}
We summarise the (local) asymptotic properties of $\tau_{\gamma_1}$ for OLS weights in \Cref{thm:lambdanullrhodist}.
\begin{restatable}[]{thm}{lambdanullrhodist}
	\label{thm:lambdanullrhodist}
	Consider model \eqref{eq:ALURT_adfreg} with $d_t = 0$ under Assumptions \ref{assum:lperrorsChangPark} and \ref{assum:aroder}. Denote $\lambda_{0,\,\rho^\star}$ the activation threshold of $y_{t-1}$ as in \Cref{def:ALURT_lambda0_definition} for some $\gamma_1>1/2,\gamma_2>0$, computed using the adaptive weights $w_1 = \lvert\widehat{\rho}\rvert^{-1}$, $w_{2,\,j}=\left\lvert\widehat{\delta}_j\right\rvert^{-1}$. Let $\tau_{\gamma_1} := T^{\gamma_1-1}\lambda_{0,\, \rho^\star}$.
	\begin{enumerate}
		\item If $\varrho^\star=1+c/T$ for some $c\in(-\infty, 0]$, then 
        \begin{align}
            \tau_{\gamma_1} \xrightarrow{d} \left\lvert \phi(1)^{-1} \frac{W_c(1)^2-1}{2\int_0^1 W_c(r)^2\mathrm{d}r}\right\rvert^{\gamma_1} \biggl\lvert\frac{1}{2}\sigma^2\phi(1)\left(W_c(1)^2-1\right)\biggr\rvert,\label{eq:taugammaNURLimit}
        \end{align}
		where $W_c(r) := \int_0^r \mathrm{exp}[c(r-s)]\mathrm{d}W(s)$ is an Ornstein-Uhlenbeck process driven by the standard Wiener process $W(s)$ on $s\in[0,1]$.  
		\item If $\lvert\varrho^\star\rvert<1$ is fixed, then $\tau_{\gamma_1} = \Theta(T^{\gamma_1})$.
	\end{enumerate}
\end{restatable}
\begin{proof}
	See \Cref{sec:ALURT_proofs}.
\end{proof}

Part 1 of \Cref{thm:lambdanullrhodist} states the limiting distribution of the (properly scaled) activation knot of $y_{t-1}$ obtained from the LARS-computed solution path of the adaptive Lasso estimator considered in \textcite{kock2016consistent} in (near) unit root settings where $\varrho^\star=1+c/T$. Setting $c=0$ obtains the null distribution under a unit root process. For $-\infty<c<0$, \vref{eq:taugammaNURLimit} gives the limit of $\tau_{\gamma_1}$ for stationary local deviations from the null. Result \eqref{eq:taugammaNURLimit} shows the limiting distribution to depend on $\gamma_1$, which controls the impact of the adaptive weight $w_1$. The parameter $\gamma_1$ thus governs the influence of the information about $\rho^\star$, obtained from an auxiliary OLS regression. Part 2 of \Cref{thm:lambdanullrhodist} states that $\tau_{\gamma_1}\to\infty$ at rate $T^{\gamma_1}$ under a fixed stationary alternative $\lvert\varrho^\star\rvert<1$.
   	
\begin{restatable}[]{rem}{adaLassocomparison}
    An advantage over the tuned adaptive Lasso estimator $\widehat{\rho}_\lambda$ is that $\tau_{\gamma_1}$ allows for inference with respect to $\rho^\star = 0$ using a tractable asymptotic distribution, as per Theorem 1. In contrast, the distributional properties of the adaptive Lasso are highly intricate. For cross-section models, \textcite{PoetscherSchneider2009} show that finite-sample distributions may be highly non-standard and asymptotic (Gaussian) distributions provide poor approximations, even for large $T$. This issue is likely exacerbated by our potentially non-stationary time series setup. A thorough comparison of the tuned $\widehat{\rho}_\lambda$ to the $\tau_{\gamma_1}$ test is an interesting point which is, however, beyond the scope of this paper.
\end{restatable}

\begin{restatable}[]{rem}{localpower}\label{rem:localpower}
    In terms of correct selection probability, $\widehat{\rho}_\lambda$ is \enquote{blind} for local alternatives $\rho^\star = O\left(1/T\right)$ if $\lambda$ is tuned for consistent selection. This drawback of $\widehat{\rho}_\lambda$ is explicitly addressed by \textcite{kock2016consistent} for an AR(1) model with i.i.d.~errors. \Cref{thm:lambdanullrhodist} reveals that this result also applies to our generalised setting: for $c\in(-\infty, 0]$, $\lambda_{0,\,\rho^\star\sim c/T} = o_p(\lambda_\mathcal{O})$. Hence, under the local alternative $\widehat\rho_{\lambda=\lambda_\mathcal{O}}=0$ w.p.~1 as $T\to\infty$, i.e., $y_{t-1}$ is asymptotically never activated at $\lambda=\lambda_\mathcal{O}$. In contrast, $\tau_{\gamma_1}$ \emph{has} asymptotic power against local alternatives, with the asymptotic power function depending on the limiting random variable in part 1 of \Cref{thm:lambdanullrhodist}. We investigate power against local alternatives in the Monte Carlo section.  
\end{restatable}

\subsection{Comparison with conditional post-selection inference for LARS fits}

We next summarise the framework of \textcite{Leeetal2016,Tibshiranietal2016}, review the LAR spacing test, and contrast the method with our testing concept. Consider the representative model $$\vy = \vmu + \boldsymbol{\epsilon}, \quad\boldsymbol{\epsilon} \sim N(\vzero,\mSigma_T),$$ with fixed target $\vmu$ so that $\vy\sim N(\vmu,\mSigma_T)$ and $\mSigma_T$ is assumed to be known. Our interest is in making inferences about a linear functional $\boldsymbol{\eta}'\boldsymbol{\mu}$, where $\boldsymbol{\eta}\in\mathbb{R}^T$ may depend on a selection event. \textcite{Tibshiranietal2016,Leeetal2016} show that a selection event on a LARS-computed solution path can be represented by a set of inequalities forming a polyhedral set $\left\{\vy: \mD\vy\leq\vb \right\}$ defined by a matrix $\mD$ and a vector $\vb$, where $\leq$ is understood element-wise. Both papers paraphrase the same core result
\begin{restatable}[Polyhedral selection as truncation]{lem}{plolyhedlemma}\label{lem:polyhedlemma}
For any $\mSigma_T$, $\veta$ with $\sigma^2_{\veta} := \veta'\mSigma_T\veta>0$,
\begin{align*}
    \mD\vy\leq\vb \quad \Longleftrightarrow \quad \mathcal{V}^-(\vy)\leq\boldsymbol{\eta}'\vy\leq\mathcal{V}^+(\vy), \, \mathcal{V}^0(\vy)\geq0,
\end{align*}
where
\begin{align*}
	\mathcal{V}^-(\vy) :=&\, \max_{j:\alpha_j<0}\frac{b_j-(\mD\vy)_j + \alpha_j\veta'\vy}{\alpha_j},\\
	\mathcal{V}^+(\vy) :=&\, \min_{j:\alpha_j>0}\frac{b_j-(\mD\vy)_j + \alpha_j\veta'\vy}{\alpha_j},\\
	\mathcal{V}^0(\vy) :=&\, \min_{j:\alpha_j = 0} \left(b_j - (\mD\vy)_j\right).
\end{align*}   
	and $\valpha := \mD\mSigma_T\veta/\veta'\mSigma_T\veta$. The triplet  ($\mathcal{V}^-(\vy)$, $\mathcal{V}^+(\vy)$, $\mathcal{V}^0(\vy))$ is independent of $\boldsymbol{\eta}'\vy$.
\end{restatable}
\begin{proof}
    See \textcite{Leeetal2016}
\end{proof}

\Cref{lem:polyhedlemma} states that the selection event $\left\{\mD\vy\leq\vb \right\}$ is equivalent to $\boldsymbol{\eta}'\vy$ falling in the range $\left\{\mathcal{V}^-(\vy)\leq\boldsymbol{\eta}'\vy\leq\mathcal{V}^+(\vy)\right\}$. Inference conditional on selection can hence be conducted with the CDF of a $N(\veta'\vmu, \sigma^2_{\veta})$ variable that is truncated to $[\mathcal{V}^-(\vy),\,\mathcal{V}^+(\vy)]$,
	\begin{align}
		F_{\veta'\vmu,\, \sigma^2_{\veta}  }^{[\mathcal{V}^-(\vy),\, \mathcal{V}^+(\vy)]} \ (x) := \frac{\Phi((x-\veta'\vmu)/\sigma^2_{\veta}  )-\Phi((\mathcal{V}^-(\vy)-\veta'\vmu)/\sigma^2_{\veta}  )}{\Phi((\mathcal{V}^+(\vy)-\veta'\vmu)/\sigma^2_{\veta}  )-\Phi((\mathcal{V}^-(\vy)-\veta'\vmu)/\sigma^2_{\veta}  )},\label{eq:TGgeneral}
	\end{align}
 	using the transform 
 	\begin{align}
 		 F_{\veta'\vmu,\, \sigma^2_{\veta}  }^{[\mathcal{V}^-(\vy),\, \mathcal{V}^+(\vy)]} \, (\veta'\vy)\,\big\vert \left\{\mD\vy\leq\vb \right\} \sim U(0,1),\label{eq:polyhedralcdf}
 	\end{align}
  with $U(0,1)$ the standard uniform distribution. \textcite{Tibshiranietal2016} discuss several tests based on \eqref{eq:polyhedralcdf}. A test statistic for
 \begin{align}
 	H_0: \veta'\vmu = 0 \quad \mathrm{vs.} \quad H_1: \veta'\vmu\neq0\label{eq:TGH0}
 \end{align}
is the marginal significance level
 \begin{align}
 	\mathcal{TG} = 2 \cdot \min \left\{F_{0,\, \sigma^2_{\veta}  }^{[\mathcal{V}^-,\, \mathcal{V}^+]} \ (\veta'\vy), \, 1-F_{0,\, \sigma^2_{\veta}  }^{[\mathcal{V}^-,\, \mathcal{V}^+]} \ (\veta'\vy) \right\}.\label{eq:polyhedraltsms}
 \end{align}
 Using $\mathcal{TG}$ as a p-value gives an exact level-$\alpha$-test under the above-mentioned conditions.

For $\vmu = \boldsymbol{X}\vbeta^\star$ result \eqref{eq:polyhedraltsms} is the p-value for a test of linear constraints on the coefficients $\vbeta$, conditional on a selection event on a LARS solution path corresponding to some fixed value for the regularisation parameter $\lambda$. Denote $\mX_{\mathcal{A}_\lambda}\in\mathbb{R}^{T\times\lvert\mathcal{A}_\lambda\rvert}$ the sub-matrix of $\mX\in\mathbb{R}^{p+1}$ containing the active variables at $\lambda$ with the index set $\mathcal{A}_\lambda\subset\{1,\dots,p+1\}$. An interesting specification is $\veta=(\mX_{\mathcal{A}_\lambda}^+)'\ve_j$, $j \in\mathcal{A}_\lambda$ with $\mX_{\mathcal{A}_\lambda}^+ := (\mX_{\mathcal{A}_\lambda}'\mX_{\mathcal{A}_\lambda})^{-1}\mX_{\mathcal{A}_\lambda}'$ and $\ve_j$ the suitable $j^{\mathrm{th}}$ standard basis vector. Using $\mathcal{TG}$, we may test
    \begin{align}
 	  H_0: \veta'\vmu = \ve_j' (\mX_{\mathcal{A}_\lambda}'\mX_{\mathcal{A}_\lambda})^{-1}\mX_{\mathcal{A}_\lambda}'\mX\vbeta^\star = 0, \quad\mathrm{vs.}\quad H_1: \veta'\vmu\neq0,
    \end{align}
i.e., the significance of the partial regression coefficient of variable $j$ in the projected linear model of $\vmu = \mX\vbeta^\star$ on $\mX_{\mathcal{A}_\lambda}$, conditional on the random selection $\mathcal{A}_\lambda$ at $\lambda$.
 
\textcite{Tibshiranietal2016} suggest a special test based on the polyhedral set representation of selection events for LAR that has similarities with our activation knot tests. Their theory considers the active set $\mathcal{A}_l$ at $\lambda = \lambda_l$, the $l^\mathrm{th}$ step of LAR. Denote $\vx_{j_l}$ the variable \emph{entering} $\mathcal{A}_l$ at step $l$. They suggest setting $\veta_l:=(\mX_{\mathcal{A}_l}^+)'\ve_j$ and to test $\veta_l'\vmu = 0$ against $\veta_l'\vmu > 0$, which is
\begin{align}
	H_0:\ \ve_j'\mX_{\mathcal{A}_l}^+\vmu = 0 \quad \mathrm{vs.} \quad H_1:\ \sgn\left(\ve_j'\mX_{\mathcal{A}_l}^+\vy\right) \cdot \ve_j'\mX_{\mathcal{A}_l}^+\vmu > 0.\label{eq:TGSH0}
\end{align}
Testing $H_0:\veta_l'\vmu = 0$ amounts to testing the variable entering the active set at the $l^\mathrm{th}$ LARS step to have zero partial regression coefficient $\ve_j'\mX_{\mathcal{A}_l}^+\vmu$ against the alternative that the coefficient is non-zero and has the same sign as the sample regression coefficient $\ve_j'\mX_{\mathcal{A}_l}^+\vy$. 

\textcite{Tibshiranietal2016} add the restriction $\mSigma_T = \sigma^2 \mI_T$ to devise the feasible statistic 
\begin{align}
	\mathcal{S} := \frac{ \Phi\left(\lambda_{l-1}\frac{\nu_l}{\sigma}\right) - \Phi\left(\lambda_l\frac{\nu_l}{\sigma}\right) }{ \Phi\left(\lambda_{l-1}\frac{\nu_l}{\sigma}\right) - \Phi\left(\lambda_{l+1}\frac{\nu_l}{\sigma}\right) }, \label{eq:larspacingtest}
\end{align}
which is based on an approximate representation of the selection event at step $l$, where
\begin{equation*}
    \nu_l := \left\lVert(\mX_{\mathcal{A}_l}^+)'\vs_{\mathcal{A}_l} - (\mX_{\mathcal{A}_{l-1}}^+)'\vs_{\mathcal{A}_{l-1}} \right\rVert_2,
\end{equation*}
and $\vs_{\mathcal{A}_l}$ is the sign vector of the active variables at step $l$, see Theorem 2 of \textcite{Tibshiranietal2016}. $\mathcal{S}$ decreases monotonically in $\lambda_l - \lambda_{l+1}>0$ and tests for an insignificant spacing between the LAR activation knots $\lambda_l$ and $\lambda_{l+1}$, and is hence called the \emph{spacing test} for LAR. We stress that $\mathcal{S}$ is not readily applicable to time series regressions due to the restriction of a spherical $\mSigma_T$.

\textcite{Tibshiranietal2018} relax the rigid assumption of Gaussian errors towards a broad class of error distributions for which they prove the $\mathcal{TG}$-type statistics to remain pivotal in low-dimensional ($p<T$) cases. They also propose a plug-in estimator for $\sigma^2$ and provide simulation evidence for the validity of their asymptotic results in broader settings like heteroskedastic errors.

The above exposition shows how the underlying principles of $\tau_{\gamma_1}$ and $\mathcal{S}$ differ in testing the relevance of $y_{t-1}$ in ADF regressions: $\mathcal{S}$ tests whether the partial regression coefficient of $y_{t-1}$ is zero in a regression of $\Delta y_{t}$ on $y_{t-1}$ and the set of the $\Delta y_{t-j}$, $j\in\mathcal{A}_l$ at $\lambda_l = \lambda_{0,\,\rho^\star}^\textup{LAR}$, the LAR activation knot of $y_{t-1}$. The inference is \emph{conditional} on a randomly selected lag structure and is approximate for $T<\infty$ in non-Gaussian settings. $\mathcal{S}$ rejects the null of irrelevance of $y_{t-1}$ in a model with regressors $\mX_{\mathcal{A}_l}$ for a significant difference $\lambda_l-\lambda_{l+1}$. Hence $\mathcal{S}$ places its level-$\alpha$ critical value somewhere between $\lambda_{l-1}$ and $\lambda_{l+1}$, rendering the critical region data-dependent.

In contrast, $\tau_{\gamma_1}$ compares the scaled LARS-computed adaptive Lasso activation knot $\lambda_{0,\,\rho^\star}$ with a pre-determined quantile $\lambda_{c_\alpha}$ of its \emph{marginal} asymptotic distribution in a non-stationary model. If the \emph{stochastic order} of $\lambda_{0,\,\rho^\star}$ is larger than expected under non-stationarity, the activation knot $\lambda_{0,\,\rho^\star}$ eventually exceeds the fixed critical value $\lambda_\alpha$, causing $\tau_{\gamma_1}$ to reject the null hypothesis of $y_{t-1}$ being an irrelevant regressor. The property $\lambda_{0,\,\rho^\star = 0} = o_p(\lambda_\mathcal{O})$ ensures that the selected model at $\lambda=\lambda_{0,\,\rho^\star=0}$ contains the true model w.p.~1 as $T\to\infty$ under $H_0$, akin to the \emph{screening property} in \textcite{Meinshausenetal2012}. Consistency implies the potential extra regressors do not impact the asymptotic distribution of $\lambda_{0,\,\rho^\star}$, as stated in \Cref{thm:lambdanullrhodist}. Therefore, finite-sample inference based on $\tau_{\gamma_1}$ is also approximate but is \emph{unconditional} on the stochastic variable selection at $\lambda=\lambda_{0,\,\rho^\star=0}$. 

Since LAR is a greedy algorithm, it is complicated to assess the usefulness of the spacing test analytically, especially in time series models with stochastic regressors. However, as the solution path of LAR is similar to that of the inconsistent plain (non-adaptive) Lasso, penalising non-zero parameters in ADF regressions even as $T\to\infty$ (see \textcite{kock2016consistent}, Theorem 7), we expect poor performance of LAR as well.\footnote{Simulation results in \textcite{Reinschlussel2024} confirm the erratic performance of the BIC-tuned plain Lasso in ADF regressions.} To investigate the performance of the spacing test for solution paths similar to those of a consistent adaptive Lasso estimator, we consider a variant of LAR based on regressors that are weighted as implied by \eqref{eq:ALURT_adaptive_lassooptim} in the Monte Carlo section. We refer to this estimator as \emph{adaptive LAR}, implemented like the adaptive Lasso but using the LARS algorithm in LAR mode.

\begin{restatable}[]{rem}{larspacing}
    \label{rem:aLARspacing}
    Notably, the adaptive LAR features the oracle property, placing both $\lambda_{0,\,\rho^\star=0}^\mathrm{LAR}$ and the spacing test's critical value $\lambda_\alpha^\mathcal{S}$ below $\lambda_{\mathcal{O}}$ as $T\to\infty$. By consistency of the coefficient estimators in this interval on the solution path, the true innovations $\varepsilon_t$ are asymptotically recovered and meet the fundamental condition $\boldsymbol{\epsilon} \sim (\vzero, \sigma^2 \mI_T)$. Relaxing the MDS condition on $\varepsilon_t$ in \Cref{assum:lperrorsChangPark} to independence ensures that all conditions for the spacing test are asymptotically satisfied at $\lambda_\alpha^\mathcal{S}$. Therefore, we conjecture that the spacing test applied to the \emph{adaptive} LAR is asymptotically valid in similar applications which violate the conditions on $\mSigma_T$ for the original LAR-based test. We revisit this point in the Monte Carlo section.
\end{restatable}

\subsection{An enhanced test using auxiliary information}

In \textcite{Reinschlussel2024}, we suggest a modified weight for $y_{t-1}$ in the adaptive Lasso estimator \eqref{eq:ALURT_adfreg} that aims for improved discrimination between stationary and non-stationary models under (consistent) tuning of $\lambda$. The idea is to enhance the adaptive weight $w_1$ with additional information on $\rho^\star$: the enriched weight $\Breve{w}_1 = \lvert\widehat{\rho}/J_\alpha\rvert^{-1}$ scales $w_1$ by a simulated statistic $J_\alpha$ which exploits different orders in probability in balanced and unbalanced regressions.\footnote{See \textcite{HerwartzSiedenburg2010} for details on $J_\alpha$. See \Cref{algo:ALURT_wbrewedet} in \Cref{sec:ALURT_dcpw} for details on the computation of $\Breve{w}_1$.} The merits of information enrichment result from the stochastic order of the corresponding $\lambda_{0}$, denoted $\Breve{\lambda}_{0,\,\rho^\star}$. The key is that both $\lambda_{0,\,\rho^\star}$ and $\Breve{\lambda}_{0,\,\rho^\star}$ are $O_p(T^{1-\gamma_1})$ under non-stationarity, but for fixed stationary alternatives $\rho^\star\in(-2,0)$, $$\lambda_{0,\,\rho^\star\in(-2,0)} = o\left(\Breve{\lambda}_{0,\,\rho^\star\in(-2,0)}\right),$$ implying beneficial properties for the resulting information-enriched adaptive Lasso estimator (ALIE). In particular, we demonstrate that ALIE has higher power under consistent tuning. Our simulation evidence also indicates that $\Breve{w}_1$ improves on the simple OLS-based weight regarding the activation rate of $y_{t-1}$ in small samples when $\rho^\star = 0$. 

It is possible to transfer these benefits to the $\tau_{\gamma_1}$ test if computed using $\Breve{w}_1$ instead of $w_1$. We summarise properties of the resulting modified test statistic, denoted $\Breve{\tau}_{\gamma_1}$, in the following corollary to \Cref{thm:lambdanullrhodist}.
 
 \begin{restatable}[]{cor}{brevelambdanullrhodist}
	\label{cor:brevelambdanullrhodist}
	Consider the assumptions of \Cref{thm:lambdanullrhodist}. Denote $\Breve{\lambda}_{0,\,\rho^\star}$ the activation threshold of $y_{t-1}$ as in \Cref{def:ALURT_lambda0_definition}, computed using the adaptive weights $\Breve{w}_1 = \lvert\widehat{\rho}/J_\alpha\rvert^{-1}$ and $w_{2,\,j}=\left\lvert\widehat{\delta}_j\right\rvert^{-1}$ for $\gamma_1 > 1/2$, $\gamma_2>0$. Let $\Breve{\tau}_{\gamma_1} := T^{\gamma_1-1}\Breve{\lambda}_{0,\, \rho^\star}$.
	\begin{enumerate}
		\item If $\varrho^\star=1+c/T$ for some $c\in(-\infty, 0]$, then $$\Breve{\tau}_{\gamma_1} = \frac{\tau_{\gamma_1}}{J_\alpha^{\gamma_1}} \xrightarrow{d} \frac{1}{J_{\alpha,\,c}^{\gamma_1}}\left\lvert \phi(1)^{-1} \frac{W_c(1)^2-1}{2\int_0^1 W_c(r)^2\mathrm{d}r}\right\rvert^{\gamma_1} \left\vert\frac{1}{2}\sigma^2\phi(1)\left(W_c(1)^2-1\right)\right\rvert,$$
		where $J_{\alpha,\,c}$ denotes the $c$-dependent weak limit of $J_\alpha$. All other quantities are defined in \Cref{thm:lambdanullrhodist}. 
		\item If $\lvert\varrho^\star\rvert<1$ is fixed, then $\Breve{\tau}_{\gamma_1} = \Theta(T^{2\gamma_1})$.
	\end{enumerate}
\end{restatable}
\begin{proof}
	See \Cref{sec:ALURT_proofs}.
\end{proof}
\Cref{cor:brevelambdanullrhodist} shows how the information-enriched adaptive weight suggested in \textcite{Reinschlussel2024} alters the asymptotic distribution of the scaled activation knot of $y_{t-1}$. Part 1 of \Cref{cor:brevelambdanullrhodist} states that $\Breve{\tau}_{\gamma_1} = \tau_{\gamma_1}/J_\alpha^{\gamma_1}$ and so the modified weight $\Breve{w}_1$ results in a scaling of the limiting random variable in part 1 of \Cref{thm:lambdanullrhodist} under non-stationarity or local alternatives. From Part 2 of \Cref{cor:brevelambdanullrhodist}, we deduce that $\Breve{\tau}_{\gamma_1}$ diverges with the squared rate of $\tau_{\gamma_1}$ under a fixed stationary alternative, suggesting a power advantage through information enrichment for all admissible choices of $\gamma_1$.

\subsection{Implementation}

\begin{figure}[H]
\centering
\caption{Simulated null distributions of the adaptive Lasso activation knot tests}
\label{fig:Tnulldist}
\vspace{.25cm}
\includegraphics[width=.9\textwidth]{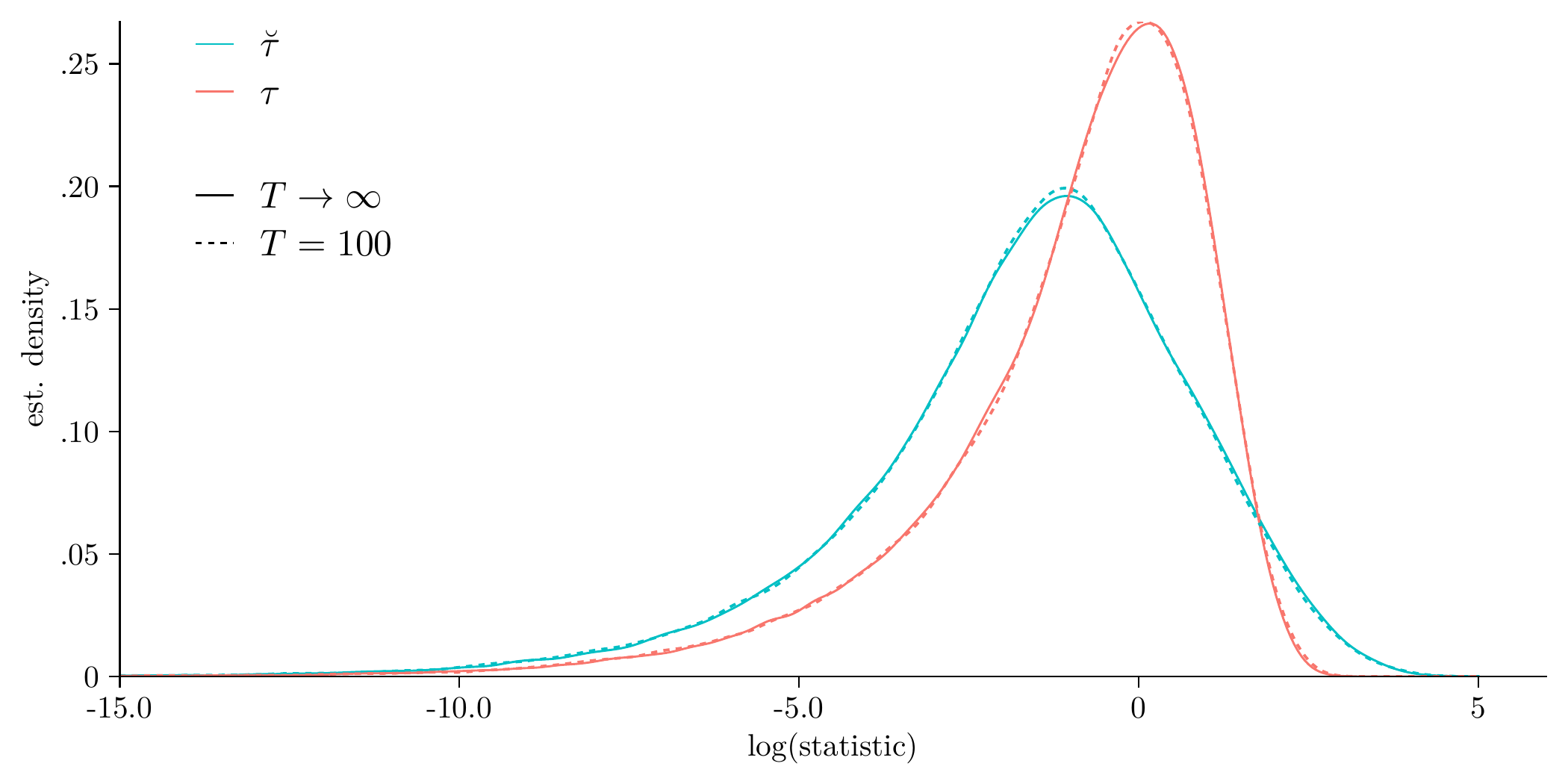}	
\vspace{.25cm}
\begin{minipage}{\textwidth}
	\scriptsize\textit{Notes:} Gaussian KDEs on the natural log scale using the \textcite{Silverman1986} rule. Dashed lines show finite-sample distributions for DGP \eqref{eq:ALURT_thedgp} with $\boldsymbol{\theta}=0$, $\varrho = 1$, and $u_t\sim i.i.d.\,N(0,1)$. $\lambda_{0,\,\rho^\star}$ and $\Breve{\lambda}_{0,\,\rho^\star}$ are taken from the \texttt{lars} computed solution path for the adaptive Lasso with $\gamma_1=\gamma_2=1$ in model \eqref{eq:ALURT_adfreg}. Asymptotic distributions (solid lines) are computed using discrete approximations of the limiting r.v.s in \eqref{eq:gamma1limit} for $10^4$ steps and with $\sigma^2=1$. $10^5$ replications.
\end{minipage}
\end{figure}

Computing the Lasso solution path for \eqref{eq:ALURT_adaptive_lassooptim} is straightforward, e.g., using the \texttt{lars} package \parencite{pkg-lars} for the statistical programming language R \parencite{R}. The tests generally require choosing $\gamma_1$ and accommodation for the LRV component $\sigma^2\phi(1)$ to obtain nuisance-parameter free limit distributions. A common choice in applications of the adaptive Lasso is $\gamma_1 = \gamma_2 = 1$. This is convenient, since then 
	\begin{align}
		\tau_{\gamma_1=1} \xrightarrow{d} \sigma^2 \frac{\left(W_c(1)^2-1\right)^2}{4\int_0^1 W_c(r)^2\mathrm{d}r},\qquad \Breve{\tau}_{\gamma_1=1} \xrightarrow{d} \frac{\sigma^2}{J_{\alpha,\,c}}\frac{\left(W_c(1)^2-1\right)^2}{4\int_0^1 W_c(r)^2\mathrm{d}r},\label{eq:gamma1limit}
	\end{align}
	by part 1 of \Cref{thm:lambdanullrhodist} and \Cref{cor:brevelambdanullrhodist}, respectively. Notably, the limiting random variables in \eqref{eq:gamma1limit} do not depend on $\phi(1)$ and thus, the asymptotic impact of the error process $u_t$ reduces to scaling by the innovation variance $\sigma^2$. Hence, we only need to adjust for $\sigma^2$ to obtain nuisance-parameter-free limit distributions. For consistent estimation of $\sigma^2$ we suggest using
	\begin{align}
		\widehat\sigma^2 := \frac{1}{T-p-1} \sum_t \widehat{\varepsilon}_{p,\,t}^{\,2}, \label{eq:s2}
	\end{align}
	where the $\widehat{\varepsilon}_{p,\,t}$ are residuals from the auxiliary ADF($p$) regression for computing the penalty weights. We henceforth refer to these test statistics as $\tau := \tau_{\gamma_1 = 1} / \widehat\sigma^2$ and $\Breve{\tau} := \Breve{\tau}_{\gamma_1 = 1} / \widehat\sigma^2$.
	
	Kernel density estimates (KDEs) of the asymptotic null distributions of $\tau$ and $\Breve{\tau}$ for $c=0$ are shown (by the solid lines) in \Cref{fig:Tnulldist}. Given the high kurtosis of the distributions, we transform the simulated statistics to the natural log scale for ease of interpretation. We compare these with KDEs of the sample distributions for $T=100$ in a non-stationary ADF model with $p=0$. Both limit distributions seem well approximated by their corresponding finite-sample distribution for $p=0$. Simulated critical values of $\tau$ and $\Breve{\tau}$ are presented in \Cref{tab:cvs}.

\begin{table}[tbp]
    \centering
    \caption{Simulated critical values of adaptive Lasso activation knot tests}
    \label{tab:cvs}
    \vspace{.25cm}
    \setlength{\tabcolsep}{28pt}
    \renewcommand{\arraystretch}{1.2}
    \resizebox{\textwidth}{!}{
        \input{tabs/Tab_CVs.tex}
    }
    \begin{minipage}{\textwidth}
        \vspace{.25cm}
        \scriptsize\textit{Notes:} Critical values are computed based on the adaptive Lasso solution to \eqref{eq:ALURT_adfreg} with $d_t=0$ and $p=0$ for Gaussian random walks. $\Breve{\tau}$ is computed using $J_\alpha$ with $\alpha=.1$ and $\sigma^2$ estimated by $\widehat\sigma^2$. $5\cdot10^5$ replications.
    \end{minipage}
\end{table}

\subsection{Deterministic components}
\label{sec:FDD}

While we so far assumed that $d_t = 0$, it is interesting to consider empirically relevant scenarios where the DGP is allowed to include a constant term or a linear time trend. Following \textcite{Reinschlussel2024}, we consider first-difference detrending (FD), see \textcite{SchmidtPhillips1992}. While adjusting for a constant by FD does not affect the limiting random variable of $\tau$ in \eqref{eq:gamma1limit}, the trend-adjusted data are
	\begin{align*}
		y_t^{\textup{FD}} := y_t - y_1 - \frac{t}{T}(y_T - y_1)
	\end{align*}
	which results in $W_c(r)$ being replaced by $V(r) := W_c(r) - rW_c(1)$. The latter carries over to the limit of $\Breve{\tau}$. Note that the limit of $\Breve{\tau}$ is altered under demeaning as well since $J_\alpha$ is computed using OLS regressions adjusting for deterministic components, see \Cref{algo:ALURT_wbrewedet} in \Cref{sec:ALURT_dcpw}. We provide simulated finite-sample critical values for statistics computed on adjusted data in \Cref{tab:CVsFDD}. 
    We investigate performance under detrending in the Monte Carlo studies in \Cref{sec:ALURT_ctutnif}.

\section{Monte Carlo evidence}
\label{sec:alurt_mce}

We next present two simulation studies. In \Cref{sec:alurt_cwsbi}, we compare methods for inference on $\rho^\star = 0$ using the knots on LARS-computed solution paths. Besides the new tests $\tau$ and $\breve{\tau}$, we consider the spacing test for plain and adaptive LAR, and the adaptive Lasso estimator from \textcite{kock2016consistent} tuned with BIC for consistent model selection ($\mathrm{AL}_\mathrm{BIC}$), as well as its information-enriched version ($\mathrm{ALIE}_\mathrm{BIC}$) introduced in \textcite{Reinschlussel2024}.

We consider AR DGPs where $\lvert\mathcal{M}\rvert < \infty$ so that selecting all relevant variables is feasible for $T<\infty$. Such a setting is useful in several regards. First, our theoretical results in \Cref{sec:alurt_tfnutlsp} suggest that the finite-$T$ performance of $\tau$ and $\Breve{\tau}$ hinges on $\widehat{\mathcal{M}}\supseteq\mathcal{M}$, i.e., the selection of the all relevant variables, at $\lambda_{0,\,\rho^\star}$ and $\Breve{\lambda}_{0,\,\rho^\star}$ in conjunction with sufficient shrinkage of the coefficients of irrelevant variables. A similar reasoning applies to $\mathrm{AL}_\mathrm{BIC}$ and $\mathrm{ALIE}_\mathrm{BIC}$ at $\lambda_\mathrm{BIC}$. Second, finite-order AR models allow us to investigate our conjecture from \Cref{rem:aLARspacing} that the spacing tests deliver asymptotically valid inference when applied to \emph{adaptive} LAR in ADF models that can be consistently estimated, contrary to the (plain) LAR spacing test of \textcite{Tibshiranietal2016}. Another aspect of the study is to gauge the improvement information enrichment provides for the testing paradigm of the spacing test.

Given the apparent relation of the proposed activation knot tests to methods based on unpenalised OLS estimation of model \eqref{eq:ALURT_adfreg}, \Cref{sec:ALURT_ctutnif} investigates how $\tau$ and $\breve{\tau}$ rank among established unit root tests in terms of asymptotic efficiency. For this, we assess the methods' local power properties in a near-integration framework where $\varrho=c/T$, $c\in(-\infty, 0]$. As benchmarks, we consider the augmented Dickey-Fuller t-statistic $\textup{ADF}^{\textup{GLS}}$ of \textcite{Elliottetal1996}, the modified Phillips-Perron statistic $\textup{MZ}_t$ by \textcite{NgPerron2001} and the $J_\alpha$ statistic of \textcite{HerwartzSiedenburg2010} used in computing $\breve{\tau}$.

\subsection{Comparison with selection-based inference}
\label{sec:alurt_cwsbi}

\begin{figure}[t]
\centering
\caption{Empirical CDFs of p-values for activation knot tests}
\label{fig:pecdf}
\vspace{.25cm}
\includegraphics[width=.94\textwidth]{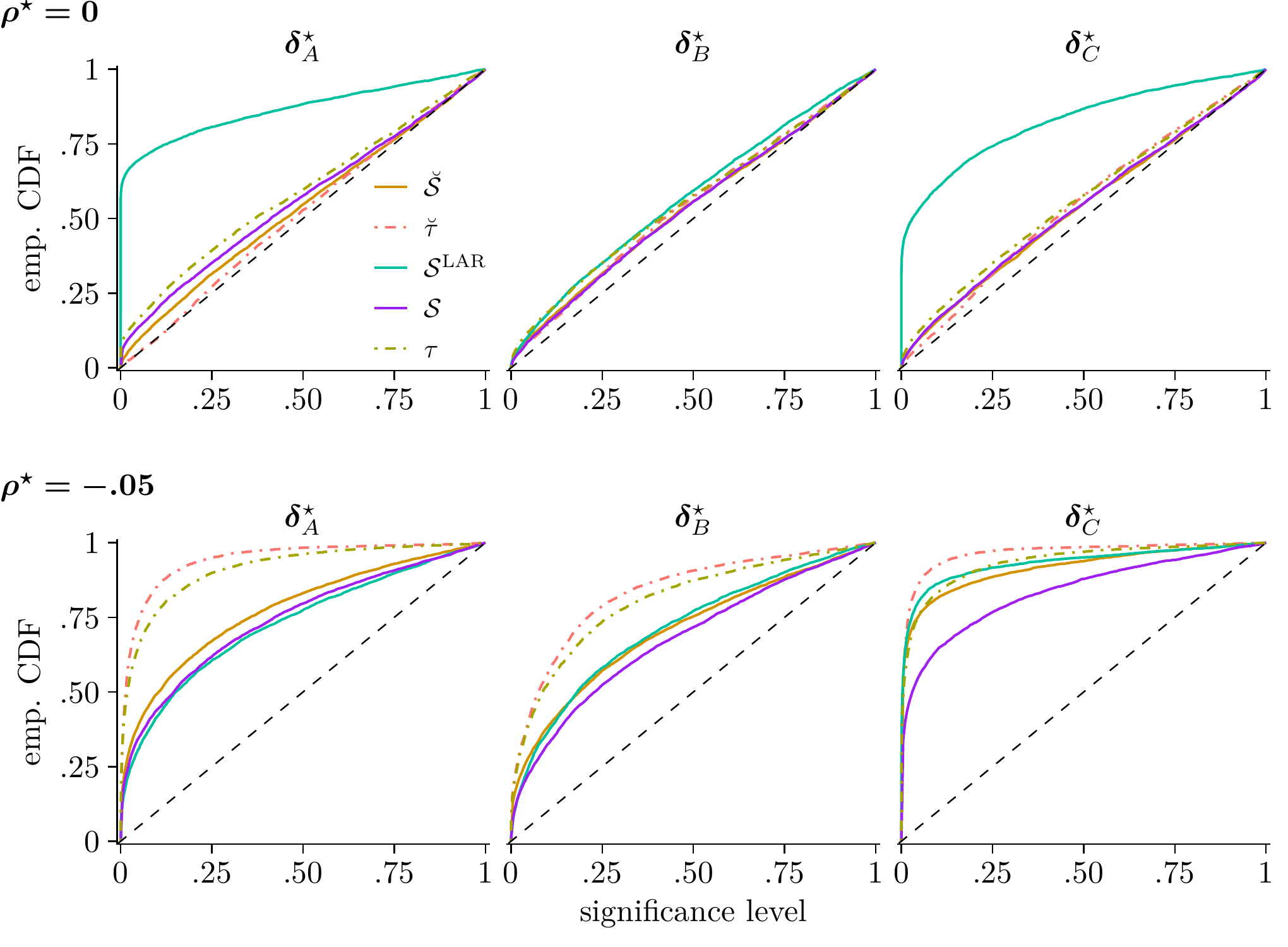}	
\vspace{.25cm}
\begin{minipage}{\textwidth}
	\scriptsize\textit{Notes:} DGP \eqref{eq:ALURT_adfdgp} with $T=75$. $\vdelta_A^\star = (.4, .3, .2, 0, 0, 0, -.2, 0, 0, .2)'$, $\vdelta_B^\star =(-.4, 0, 7)$, and $\vdelta_C^\star = .8$. Curves show empirical CDFs of observed p-values. The dashed lines reference an expected $U(0,1)$ CDF under $H_0$. Model \eqref{eq:ALURT_adfreg} with $d_t=0$ and $p = 10$. 5000 replications.
\end{minipage}
\end{figure}

We generate the data using the recursion
\begin{equation}
    y_t(1-B(L)) = v_t, \qquad v_t\overset{i.i.d.}{\sim}N(0,1), \qquad t=1,\dots,T+50, \label{eq:ALURT_adfdgp}
\end{equation}
with
\begin{align*} 
    B(L)=\,(\rho^\star + \delta_1^\star) L+ (\delta_2^\star - \delta_1^\star) L^2 + \dots + (\delta^\star_{k^\star-1}-\delta^\star_{k^\star}) L^{k^\star} - \delta^\star_{k^\star} L^{k^\star+1},
\end{align*}
and $k^\star+1$ zero initial values where $k^\star:=\dim(\vdelta^\star)$. The first 50 samples are discarded. DGP \eqref{eq:ALURT_adfdgp} has an ADF representation with coefficients $(\rho^\star, {\vdelta^\star}')'$.  We distinguish between the unit root case $\rho^\star=0$ and the fixed alternative $\rho^\star = -.05$. Three different configurations of $\vdelta^\star$ are considered: we assess the performance for a challenging sparsity pattern using $\vdelta_A^\star := (.4, .3, .2, 0, 0, 0, -.2, 0, .2)'$. $\vdelta_B^\star := (-.4, 0, .7)'$ is a low power setting for unit root tests, and $\vdelta_C^\star:= .8$ yields a short AR process with slowly decaying auto-correlation. These settings are used in \textcite{Reinschlussel2024} to compare BIC-tuned adaptive Lasso estimators based on $w_1$ or $\breve{w}_1$ in their ability to discriminate stationary and non-stationary data. 

The LAR spacing test $\mathcal{S}^\mathrm{LAR}$ as defined in \eqref{eq:larspacingtest}, termed the \emph{modified spacing test}, is computed with the \texttt{selectiveInference} R package \parencite{Tibshiranietal2019}. We further report the adaptive LAR variants suggested in \Cref{rem:aLARspacing} based on OLS-only weights ($\mathcal{S}$) and with the information-enriched weight $\Breve{w}_1$ for $y_{t-1}$ ($\Breve{\mathcal{S}}$). These are computed by the same algorithm as $\mathcal{S}^\mathrm{LAR}$ but follow from LAR estimates based on the scaled regressors
\begin{align*}
    \overset{\bm\sim}{\mX} := \left(w_1^{-\gamma_1}\vy_{t-1},\,  w_{2,\,1}^{-\gamma_2}\Delta\vy_{t-1},\, \dots,\,  w_{2,\,p}^{-\gamma_2}\Delta\vy_{t-p}\right). 
\end{align*}
To implement the spacing tests, we estimate $\sigma^2$ by \cref{eq:s2} using OLS residuals from the ADF regression for computing the penalty weights.

We first simulate data for $T=75$, a usual sample size in macroeconomic applications, and estimate model \eqref{eq:ALURT_adfreg} with fixed lag length $p=10$ and $d_t=0$. \Cref{fig:pecdf} compares the empirical CDFs of the p-values with the reference quantiles of a standard uniform distribution under $H_0$. 
All tests are somewhat anti-conservative under the null (top panel) in all scenarios. Outcomes indicate that the finite-sample distribution of $\mathcal{S}^\mathrm{LAR}$ is a poor approximation of the $U(0,1)$ distribution and may entail unacceptable upward size distortions. This observation is consistent with the theory behind the spacing test ruling out serial correlation in the dependent variable. An exception is $\vdelta_B^\star$ where $\mathcal{S}^\mathrm{LAR}$ performs comparably to the other tests. Differences among the latter are most pronounced for $\vdelta_A^\star$, where $\tau$ performs worst, followed by the spacing tests for adaptive LAR. The information-enriched statistic $\Breve{\tau}$ performs best here and has an almost exact size for typical significance levels.

Corroborating Remark 4, the results show that the adaptive LAR tests $\mathcal{S}$ and $\Breve{\mathcal{S}}$ perform much better than $\mathcal{S}^\mathrm{LAR}$ as their empirical distributions for $\vdelta_A^\star$ and $\vdelta_C^\star$ are more closely approximated by the theoretical null distributions than $\tau$. Furthermore, we find that $\Breve{w}_1$ mitigates the size distortions of $\Breve{\mathcal{S}}$ compared to $\mathcal{S}$, a direct effect of information enrichment.

All methods have power under the alternative (bottom panel), as indicated by empirical CDFs well above the dashed $45^{\circ}$ line. We again observe a positive effect of information enrichment for the adaptive LAR spacing tests, leading to a higher power of $\Breve{\mathcal{S}}$ than $\mathcal{S}$, which is most pronounced for $\vdelta_C^\star$. Analogous outcomes are seen for $\Breve{\tau}$ and $\tau$. In particular, the comparison with the null results suggests that methods based on the modified penalty weight are superior. Furthermore, the discrepancies in CDFs are strong evidence that the adaptive Lasso activation threshold tests, especially $\Breve{\tau}$, have higher power than the spacing tests. 

\begin{table}[htbp]
    \centering
    \caption{Empirical size and power of LARS solution path tests}
    \label{tab:spacingresults}
    \vspace{.25cm}
    \renewcommand{\arraystretch}{.88}
    \setlength{\tabcolsep}{14pt}
    \resizebox{\textwidth}{!}{
        \input{tabs/Tab_spacingVstau.tex}
    }
    \begin{minipage}{\textwidth}
        \vspace{.25cm}
    	\scriptsize\textit{Notes:} DGP \eqref{eq:ALURT_adfdgp}. $\vdelta_A^\star = (.4, .3, .2, 0, 0, 0, -.2, 0, 0, .2)'$, $\vdelta_B^\star = (-.4, 0, .7)'$, $\vdelta_C^\star = .8$. All methods are applied to model \eqref{eq:ALURT_adfreg} with $p=10$ fixed. $\mathcal{S}^\mathrm{LAR}$, $\mathcal{S}$, and $\Breve{\mathcal{S}}$ are LAR spacing tests for the irrelevance of $y_{t-1}$ without adaptively weighted regressors, with all OLS weights and modified weights. $\mathrm{AL}_\mathrm{BIC}$ and $\mathrm{ALIE}_\mathrm{BIC}$ are BIC-tuned adaptive Lasso estimators. $\tau$ and $\Breve{\tau}$ are activation knot tests for the solution path to \eqref{eq:ALURT_adaptive_lassooptim}. The nominal level is 5\%. Power is size-adjusted for all tests but $\mathcal{S}^\textup{LAR}$. Adaptive Lasso solution paths are computed with $\gamma_1=\gamma_2=1$. $5000$ replications.
    \end{minipage}
\end{table}

\Cref{tab:spacingresults} provides a detailed comparison of the tests at 5\% for sample sizes $T\in\{50,75,\allowbreak 100, \allowbreak 150,250, 500\}$. We additionally report outcomes for $\mathrm{AL}_\mathrm{BIC}$ and $\mathrm{ALIE}_\mathrm{BIC}$. Results for non-stationary data (top panel) confirm the evidence from \Cref{fig:pecdf}, indicating that the tests are affected by upward size distortions for small samples, which attenuate with increasing $T$. In this respect, $\Breve{\tau}$ controls size best among the hypothesis tests across all scenarios, although it is somewhat conservative for $\vdelta^\star_A$. Drastic improvements by scaling the regressors with adaptive weights are seen by comparing the spacing tests $\mathcal{S}$ and $\Breve{\mathcal{S}}$ with $\mathcal{S}^\textup{LAR}$: even for large $T$, the plain spacing test $\mathcal{S}^\mathrm{LAR}$ remains too anti-conservative to be useful as a unit root test. We note that small sample size distortions are less pronounced across the adaptive variants and diminish quickly as $T$ grows. This result not only confirms our theory on $\tau$, but further strengthens our conjecture in \Cref{rem:aLARspacing} that $\mathcal{S}$ and $\Breve{\mathcal{S}}$ could have a correct size below $\lambda_\mathcal{O}$ as $T\to\infty$. It seems likely that the spacing test can be tweaked for asymptotic validity in even more general settings than envisaged in \textcite{Tibshiranietal2016,Tibshiranietal2018}, provided all coefficients are estimated consistently---a necessary condition for $\tau$ and $\Breve{\tau}$ as well.

For stationary data (bottom panel), we report size-adjusted power at 5\% for $\tau$, $\Breve{\tau}$, $\mathcal{S}$, and $\Breve{\mathcal{S}}$, and present raw rejection rates for $\mathcal{S}^\mathrm{LAR}$, since size-adjusted rates are mostly trivial given its empirical size. Columns for $\mathrm{AL}_\mathrm{BIC}$ and $\mathrm{ALIE}_\mathrm{BIC}$ show raw activation rates of $y_{t-1}$, as it is not possible to readily size adjust the true-positive rate of these consistent classifiers.\footnote{We consider detecting a stationary time series a \emph{positive}.} The non-adaptive $\mathcal{S}^\mathrm{LAR}$ is again found to be inferior to the competitors, as its raw power may be significantly below the size-adjusted power of the other methods. On the other hand, $\mathcal{S}$ and $\Breve{\mathcal{S}}$ have less size-adjusted power than $\tau$ and $\Breve{\tau}$, despite the comparable size. As before, information enrichment proves to be beneficial in the sense that $\Breve{\mathcal{S}}$ and $\Breve{\tau}$ have higher power than their unmodified counterparts across all settings. Thus $\Breve{\tau}$ is preferable to $\tau$, especially in scenario $\vdelta_A^\star$, where $\tau$ has low power for small $T$. Although a direct comparison is not feasible, the true positive rate of $\mathrm{ALIE}_\mathrm{BIC}$ is roughly comparable to the power of $\Breve{\tau}$ given similar empirical size, see setting $\vdelta_B^\star$ for $T=100$. This observation indicates similar discriminatory power of decision rules based on consistent tuning or hypothesis testing. Further research here seems worthwhile but is beyond the scope of this paper.

\subsection{Comparison to unit root tests in a near-integration framework}
\label{sec:ALURT_ctutnif}

We next benchmark $\tau$ and $\Breve{\tau}$ against $\mathrm{MZ}_t$, $\mathrm{ADF}^\mathrm{GLS}$ and $J_\alpha$ in a near-integration setting where $\varrho=1+c/T$ with $c\in(-\infty, 0]$. The consistently tuned Lasso estimators are omitted as they do not have asymptotic power against alternatives in $1/T$ neighbourhoods. 

$\textup{ADF}^{\textup{GLS}}$ and the $\textup{MZ}_t$ are computed for GLS-adjusted data. Following the suggestions of \textcite{HerwartzSiedenburg2010}, $J_\alpha$ is calculated with $\alpha = .1$ and $R=150$, after adjusting for deterministic components using OLS. Both $J_\alpha$ and $\textup{MZ}_t$ require an estimate of the LRV $\omega^2$. Following \textcite{NgPerron2001} and \textcite{HerwartzSiedenburg2010}, we employ the autoregressive spectral density estimator at frequency zero,
	\begin{align}
    	\widehat{\omega}^2_{\textup{AR}}(k) := \frac{\widehat{\sigma}_k^2}{(1 - \sum_{j=1}^k \widehat{\delta}_j)^2}, \qquad \widehat{\sigma}^2_k := (T-k)^{-1} \sum_{t=k+1}^T (\widehat{e}_{k,\,t}^d)^2, \label{eq:ALURT_s2ar}    
	\end{align}
	as proposed by \textcite{PerronNg1998}. The $\widehat{e}_{k,\,t}^d$ are OLS residuals from the regression
	\begin{align}
		\Delta y_{t}^{d} = \xi_0 y_{t-1}^d + \sum_{j=1}^k \xi_j \Delta y_{t-j}^d + e_{k,\,t}^d,\label{eq:ALURT_adfregGLS} 	
	\end{align}
	 based on the GLS-adjusted data $y_{t}^{d}$ (if applicable). As suggested by \textcite{NgPerron2001}, we estimate $k$ as $\widehat{k}:= \argmin_{0\leq k\leq k_{\max}} \textup{MAIC}(k)$ with
	 \begin{align}
        \textup{MAIC}(k) := \log(\tilde{\sigma}^2_k) - 2 \frac{\tau_T(k)+k}{T-k_{\max}},\label{eq:maic}
    \end{align}
    the modified Akaike information criterion (AIC). The maximum lag order is $k_{\max} = \lfloor 12 (T/100)^{.25} \rfloor$ and $\tilde\sigma^2_k := (T-k_{\max})^{-1}\sum_{t=k+1}^T (\widehat{e}_{k,\,t}^d)^2$ is a deviance estimate. The MAIC differs from the AIC by the stochastic adjustment term $\tau_T(k) := (\tilde{\sigma}_k^2)^{-1}\linebreak[1]\widehat{\xi}_0^2\sum_{k_{\max}+1}^T(y_{t-1}^d)^2$ in the penalty function. $\textup{MZ}_t$ and $\textup{ADF}^{\textup{GLS}}$ are implemented based on the regression \eqref{eq:ALURT_adfregGLS} with $k$ selected by MAIC. For $\tau$ and $\breve{\tau}$, we compute the Lasso solution paths to \eqref{eq:ALURT_ALestimator} with $p = \widehat{k}$ and consider FD-detrended data, cf. \Cref{sec:FDD}.

\begin{figure}[t]
\centering
\caption{Local asymptotic power functions of Lasso-based and classical unit root tests}
\label{fig:penv_tau}
\vspace{.25cm}
\includegraphics[width=\textwidth]{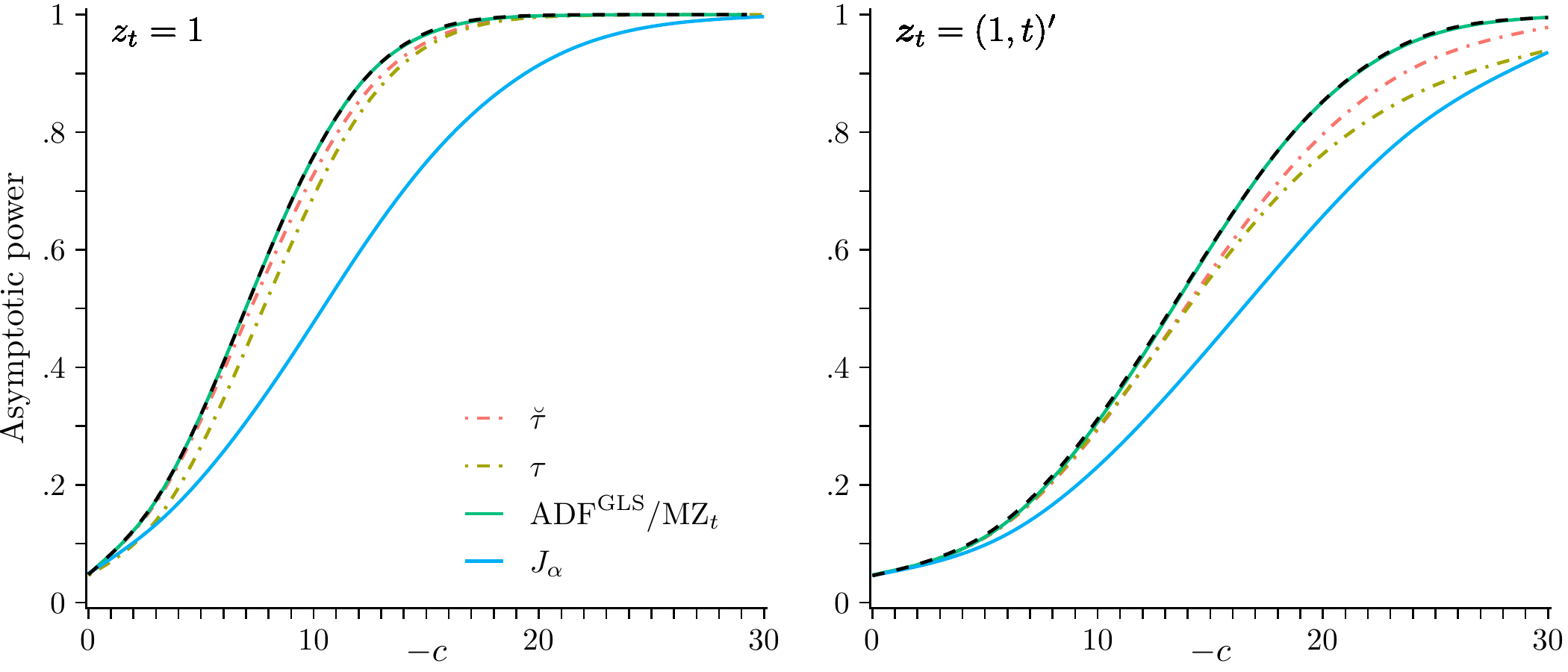}	
\vspace{.25cm}
\begin{minipage}{\textwidth}
    \scriptsize\textit{Notes:} 5\% nominal level. The dashed black lines show the Gaussian asymptotic power envelopes of unit root tests for DGP \eqref{eq:ALURT_thedgp} with a constant (left panel) and a linear time trend (right panel). Power functions of the tests are obtained by discrete approximations of $W_c(r)$ with $10^4$ steps and $10^5$ replications. $J_{\alpha}$ with $\alpha = .1$ and $R=150$ is simulated for a Gaussian AR(1) DGP and used in approximating the power function of $\Breve{\tau}$. 
\end{minipage}
\end{figure}

To analyse asymptotic efficiency, we compare the tests' asymptotic power in a Gaussian AR model with local-to-unity parameter $c\in[0,-1,-2,\dots,-30]$ and zero initial condition. The asymptotic local power functions in \Vref{fig:penv_tau} obtain from $10^5$ draws of discrete approximations of the corresponding limiting random variables with $10^4$ steps. The dashed lines show the Gaussian asymptotic power envelope for unit root tests adjusting for the specified deterministic component. For a constant ($z_t = 1$, left panel) and a linear trend ($\vz_t = (1, t)'$, right panel), the asymptotic local power functions of $\mathrm{ADF}^\mathrm{GLS}$ and $\mathrm{MZ}_t$ are virtually indistinguishable from the Gaussian power envelope by the construction of the tests. $\tau$ and $\Breve{\tau}$ track the power envelope, but show pronounced downward deviations for $\vz_t = (1, t)'$ at smaller $c$. The results indicate that information enrichment is effective: $\Breve{\tau}$ is often at least as powerful and outperforms $\tau$ for $z_t = 1$ and for processes with $c<-15$ when $\vz_t = (1, t)'$.

To investigate finite-sample powers in $1/T$ neighbourhoods under correlated errors, we generate time series as
\begin{equation}
    y_t = (1+c/T) y_{t-1} + v_t,\quad t=-49,\dots,-1,0,1,\dots,T, \label{eq:arerrorsdgp}
\end{equation} 
with starting value 0 and $c\in[0,-1,-2,\dots,-30]$ for sample sizes $T\in\{75,250\}$, discarding the first 50 samples. The errors $v_t$ are generated by the ARMA recursion
\begin{equation*}
	 v_t = \varphi v_{t-1} + \vartheta \epsilon_{t-1} + \epsilon_t, \quad\epsilon_t\overset{i.i.d.}{\sim}N(0,1).
\end{equation*}
To model different degrees of persistence in the errors, we set $\varphi,\vartheta\in\{-.8, -.4,\allowbreak 0, .4, .8\}$, considering only pure AR and MA processes.

\begin{figure}[H]
\centering
\caption{Size-adjusted local power -- AR errors, constant}
\label{fig:LPF_AR_c}
\includegraphics[width=.9\textwidth]{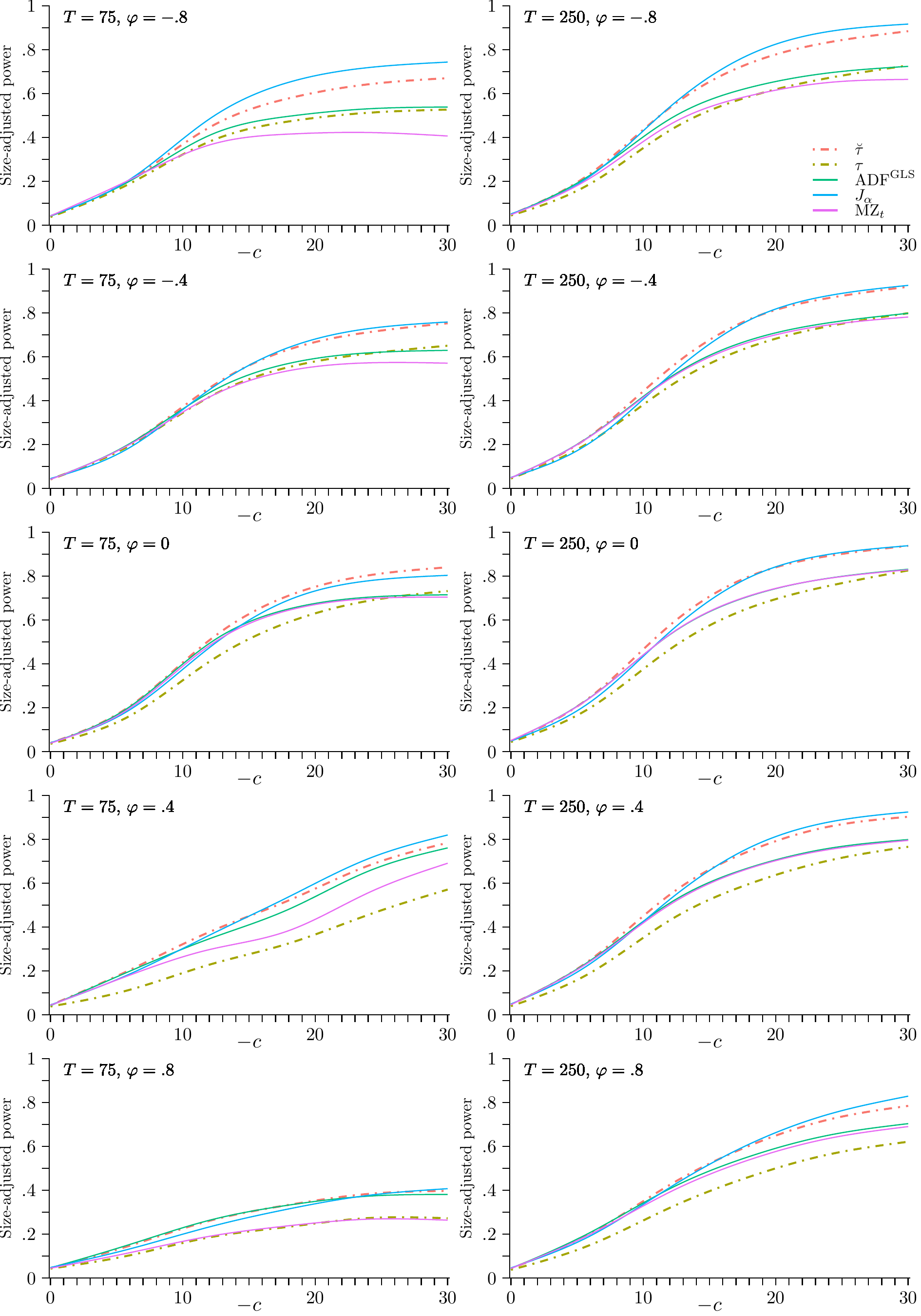}	
\begin{minipage}{\textwidth}
        \linespread{0.1}\selectfont
        \scriptsize\textit{Notes:} DGP \eqref{eq:arerrorsdgp} with $\vartheta=0$. Model \eqref{eq:ALURT_adfreg} with $d_t = 0$. $\tau$ and $\Breve{\tau}$ are activation knot tests based on FD-adjustment. $\textup{ADF}^\textup{GLS}$ and $\textup{MZ}_t$ are GLS-adjusted ADF and Phillips-Perron tests. $J_\alpha$ with $\alpha = .1$ is computed for OLS-adjusted data. Lag orders are selected by MAIC with $k_{\max}=\lfloor12\cdot(T/100)^{.25}\rfloor$. Lines show $5^{th}$-degree natural cubic spline estimates of the size-adjusted local power functions at 5\%. 5000 replications.
\end{minipage}
\end{figure}

\begin{figure}[H]
\centering
\caption{Size-adjusted local power -- AR errors, linear time trend}
\label{fig:LPF_AR_ct}
\vspace{.25cm}
\includegraphics[width=.9\textwidth]{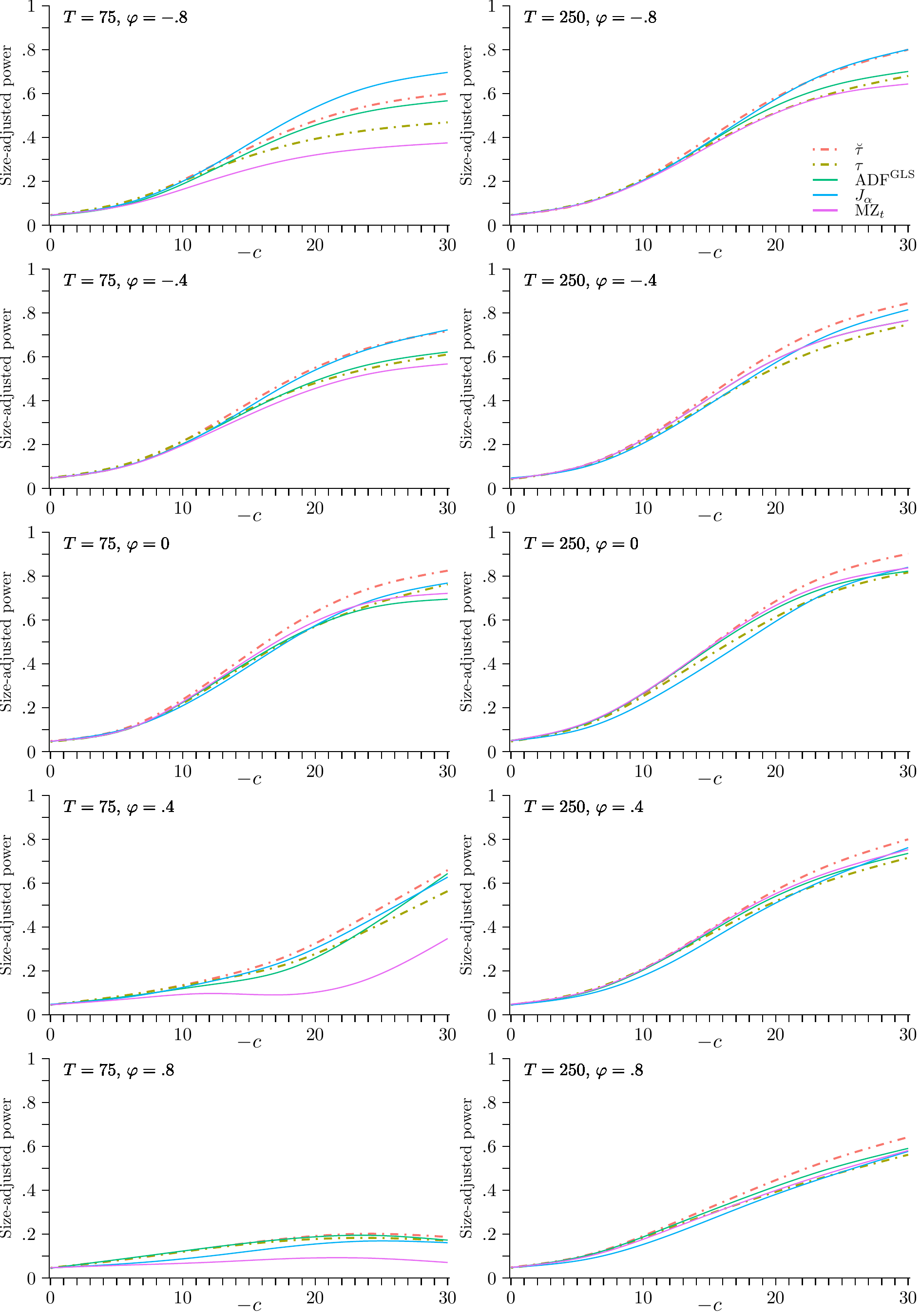}	
\vspace{.25cm}
\begin{minipage}{\textwidth}
    \scriptsize\textit{Notes:} See \Cref{fig:LPF_AR_c}.
\end{minipage}
\end{figure}

\begin{figure}[H]
\centering
\caption{Size-adjusted local power -- MA errors, constant}
\label{fig:LPF_MA_c}
\vspace{.25cm}
\includegraphics[width=.9\textwidth]{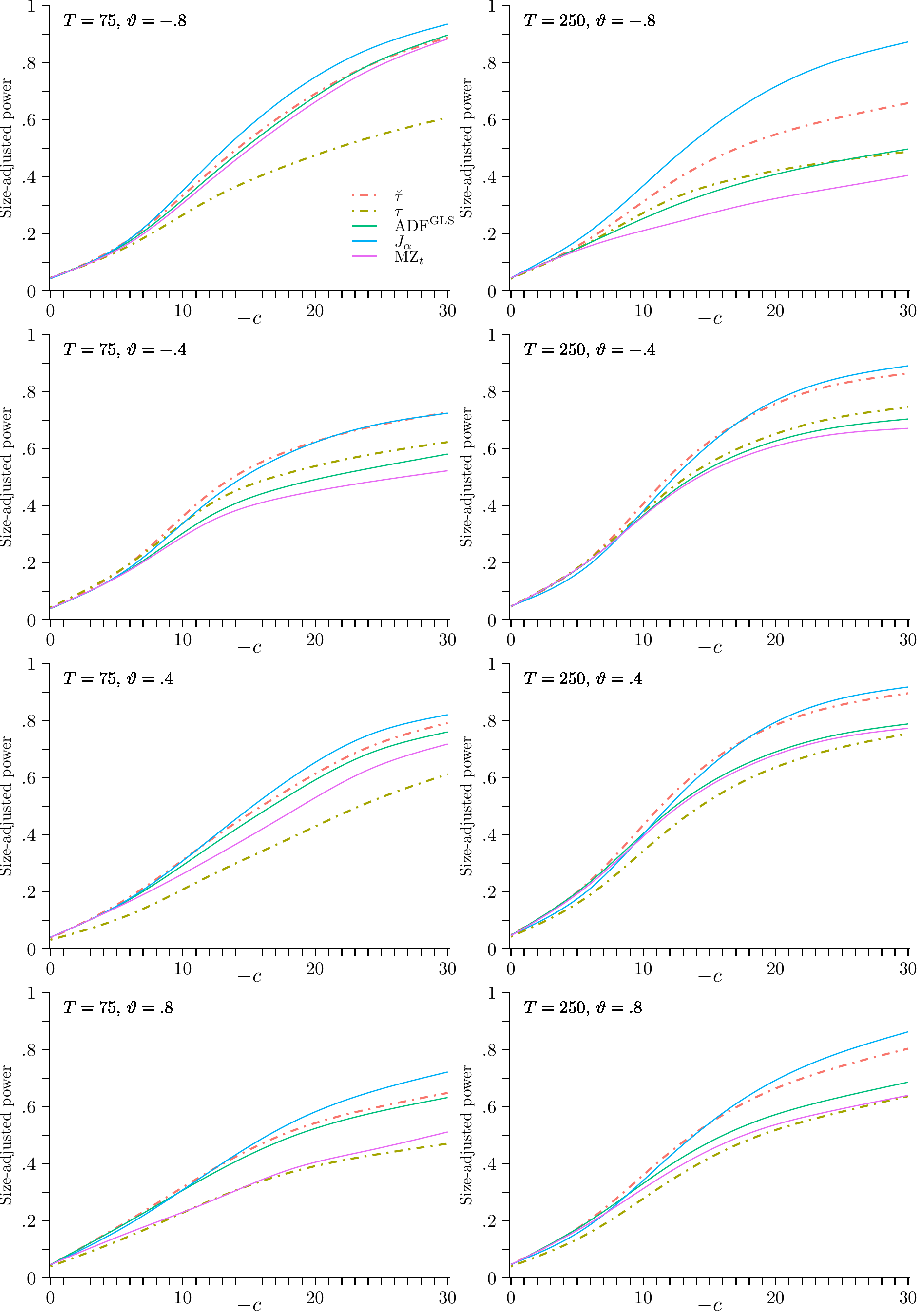}	
\vspace{.25cm}
\begin{minipage}{\textwidth}
    \scriptsize\textit{Notes:} DGP \eqref{eq:arerrorsdgp}, $v_t=\vartheta\epsilon_{t-1} + \epsilon_t$, $\epsilon_t\sim\,i.i.d.\,N(0,1)$. See \Cref{fig:LPF_AR_c} for details.
\end{minipage}
\end{figure}

\begin{figure}[H]
\centering
\caption{Size-adjusted local power -- MA errors, linear time trend}
\label{fig:LPF_MA_ct}
\vspace{.25cm}
\includegraphics[width=.9\textwidth]{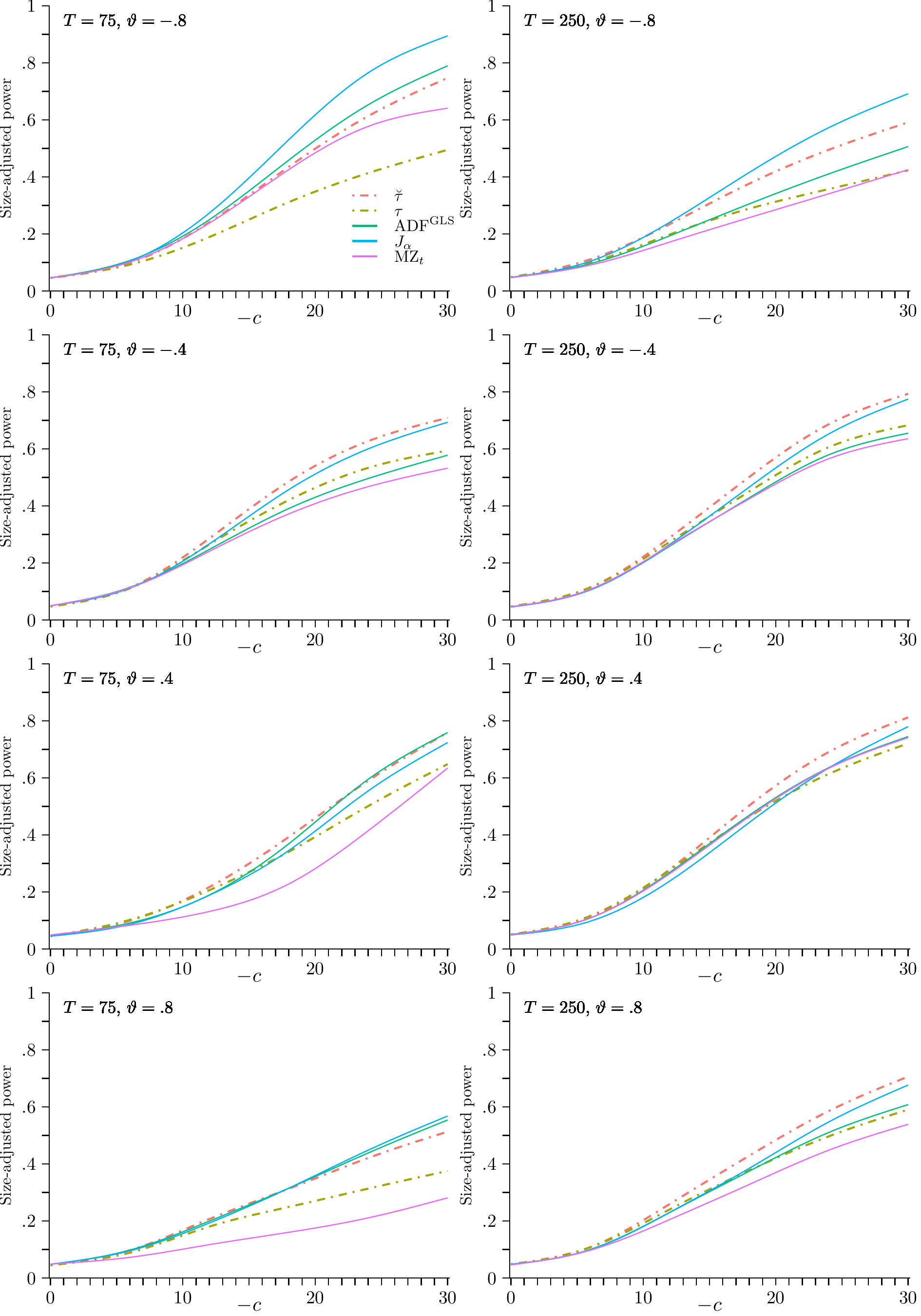}	
\vspace{.25cm}
\begin{minipage}{\textwidth}
    \scriptsize\textit{Notes:} See \Cref{fig:LPF_MA_c}.
\end{minipage}
\end{figure}

\Cref{fig:LPF_AR_c,fig:LPF_AR_ct} show size-adjusted powers under AR errors for demeaned and detrended data. The results mirror the insight from the asymptotic local power function in \Cref{fig:penv_tau} that all tests have similar power for small $c$, which is lower under detrending and for highly correlated errors ($\varphi = .8$). Depending on $\varphi$, we find substantial discrepancies for larger $c$. These are most pronounced at $\varphi = -.8$, where $J_\alpha$ performs best. Overall, $\Breve{\tau}$ tracks $J_\alpha$, dominating $\tau$ due to the enrichment via $\Breve{w}_1$. While there is no clear ranking across $c$ and $\varphi$, the size-adjusted power of $\Breve{\tau}$ is often significantly higher than for $\mathrm{ADF}^\mathrm{GLS}$ and $\mathrm{MZ}_t$.

We report similar results for MA errors in \Cref{fig:LPF_MA_c,fig:LPF_MA_ct}. Striking deviations between the tests arise for detrending under large negative MA coefficients at $T=75$, which is one of the few scenarios where $\mathrm{ADF}^\mathrm{GLS}$ surpasses $\Breve{\tau}$. It is, however, well-known that unit root tests suffer from upward size distortions in these scenarios (\cite[cf.][]{NgPerron2001}), making it difficult to recommend one test over the others. We examine this point below.

To investigate the dependence on the sample size, we further generate data using \eqref{eq:arerrorsdgp} for fixed $\rho^\star$ and $T\in\{50,75,100,150,250,500\}$ and again test at $5\%$. We set $c=0$ to generate non-stationary data. \Cref{tab:size_AR_c,tab:size_AR_ct} in \Cref{sec:ALURT_asr} present the empirical sizes of the tests for AR errors. The activation knot tests are close to the nominal level across parameter combinations and can compete with established unit root tests. Negative $\varphi$ are exemptions where $\tau$ and $\Breve{\tau}$ are slightly oversized. This behaviour is more pronounced under detrending but vanishes quickly with $T$.

We set $c=-7$ and $c=-13.5$ as local alternatives in the constant and in the trend-adjusted case, respectively. Size-adjusted powers for AR errors are presented in the \Cref{tab:sapower_AR_c,tab:sapower_AR_ct} in \Cref{sec:ALURT_asr}. The results confirm that $\Breve{\tau}$ has higher rejection rates than the $\tau$ test for all $\varphi$ considered and across $T$. This observation further corroborates our conjecture that power enhancements of the consistently tuned adaptive Lasso from the modified penalty weight $\Breve{w}_1$ reported in \textcite{Reinschlussel2024} carry over to $\Breve{\tau}$. Another indication of this feature is that the $\Breve{\tau}$ test often exceeds the power of $J_\alpha$, as indicated by the asymptotic results in \Cref{fig:penv_tau}. $\Breve{\tau}$ hence comes closer to the theoretical upper bound of 50\% local asymptotic power achievable by an optimal unit root test for the alternatives with $c=-7$ and $c=-13.5$. Power gains not vanishing for samples of size $T=1000$ substantiate this result. In this respect, $\Breve{\tau}$ benchmarks comparably to the $\textup{ADF}^\textup{GLS}$ test and sometimes even outperforms it.

Similar outcomes are obtained in case of MA errors for which we report empirical sizes in  \Cref{tab:size_MA_c,tab:size_MA_ct} and size-adjusted powers in \Cref{tab:sapower_MA_c,tab:sapower_MA_ct} in \Cref{sec:ALURT_asr}. Noteworthy here is a weak spot of the activation knots test for negative MA coefficients under the null. All tests suffer from significant upward size distortions in such scenarios. However, the type I error rates for the activation knot tests are higher than for the other tests and need larger $T$ to diminish. This effect is strongly pronounced for detrending, although $\Breve{\tau}$ does better than $\tau$. Research on the causes of this phenomenon is certainly of interest.

\section{Application to German groundwater levels}
\label{sec:alurt_atggl}

At present, over seventy per cent of Germany's water demand for irrigation and drinking water production is met by groundwater reserves, which often are the sole regional source of water supply, particularly in the northeastern federal states \parencite{Umweltbundesamt2018}. Climate change and anthropogenic influences are suspected to exert a negative impact on regional groundwater supplies in recent decades. Deep learning forecasts across possible climate pathways indicate further substantial declines until the end of the century for the whole country, see \textcite{Wuenschetal2022}. 

Reductions in groundwater levels may result in several economic ramifications, including increased expenditures for municipal drinking water supply, agricultural production drops and profitability declines due to elevated costs for irrigation, and the prospect of production disruptions within the industrial sector, to name a few \parencite{Wuenschetal2022}. Managing water resources will hence pose challenges for German policymakers. In particular, it is essential to enact countermeasures early to mitigate the risk of shortages caused by seasonal demand peaks during prolonged dry periods in the following decades, cf. \textcite{Scheihingetal2022}.

\subsection{Data and models}

For the empirical analysis, we examine six groundwater level time series recorded at official monitoring stations in the federal state of North Rhine-Westphalia (NRW). We consider stations for which particularly long samples are available, with low anthropogenic influence. These stations are located in the municipalities of Bokel (city of Rietberg), Hamminkeln, Niederbauer, Ratingen, Stellerdamm (city of Rahden) and Wachtendonk. Positions within Germany are shown in \Cref{fig:gwstationsmap}.

\begin{figure}[t]
\centering
\caption{Evolution of groundwater levels in NRW, Germany}
\label{fig:groundwaterlevels}
\vspace{.25cm}
\includegraphics[width=\textwidth]{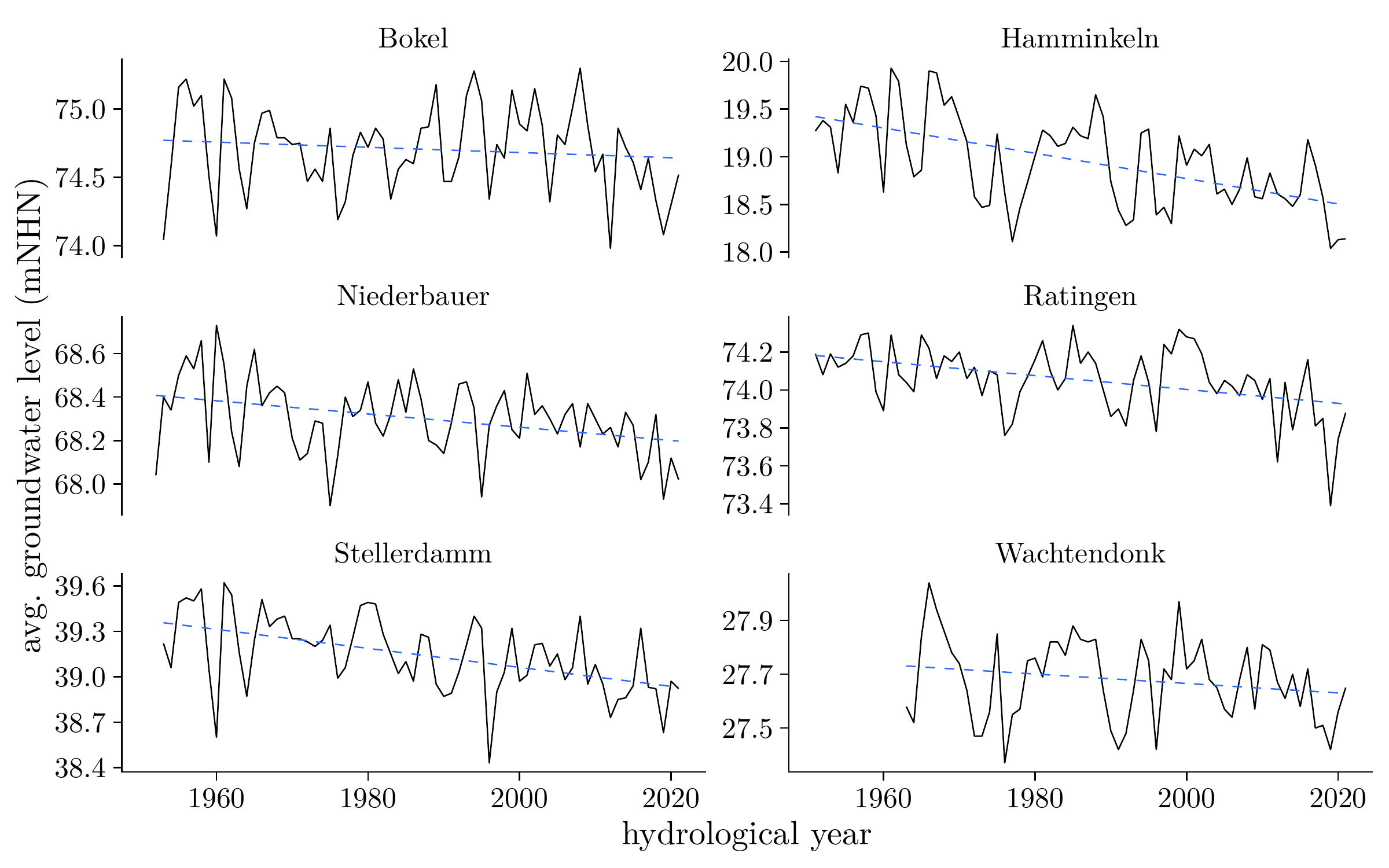}	
\vspace{.25cm}
\begin{minipage}{\textwidth}
	\scriptsize\textit{Notes:} A hydrological year is from November 1 to October 31. Levels are the height above sea level in meters (mNHN), measured as the difference in elevation between the measuring point in the German elevation reference system and the groundwater level. Dashed lines show linear trend estimates.
\end{minipage}
\end{figure}

We use the average groundwater level of the hydrological year, measured as the height above sea level in meters (mNHN), as the indicator. The data are freely available from the NRW State Office for Nature, Environmental and Consumer Protection \parencite{LANUV2023}. The evolution of groundwater levels and estimated linear trends are shown in \Cref{fig:groundwaterlevels}. This preliminary analysis indicates downward trends in all regions.

We apply the procedures to data adjusted for a constant or a constant and a linear time trend. Regarding the DGP \eqref{eq:ALURT_thedgp}, these appear reasonable specifications: In the case of a constant $\theta_1$, we test the null of groundwater levels following a purely stochastic trend against a stationary process with a non-zero mean. When adjusting for a linear trend component $\vz_t = (1, t)'$ with coefficients $\vtheta = (\theta_1, \theta_2)'$, we test the null of a random walk (possibly with drift) against stationary fluctuations around a deterministic time trend.
	
Given the low power of univariate procedures in small samples, a panel unit root test would be appropriate for a thorough analysis. However, the example serves to illustrate the differences between the methods in empirical applications. 

\subsection{Empirical results}

\Cref{tab:gwoutcomes} presents the results. In addition to the five tests considered, we also report the classification result obtained from the BIC-tuned adaptive Lasso estimators with all-OLS weights in column $\tau$ and weight $\Breve{w}_1 = (\widehat\rho/J_\alpha)^{-1}$ in column $\breve{\tau}$. Grey shading indicates classification as stationary. Evidence for a stationary process (top panel) is inconclusive. None of the methods rejects the null hypothesis at common significance levels for the sites Bokel, Ratingen, and Stellerdamm. Evidence is mixed for Hamminkeln and Niederbauer. For the former, only $\breve{\tau}$ rejects at 5\% among the hypothesis tests. For Niederbauer, the activation knot tests and $\textup{ADF}^\textup{GLS}$ are significant, whereby marginal significance levels (in parentheses) range from three to nine per cent. The tuned adaptive Lasso estimators classify water levels at Niederbauer and Hamminkeln as stationary. An exception is Wachtendonk, where all procedures reject the null of a random walk.

\begin{table}[t]
\centering
\caption{Inference on stochastic trends in German groundwater level time series}
\label{tab:gwoutcomes}
\vspace{.25cm}
\setlength{\tabcolsep}{14pt}
\renewcommand{\arraystretch}{.6}
\resizebox{\textwidth}{!}{
\begin{tabular}{llccccccc}
\toprule
& location & $T$ & $\tau$ & $\breve{\tau}$  &  $\textup{ADF}^{\textup{GLS}}$ & $\textup{MZ}_t$ & $J_\alpha$ &\\
\midrule
\addlinespace[2ex]

\multicolumn{8}{l}{$z_t = 1$ (\textit{constant})}\\[2ex]
& Bokel & \multirow{2}{*}{69} & 0.90  & .79    & $-0.94$  & $-0.38$  & 1.10 \\
& (1953--2021)	&& (.40) & (.33)  & (.48)  & (.69)  & (.31) \\[2ex]

& Hamminkeln & \multirow{2}{*}{71} & \cellcolor{gray!25} 2.23  & \cellcolor{gray!25} $3.17^*$ & $-1.44$ & $-1.26$ & .81\\
& (1951--2021) &&  (.16) & (.10)  & (.23) & (.27) & (.14)\\[2ex]

& Niederbauer & \multirow{2}{*}{80} & \cellcolor{gray!25} $5.00^{**}$ & \cellcolor{gray!25} $4.83^*$ & $-1.94^*$ & $-.65$ & 1.03\\  
& (1952--2021) && (.03) 	  & (.06) 	 & (.09) 	 & (.58)  & (.25)\\[2ex]

& Ratingen & \multirow{2}{*}{71}	& .01   & .01   & $-.81$ & $-1.01$ & $.89$\\
& (1951--2021) && (.93) & (.90) & (.54)  & (.40)   & (.18)\\[2ex]

& Stellerdamm & \multirow{2}{*}{69} & .08  & .04  & -.56  & -.40   & 1.68\\
& (1953--2021) && (.82) & (.83) & (.66) & (.68) & (.59)\\[2ex]

& Wachtendonk & \multirow{2}{*}{59} & \cellcolor{gray!25} $11.91^{***}$ & \cellcolor{gray!25} $45.48^{***}$ & $-3.56^{***}$ & $-3.07^{***}$ & $.26^{***}$\\
& (1963--2021) && (<1e-6) & (3e-4) & (<1e-6) & (1e-4) & (3e-4)\\[4ex]

\multicolumn{8}{l}{$\vz_t=(1,t)'$ (\textit{linear time trend})}\\[3ex]

& Bokel & \multirow{2}{*}{69}& 2.60   & 1.70   & $-0.86$ & $-0.49$ & 1.40\\
& (1953--2021) && (.39)  & (.53)  & (.95)   & (.98)   & (.70)\\[2ex]

& Hamminkeln & \multirow{2}{*}{71} & \cellcolor{gray!25} $16.95^{***}$ & \cellcolor{gray!25} $61.47^{***}$ & $-5.10^{***}$ & $-3.66^{***}$ & $.28^{***}$\\
& (1951--2021) && (6.9e-4) & (1.6e-4) & (9.4e-5) & (1.8e-4) & (1.6e-4)\\[2ex]

& Niederbauer & \multirow{2}{*}{80} & \cellcolor{gray!25} 2.40   & \cellcolor{gray!25} 6.53   & $-2.08$ & $-1.60$ & $.50^{**}$\\  
& (1952--2021) && (.44)  & (.13)  & (.37)   & (.59)   & (.02)\\[2ex]

& Ratingen & \multirow{2}{*}{71} & \cellcolor{gray!25} $31.69^{***}$ & \cellcolor{gray!25} $87.28^{***}$ & $-4.46^{***}$ & $-3.65^{***}$ & $0.36^{***}$\\
& (1951--2021) && (8e-6) & (3.6e-5) & (8.3e-4) & (1.9e-4) & (.0020)\\[2ex]

& Stellerdamm & \multirow{2}{*}{69} & \cellcolor{gray!25} $37.57^{***}$ & \cellcolor{gray!25} $208.72^{***}$ & $-6.64^{***}$ & $-3.97^{***}$ & $.18^{***}$ \\
& (1953--2021) && (<1e-6) & (<1e-6) & (<1e-6) & (2e-6) & (<1e-6)\\[2ex]

& Wachtendonk & \multirow{2}{*}{59} & \cellcolor{gray!25} $14.60^{***}$ & \cellcolor{gray!25} $40.95^{***}$ & $-3.85^{***}$ & $-3.25^{***}$ & $.36^{***}$\\
& (1963--2021) && (<1e-6) & (8.1e-4) & (.01) & (1.8e-4) & (8.1e-4)\\[1.5ex]
\bottomrule
\end{tabular}
}
\begin{minipage}{\textwidth}
        \vspace{.25cm}
        \scriptsize\textit{Notes:} Data are adjusted for the stated deterministic component. AL estimators in columns $\tau$ and $\breve{\tau}$ are computed using the FD approach in \textcite{SchmidtPhillips1992}. Grey shading indicates classification as stationary by the corresponding BIC-tuned AL estimator. Values in parenthesis are marginal significance levels obtained from simulation. For $\textup{MZ}_t$ and $J_\alpha$ the LRV is estimated by $\widehat{\omega}^2_{\textup{AR}}$. All lag orders are computed by MAIC with maximum lag order $p=\lfloor12\cdot(100/T)^{.25}\rfloor$.
    \end{minipage}
\end{table}

Evidence favouring trend-stationary processes (bottom panel) is substantial across measuring sites. All tests reject the null of a stochastic trend in groundwater levels at significance levels below 1\% for the sites Hamminkeln, Ratingen, Stellerdamm and Wachtendonk. For Niederbauer, $J_\alpha$ is the only hypothesis test with a significant outcome at 5\%. The tuned adaptive Lasso estimators classify measurements at Niederbauer as trend-stationary, whereas the corresponding activation knot tests do not. As for the case with a constant, none of the procedures indicate a (trend-)stationary process for Bokel. Overall, these results substantiate the conjecture that groundwater levels are stationary around a deterministic trend rather than following a stochastic trend. 

The empirical results mirror the Monte Carlo results, which show that the activation knot tests track the established unit root tests, often yielding similar results. We further find that information enrichment can be helpful in empirical applications: $\breve{\tau}$ rejects the null hypothesis of a random walk the most often among the hypothesis tests. In the case with a constant, $J_\alpha$ has the second lowest p-value for Hamminkeln, and its impact via the adaptive penalty weight leads to a rejection based on $\breve{\tau}$ at 10\% while $\tau$ has p-value 16\%. In the linear trend case for Niederbauer, a comparison with $\tau$ shows that $\breve{\tau}$ is pushed markedly towards the margin of rejection by the adaptive weight.

\section{Conclusion and outlook}
\label{sec:alurt_cao}
Conventional methods for inference based on the (adaptive) Lasso and its solution path require tuning, sophisticated post-selection procedures or conditioning on random hypotheses. Tuning-based methods, in particular, may suffer from uncontrolled classification error rates or a lack of power against local alternatives.

Focusing on the potentially non-stationary regressor $y_{t-1}$ in ADF regressions, we studied the distribution of its activation knot $\lambda_{0,\,\rho^\star}$ on the adaptive Lasso's solution path and used an appropriate scaling to devise a significance test for the relevance of $y_{t-1}$. The novel test $\tau_{\gamma_1}$ exploits that the oracle property of the adaptive Lasso implies different stochastic orders of $\lambda_{0,\,\rho^\star}$ for stationary and non-stationary data. Our proposed implementation $\tau$ follows naturally from the output of the algorithm \parencite{Efronetal2004} as implemented in the R package \texttt{lars} \parencite{pkg-lars} and is straightforward to compute.

The new test is compared with testing in the post-selection framework of \textcite{Leeetal2016,Tibshiranietal2016} where inference is conditional on $\widehat{\mathcal{M}}$ which is subject to the stochastic variable selection process. We contrast $\tau$ in detail with the spacing test of \textcite{Tibshiranietal2016}, which tests whether the partial regression coefficient of $y_{t-1}$ at its LAR activation knot is zero, conditional on the lagged dependent variables that are currently in the model. Instead, $\tau$ performs inference based on the asymptotic null distribution of $\lambda_{0,\,\rho^\star=0}$, which is tractable if the adaptive Lasso can consistently estimate the coefficients.

Noting that the $\tau$ test is significantly influenced by the adaptive penalty weight of $y_{t-1}$ in the adaptive Lasso loss function, we propose a modified version of the new test, using the enhanced weight $\Breve{w}_1$. This modified test $\Breve{\tau}$ is based on the information-enrichment principle considered in \textcite{Reinschlussel2024} for improving the classification performance of tuned adaptive Lasso estimators in ADF models. In addition to the OLS coefficient estimate of $y_{t-1}$, $\Breve{\tau}$ incorporates a second identification principle regarding the relevance of the regressor to compute the penalty weight. Analogously to the construction of the ALIE estimator proposed by \textcite{Reinschlussel2024}, we make use of the $J_\alpha$ statistic of \textcite{HerwartzSiedenburg2010} which exploits distinct orders in probability of the OLS estimator in time series regressions where the degree of integration differs.

Our Monte Carlo comparison of LARS-computed tests shows that $\tau$ and $\Breve {\tau}$ are significantly more reliable than the spacing test for LAR, which is often heavily oversized and has considerably lower power. We also found power shortfalls but mitigated size distortions for two suggested adaptive LAR spacing tests. Information enrichment with $\Breve{w}_1$ curbs size distortions and yields a higher size-adjusted power for the spacing test and $\Breve {\tau}$, which performs best. These results suggest that performance gains through information enrichment are not confined to tuning-based inference \parencite[cf.][]{Reinschlussel2024}, but extend well to hypothesis testing. While $\tau$ and $\Breve {\tau}$ have similar precision under the null, we showed that $\Breve {\tau}$ is superior to $\tau$ in a local power analysis. Both methods keep up with or outperform established unit root tests under local alternatives in $1/T$ neighbourhoods for ARMA DGPs. A weak spot of the new tests is large negative MA coefficients, which entail slowly vanishing upward size distortions that are pronounced under detrending---a feature that also affects established unit root tests.

Applying the tests to German groundwater levels, we found evidence of declines along deterministic trends over the last seven decades for five out of six monitoring sites. The outcomes mostly agree with results from other unit root tests and consistently tuned adaptive Lasso estimators. However, we find differences in outcomes that reflect some of the observed characteristics in the Monte Carlo simulations, pointing to the benefits of information enrichment.

There are several directions for further research. First, augmenting the ADF regression with other (exogenous) covariates similar to ARX regressions could improve power for stationary data with a low signal-to-noise ratio. Assuming the coefficients of additional covariates can be consistently estimated, this would be a straightforward extension, not incurring any nuisance parameters. An extension of the test to saturated models with $p \geq T$ seems also worth investigating. The central issue is that we require consistent coefficient estimates for all covariates, including those erroneously activated. A path forward may follow the post-double-selection procedure \parencite{Bellonietal2014}. Such an approach would enable practitioners to apply our test to high-dimensional regressions where the number of (potential) covariates grows faster than our assumptions allow.

Furthermore, using $\tau$ instead of cross-validation or the AIC to select $\lambda$ could aid the detection of small effects or raise power against local alternatives in post-selection inference. A significant advantage of using $\tau$ rather than tuning is reliable control of the classification error rate in conservative model selection. Comparing the performance of popular conservative tuning criteria with $\tau$ would be an interesting exercise easy to implement within the framework of stationary time series regressions since all $\lambda_{0,\,\delta_j^\star=0}$-distributions are identical. Using $\tau$ to tune cross-section regressions is just as appealing, but requires knowledge about all $\lambda_{0,\,\beta_j^\star=0}$-distributions, which may vary across the covariates.

Finally, it would be interesting to investigate whether our test principle transfers to cross-section regression, e.g., for causal analysis. Similar to the bootstrap for the spacing test presented in \textcite{Tibshiranietal2018}, advances in the bootstrap literature summarised in \textcite{Chernozhukov2023} could help recover the null distribution in such scenarios, possibly even in high-dimensional regressions. A multiplier (or wild) bootstrap could also be useful to improve the precision of the tests under detrending with MA errors and provide robustness to non-stationary volatility. We are currently investigating this idea.

%% file: tabs/Tab_CVs.tex
\begin{tabular}{lcccccc}
\toprule
\multicolumn{1}{c}{ } & \multicolumn{3}{c}{$\tau$} & \multicolumn{3}{c}{$\Breve{\tau}$} \\
\cmidrule(l{3pt}r{3pt}){2-4} \cmidrule(l{3pt}r{3pt}){5-7}
$T$ & 1\% & 5\% & 10\% & 1\% & 5\% & 10\%\\
\midrule
50 & 7.43 & 4.30 & 3.07 & 15.87 & 5.73 & 3.06\\
75 & 7.23 & 4.22 & 3.03 & 15.94 & 5.81 & 3.11\\
100 & 7.18 & 4.23 & 3.03 & 16.55 & 5.93 & 3.15\\
150 & 7.06 & 4.18 & 3.00 & 16.32 & 5.94 & 3.17\\
250 & 7.03 & 4.15 & 2.99 & 16.60 & 5.98 & 3.20\\
500 & 7.00 & 4.13 & 2.97 & 16.65 & 6.01 & 3.20\\
1000 & 6.97 & 4.13 & 2.97 & 16.78 & 6.04 & 3.22\\
\bottomrule
\end{tabular}

%% file: tabs/Tab_spacingVstau.tex
\begin{tabular}{llccccccc}
\toprule
\multicolumn{3}{c}{ } & \multicolumn{3}{c}{$w_1=1/\lvert\widehat\rho\rvert,\ w_{2,\,j} = 1/\lvert\widehat\delta_j\rvert$} & \multicolumn{3}{c}{$\Breve{w}_1=\lvert J_\alpha/\widehat\rho\rvert,\ w_{2,\,j} = 1/\lvert\widehat\delta_j\rvert$} \\
\cmidrule(l{3pt}r{3pt}){4-6} \cmidrule(l{3pt}r{3pt}){7-9}
 & $T$ & $\mathcal{S}^\mathrm{LAR}$ & $\tau$ & $\mathcal{S}$ & $\mathrm{AL}_\mathrm{BIC}$ & $\Breve{\tau}$ & $\Breve{\mathcal{S}}$ & $\mathrm{ALIE}_\mathrm{BIC}$\\
\midrule
\addlinespace[.6em]
\multicolumn{1}{l}{\textbf{$\boldsymbol{\rho^\star = 0}$}}\\
\addlinespace[0.6em]\addlinespace[0.3em]
\multicolumn{9}{l}{\textbf{$\vdelta_A^\star$}}\\
 & 50 & .710 & .256 & .227 & .224 & .086 & .157 & .109\\
 & 75 & .696 & .166 & .137 & .140 & .049 & .094 & .058\\
 & 100 & .692 & .129 & .112 & .103 & .037 & .079 & .043\\
 & 150 & .689 & .100 & .080 & .070 & .030 & .064 & .035\\
 & 250 & .678 & .075 & .066 & .051 & .031 & .059 & .035\\
 & 500 & .678 & .068 & .057 & .042 & .043 & .060 & .044\\
\addlinespace[0.3em]
\multicolumn{9}{l}{\textbf{$\vdelta_B^\star$}}\\
 & 50 & .154 & .178 & .137 & .135 & .118 & .149 & .127\\
 & 75 & .111 & .123 & .090 & .068 & .090 & .101 & .079\\
 & 100 & .105 & .101 & .079 & .043 & .072 & .080 & .055\\
 & 150 & .097 & .084 & .073 & .026 & .072 & .075 & .044\\
 & 250 & .097 & .064 & .053 & .011 & .054 & .065 & .021\\
 & 500 & .096 & .059 & .058 & .006 & .052 & .064 & .014\\
\addlinespace[0.3em]
\multicolumn{9}{l}{\textbf{$\vdelta_C^\star$}}\\
 & 50 & .590 & .174 & .138 & .100 & .088 & .142 & .074\\
 & 75 & .534 & .125 & .101 & .049 & .071 & .098 & .049\\
 & 100 & .520 & .094 & .081 & .032 & .059 & .088 & .034\\
 & 150 & .523 & .086 & .069 & .019 & .059 & .070 & .026\\
 & 250 & .502 & .066 & .061 & .010 & .050 & .067 & .017\\
 & 500 & .491 & .059 & .054 & .003 & .051 & .061 & .009\\
\addlinespace[.6em]
\multicolumn{1}{l}{\textbf{$\boldsymbol{\rho^\star = -.05}$}}\\
\addlinespace[0.6em]\addlinespace[0.3em]
\multicolumn{9}{l}{\textbf{$\vdelta_A^\star$}}\\
 & 50 & .310 & .002 & .004 & .394 & .342 & .113 & .495\\
 & 75 & .312 & .041 & .118 & .540 & .743 & .277 & .714\\
 & 100 & .367 & .309 & .268 & .672 & .911 & .405 & .861\\
 & 150 & .435 & .857 & .465 & .872 & .991 & .548 & .979\\
 & 250 & .541 & .997 & .716 & .988 & 1 & .619 & 1\\
 & 500 & .736 & 1 & .900 & 1 & 1 & .807 & 1\\
\addlinespace[0.3em]
\multicolumn{9}{l}{\textbf{$\vdelta_B^\star$}}\\
 & 50 & .216 & .110 & .096 & .255 & .151 & .117 & .311\\
 & 75 & .254 & .193 & .156 & .252 & .251 & .184 & .355\\
 & 100 & .294 & .327 & .197 & .281 & .419 & .276 & .432\\
 & 150 & .414 & .537 & .308 & .353 & .634 & .432 & .571\\
 & 250 & .557 & .916 & .652 & .644 & .955 & .816 & .855\\
 & 500 & .669 & .999 & .956 & .976 & 1 & .985 & .998\\
\addlinespace[0.3em]
\multicolumn{9}{l}{\textbf{$\vdelta_C^\star$}}\\
 & 50 & .652 & .228 & .255 & .448 & .544 & .376 & .604\\
 & 75 & .808 & .547 & .443 & .598 & .814 & .673 & .810\\
 & 100 & .907 & .800 & .643 & .756 & .950 & .865 & .925\\
 & 150 & .971 & .972 & .886 & .942 & .995 & .970 & .994\\
 & 250 & .995 & 1 & .983 & .999 & 1 & .962 & 1\\
 & 500 & 1 & 1 & 1 & 1 & 1 & .892 & 1\\
\bottomrule
\end{tabular}

%% file: C_appendix.tex
\subsection{Proofs}
\label{sec:ALURT_proofs}

\begin{proof}[\textbf{Proof of \Cref{thm:lambdanullrhodist}}]
\noindent
\begin{enumerate}[leftmargin=*]
    \item We divide the proof to result 1 of \Cref{thm:lambdanullrhodist} into three parts. In part (a), we show that the adaptive Lasso estimator $\widehat{\vbeta}_\lambda$ for sequences $\lambda$ in the interval $[0, \lambda_\mathcal{O}]$ is consistent for $\vbeta_0^\star$, the \emph{fixed} limit of $\vbeta^\star$ as $T\to\infty$ under \Cref{assum:lperrorsChangPark,assum:aroder}. The notion of $\vbeta_0^\star$ allows $\vbeta^\star$ to depend on $T$ to cover local-to-unity processes. Because $\lambda_{0,\,\rho^\star\sim\,c/T} \in [0, \lambda_\mathcal{O}]$, $\widehat{\vbeta}_\lambda$ must be consistent for any $\lambda = \lambda_{0,\,\rho^\star\sim\,c/T}$ with $c\in(-\infty, 0]$. Part (b) uses this property and analyses the asymptotic behaviour of $ \lambda_{0,\,\rho^\star\sim\,c/T}$, appealing to results from \textcite{Aylaretal2019} who generalise the asymptotic theory for OLS estimators of model~\eqref{eq:ALURT_adfreg} under lag truncation introduced by \textcite{ChangPark2002} to processes with local alternatives. The limits derived in part (b) allow us to identify the limit distribution of $\tau_{\gamma_1}$, which is summarised in part (c).
    
   \begin{enumerate}[leftmargin=*]
        \item We begin by proving the consistency of $\widehat{\vbeta}_\lambda$ in model \eqref{eq:ALURT_adfreg} under \Cref{assum:lperrorsChangPark,assum:aroder} if the penalty parameter $\lambda$ satisfies $\lambda = O(\lambda_\mathcal{O})$. Denote $\mX$ the $(T-p-1) \times (p+1)$ design matrix of model \eqref{eq:ALURT_adfreg} and consider the partition $\mX = (\mX_\mathcal{M}\,\vdots\, \mX_{\neg\mathcal{M}})$ with the sub-matrix $\mX_\mathcal{M}$ containing the relevant regressors $\vx_i\in\mathcal{M}$ and $\mX_{\neg\mathcal{M}}$ the irrelevant ones. The DGP coefficient vector $\vbeta^\star = (\vbeta^{\star'}_\mathcal{M}, \vbeta^{\star'}_{\neg\mathcal{M}})' = (\vbeta^{\star'}_\mathcal{M}, \vzero')'$ is partitioned analogously. We rewrite the adaptive Lasso loss function \eqref{eq:ALURT_adaptive_lassooptim} accordingly as
	\begin{align}\label{eq:ALURT_sampleloss}
		\begin{split}
                \Psi_T\left(\dot\vbeta_\mathcal{M},\,\dot\vbeta_{\neg\mathcal{M}}\big\vert\lambda\right) =&\, \left\lVert\Delta\vy - \mX_\mathcal{M}\dot\vbeta_\mathcal{M} - \mX_\mathcal{\neg M}\dot\vbeta_{\mathcal{\neg M}} \right\rVert_2^2\\ 
                +&\, 2\lambda \left\lVert\vw_\mathcal{M}\odot\dot\vbeta_\mathcal{M}\right\rVert_1  + 2\lambda \left\lVert\vw_\mathcal{\neg M}\odot\dot\vbeta_\mathcal{\neg M}\right\rVert_1 ,
		\end{split}
	\end{align}
	where $\vw_\mathcal{M}$ and $\vw_\mathcal{\neg M}$ denote the stacked penalty weights for some $\gamma_1>1/2,\gamma_2>0$. After rescaling with $1/T$, we obtain the limit loss function
	\begin{align}\label{eq:ALURT_asymptoticLoss}
		\begin{split}
        \Psi_0\left(\dot\vbeta_\mathcal{M},\,\dot\vbeta_{\neg\mathcal{M}}\big\vert\lambda\right) 
		:=&\, \lim_{T\to\infty} \frac{1}{T} \Psi_T\left(\dot\vbeta_\mathcal{M},\,\dot\vbeta_{\neg\mathcal{M}}\big\vert\lambda\right) \\
		=&\, \lim_{T\to\infty} \frac{1}{T}\left\lVert\Delta\vy - \mX_\mathcal{M}\dot\vbeta_\mathcal{M} - \mX_\mathcal{\neg M}\dot\vbeta_{\mathcal{\neg M}} \right\rVert_2^2 \\ 
		+&\, \lim_{T\to\infty} \frac{2\lambda}{T} \left\lVert\vw_\mathcal{M}\odot\dot\vbeta_\mathcal{M}\right\rVert_1 + \lim_{T\to\infty} \frac{2\lambda}{T} \left\lVert\vw_\mathcal{\neg M}\odot\dot\vbeta_\mathcal{\neg M}\right\rVert_1 .
		\end{split}
	\end{align}
Define $\mathcal{B}$ as the set of all solutions along the Lasso path for $\lambda\in[0,\infty)$,
\begin{equation*}
    \mathcal{B} := \bigcup_{\lambda\in[0,\infty)} \biggl\{ (\widehat\vbeta_{\lambda,\,\mathcal{M}}',\,\widehat\vbeta_{\lambda,\,\neg\mathcal{M}}')' := \argmin_{\dot\vbeta_\mathcal{M},\,\dot\vbeta_{\neg\mathcal{M}}} \Psi_0\left(\dot\vbeta_\mathcal{M},\,\dot\vbeta_{\neg\mathcal{M}}\big\vert\lambda\right) \biggr\} .
\end{equation*}
Because $\mathcal{B}$ exclusively contains minimisers which by definition improve on the path's starting point $\Psi_0(\vzero, \vzero\vert\lambda)$ for $\lambda \geq \max_i \lambda_{0,\,\beta_i \neq 0}$, the limit loss function $\Psi_0(\dot\vbeta_\mathcal{M},\,\dot\vbeta_{\neg\mathcal{M}}\vert\lambda)$ is bounded for every element in $\mathcal{B}$, we have
\begin{align*}
		(\dot\vbeta_\mathcal{M}',\,\dot\vbeta_{\neg\mathcal{M}}')' \in \mathcal{B}: \quad \Psi_0\left(\dot\vbeta_{\mathcal{M}},\,\dot\vbeta_{\neg\mathcal{M}}\big\vert\lambda\right) &\leq \Psi_0\left(\vzero,\, \vzero\,\big\vert\lambda\right) \\
		&= \E(\Delta y_t^2) \\ 
            &< \infty , 
\end{align*}
where existence of $\E(\Delta y_t^2)$ is ensured by weak stationarity. Therefore, the limit loss function exists for \emph{all} solutions on the Lasso path.\footnote{Note that this property also holds for the finite-sample loss function $\Psi_T$.}

Supposing that $\lambda=O(\lambda_\mathcal{O})$ and $\dot\vbeta_\mathcal{M}\in\mathcal{B}$ hereafter,
	\begin{equation}\label{eq:unpenalisedM_ALURT}
		\plim \frac{2 \lambda}{T} \left\lVert\vw_\mathcal{M}\odot\dot\vbeta_\mathcal{M}\right\rVert_1 = \vzero, 
	\end{equation}
	since all elements of $\vw_\mathcal{M}$ are $O_p(T^{-1/2})$, $\lambda_{\mathcal{O}} = o(T^{1/2})$ by construction and the dimension of each $\vw_\mathcal{M}$ and $\vbeta_\mathcal{M}$ is bounded by $p = o(T^{1/3})$. 

The limit of the second $\ell_1$-penalty term in \eqref{eq:ALURT_asymptoticLoss} has a lower bound attained at $\vbeta_{0,\,\mathcal{\neg M}}^\star = \vzero$,
\begin{equation}\label{eq:ALURT_irrelevantzerocoef}
    \lim_{T\to\infty} \frac{2\lambda}{T} \left\lVert\vw_\mathcal{\neg M}\odot\dot\vbeta_\mathcal{\neg M}\right\rVert_1 \, \begin{cases}
        = 0, & \textup{if}\ \dot{\vbeta}_\mathcal{\neg M} = \vzero \\
	\in [0,\infty), & \textup{if}\ \dot{\vbeta}_\mathcal{\neg M} \neq \vzero
    \end{cases},
\end{equation}
with the second case's value depending on the choices for $\gamma_1$, $\gamma_2$ and the growth rate of $\lambda$.
    
Recalling that collinearity is ruled out since \Cref{assum:aroder} ensures $\mX$ to have full column rank, results \eqref{eq:unpenalisedM_ALURT} and \eqref{eq:ALURT_irrelevantzerocoef} establish a lower limit for $\Psi_0\left(\cdot,\cdot\vert\lambda\right)$,
\begin{equation}
    \inf_{\dot\vbeta_\mathcal{M},\,\dot\vbeta_\mathcal{\neg M}}\Psi_0\left(\dot\vbeta_\mathcal{M},\,\dot\vbeta_{\neg\mathcal{M}}\big\vert\lambda\right) = \Psi_0\left(\vbeta^\star_{0,\,\mathcal{M}},\,\vzero\,\big\vert\lambda\right) = \sigma^2 .
\end{equation}	
Because $\vbeta^\star_0:=(\vbeta^{\star'}_{0,\,\mathcal{M}},\vzero')'$ uniquely minimises the quadratic component of the loss function \eqref{eq:ALURT_asymptoticLoss} and attains the zero infima of the terms in \eqref{eq:unpenalisedM_ALURT} and \eqref{eq:ALURT_irrelevantzerocoef}, it must be the unique minimiser of $\Psi_0\left(\cdot,\,\cdot\vert\lambda\right)$:
\begin{alignat*}{2}
\Psi_0\left(\dot\vbeta_\mathcal{M},\,\dot\vbeta_{\neg\mathcal{M}}\big\vert\lambda\right) - \Psi_0\left(\vbeta^\star_{0,\,\mathcal{M}},\vzero\,\big\vert\lambda\right) &> 0 \quad \forall && \left( \dot\vbeta'_\mathcal{M},\,\dot\vbeta'_{\neg\mathcal{M}} \right)' \neq \left(\vbeta^{\star'}_{0,\,\mathcal{M}},\,\vzero'\right)', \\
	\Psi_0\left(\dot\vbeta_\mathcal{M},\,\dot\vbeta_{\neg\mathcal{M}}\big\vert\lambda\right) - \Psi_0\left(\vbeta^\star_{0,\,\mathcal{M}},\vzero\,\big\vert\lambda\right) &= 0 \quad \textup{iff} && \left( \dot\vbeta'_\mathcal{M},\,\dot\vbeta'_{\neg\mathcal{M}} \right)' = \left(\vbeta^{\star'}_{0,\,\mathcal{M}},\,\vzero'\right)'.
\end{alignat*}
Therefore $\vbeta^\star_0$, the fixed limit of $\vbeta^\star$ as $T\to\infty$ is \emph{identified} in $\Psi_0\left(\dot\vbeta_\mathcal{M},\,\dot\vbeta_{\neg\mathcal{M}}\vert\lambda\right)$ for $\lambda = O(\lambda_\mathcal{O})$. As $T \to \infty$, the minimum of the sample loss function \eqref{eq:ALURT_sampleloss} (epi-)converges to the limit loss function's unique minimum,
\begin{align*}
	\frac{1}{T} \Psi_T\left(\vbeta^\star_{0,\,\mathcal{M}},\,\vzero\,\big\vert\lambda\right) &= \frac{1}{T}	\left\lVert\Delta\vy - \mX_\mathcal{M}\vbeta^\star_{0,\,\mathcal{M}} \right\rVert_2^2 + \frac{2\lambda}{T} \left\lVert\vw_\mathcal{M}\odot\vbeta^\star_{0,\,\mathcal{M}}\right\rVert_1 \\
	&\xrightarrow{p} \sigma^2 + o_p(1) \\
	&= \min_{\dot\vbeta_\mathcal{M},\, \dot\vbeta_\mathcal{\neg M}} \Psi_0\left(\dot\vbeta_\mathcal{M},\,\dot\vbeta_\mathcal{\neg M}\big\vert\lambda\right) \quad \forall\, \lambda = O(\lambda_\mathcal{O}).
\end{align*}
Hence the adaptive Lasso estimator $\widehat{\vbeta}_\lambda$ is consistent for $\vbeta^\star_0$ for $\lambda = O(\lambda_\mathcal{O})$:
	\begin{align}\label{eq:pthm1a}
		\begin{split}
		\argmin_{\dot\vbeta_\mathcal{M},\, \dot\vbeta_\mathcal{\neg M}} \Psi_T\left(\dot\vbeta_\mathcal{M},\,\dot\vbeta_{\neg\mathcal{M}}\big\vert\lambda\right)  &=: \left(\widehat\vbeta_{\lambda,\,\mathcal{M}}',\,\widehat\vbeta_{\lambda,\,\neg\mathcal{M}}'\right)' \\
		&\xrightarrow{p} \left( \vbeta_\mathcal{M}^{\star'},\, \vzero^{'}  \right)' \\
		&= \argmin_{\dot\vbeta_\mathcal{M},\, \dot\vbeta_\mathcal{\neg M}} \Psi_0\left(\dot\vbeta_\mathcal{M},\,\dot\vbeta_{\neg\mathcal{M}}\big\vert\lambda\right) . 		
		\end{split}
	\end{align}
	 
\item We next analyse the properties of $\lambda_{0,\,\rho^\star\sim\,c/T}$ as $T\to\infty$. Note that $\lim_{T\to\infty}\rho^\star = 0$ for $\rho^\star\sim\,c/T$. Therefore, by definition of $\lambda_{\mathcal{O}}$,
\begin{align}
    \lambda_{0,\,\rho^\star\sim\,c/T} = o_p(\lambda_{\mathcal{O}}),
\end{align}
i.e., sequences $\lambda_{0,\,\rho^\star\sim\,c/T}$ are asymptotically restricted to the interval $[0,\lambda_\mathcal{O})$. By \eqref{eq:pthm1a}, the coefficients $\vbeta^\star$ hence are consistently estimated at all possible activation thresholds of $y_{t-1}$,
\begin{equation}\label{eq:betaconvergence_F}
    \widehat{\vbeta}_{\lambda} \xrightarrow{p} \vbeta^\star_0 \quad \forall\, \lambda = \lambda_{0,\,\rho^{\star}\sim\,c/T}. 
\end{equation}

To analyse the asymptotic behaviour of $\lambda_{0,\,\rho^\star\sim\,c/T}$, note that $\rho^\star = \phi(1)^{-1}c/T$ under the local parameterisation $\varrho^\star = 1+c/T$, $c\in(-\infty, 0]$, cf. the discussion preceding Theorem 1 in \textcite{Aylaretal2019}. Appealing to \Vref{eq:foc_lambda0_anatomy2}, we now have
\begin{align}
    \lambda_{0,\,\rho^\star} =&\, {\lvert\!\underbrace{\vphantom{\sum_{j=1}^p}\widehat{\rho}}_{=A_T}\!\rvert^{\gamma_1}} \biggl\lvert\underbrace{\vphantom{\sum_{j=1}^p}\sum_t y_{t-1}\varepsilon_{p,\,t}}_{=B_{T,\,\lambda}} + \underbrace{\sum_{j=1}^{p} \left(\delta_j^\star - \widehat{\delta}_{\lambda,\,j}\right)\sum_t y_{t-1} \Delta y_{t-j}}_{=C_{T,\,\lambda}} + \phi(1)^{-1}c/T \underbrace{\vphantom{\sum_{j=1}^p} \sum_t y_{t-1}^2}_{=: D_T} \biggr\rvert\label{eq:scaledlambanaught}\\
    \intertext{By substituting and rearranging, we get}
    \tau_{\gamma_1} =&\, T^{\gamma_1-1}\lambda_{0,\,\rho^\star}\notag\\
    =&\, \big\lvert T A_T \big\rvert^{\gamma_1} \left\lvert T^{-1} B_{T,\,\lambda} + T^{-1}C_{T,\,\lambda} + \phi(1)^{-1}c \cdot T^{-2} D_T\right\rvert\label{eq:compscaledlambanaught}
\end{align}

We next consider the individual components of \eqref{eq:compscaledlambanaught}. By results of \textcite{Aylaretal2019,Hansen1995},
\begin{align}
    T A_T \xrightarrow{d} A :=&\, \phi(1)^{-1}\left(\frac{\int_0^1W_c(r)\mathrm{d}W(r)}{\int_0^1 W_c(r)^2\mathrm{d}r}+c\right),\label{eq:DFlimit}\\
    T^{-2} D_T \xrightarrow{d} D :=&\, \sigma^2\phi(1)^2\int_0^1 W_c(r)^2\mathrm{d}r, 
\end{align}
with the Ornstein-Uhlenbeck process $W_c(r)$ defined by the stochastic differential equation $\mathrm{d}W_c(r) = cW_c(r)\mathrm{d}r+\mathrm{d}W(r)$, $r\in[0,1]$, with zero starting value. 

The components $B_{T,\,\lambda}$ and $C_{T,\,\lambda}$ depend on the adaptive Lasso solution $\widehat{\vbeta}_{\lambda}$. Since consistency of $\widehat{\vbeta}_{\lambda}$ at $\lambda=\lambda_{0,\,\rho^\star\sim\,c/T}$ is ensured per \eqref{eq:betaconvergence_F}, we have $\varepsilon_{p,\,t}\xrightarrow[]{p}\varepsilon_t$ by \textcite{Aylaretal2019} so that, by \textcite{Hansen1995},
\begin{align}
    T^{-1}B_{T,\,\lambda} \xrightarrow{d} B:=&\, \sigma^2\phi(1)\int_0^1 W_c(r)\mathrm{d}W(r).\label{eq:ndlimit}
\end{align}

Turning to $C_{T,\,\lambda}$, we note that
\begin{align*}
    T^{-1}\sum_t y_{t-1} \Delta y_{t-j}\bigg\vert_{c\,\in(-\infty,0]} = O_p(1) \quad \forall\,j\geq0,
\end{align*}
and further that \eqref{eq:betaconvergence_F} implies
\begin{align*}
    \delta_j^\star - \widehat{\delta}_{j,\,\lambda} \xrightarrow{p} 0,\quad j=1,\dots,p  \quad \forall\, \lambda = \lambda_{0,\,\rho^{\star}\sim\,c/T}.
\end{align*}
By Slutsky's Theorem, we hence have
\begin{align}
    T^{-1} C_{T,\,\lambda} \xrightarrow{p} 0  \quad \forall\, \lambda = \lambda_{0,\,\rho^{\star}\sim\,c/T}.\label{eq:CTlambdalimit}
\end{align}
Applying the results \eqref{eq:DFlimit} -- \eqref{eq:CTlambdalimit} and Slutsky's Theorem to \eqref{eq:compscaledlambanaught} shows that
\begin{align}
    \tau_{\gamma_1}\big\vert_{c\,\in(-\infty,0]} = O_p(1) \quad \text{and} \quad \lambda_{0,\,\rho^\star\sim\,c/T}\big\vert_{c\,\in(-\infty,0]} = O_p(T^{1-\gamma_1}).
\end{align}

\item To derive the limiting r.v. in \eqref{eq:taugammaNURLimit}, note that
\begin{align}
    \int_0^1 W_c(r)\mathrm{d}W(r) = \frac{1}{2} (W_c(1)^2-1) - c\cdot\int_0^1 W_c(r)^2\mathrm{d}r,
\end{align}
see \textcite{Phillips1987}. Using this result, it is straightforward to establish that
\begin{align}
    A = \phi(1)^{-1}\frac{W_c(1)^2-1}{2\int_0^1 W_c(r)^2\mathrm{d}r}, \quad
    B + \phi(1)^{-1} c \cdot D = \frac{1}{2}\phi(1)\sigma^2\left(W_c(1)^2-1\right).\label{eq:altrepresults}
\end{align}

Using the representations \eqref{eq:altrepresults} and applying the continuous mapping theorem (CMT) obtains
\begin{align*}
    \tau_{\gamma_1} \xrightarrow{d}&\, \left\lvert A\right\rvert^{\gamma_1} \left\lvert B + \phi(1)^{-1} c \cdot D \right\rvert \\
    =&\, \left\lvert \phi(1)^{-1} \frac{W_c(1)^2-1}{2\int_0^1 W_c(r)^2\mathrm{d}r}\right\rvert^{\gamma_1} \biggl\vert\frac{1}{2}\phi(1)\sigma^2\left(W_c(1)^2-1\right)\biggr\rvert,
\end{align*}
which completes the proof of part 1. 
\end{enumerate}
    \item Result 2 follows directly from the argument of part 2 of Proposition 2 of Arnold and Reinschlüssel (2023).\qedhere
\end{enumerate}
\end{proof}

\begin{proof}[\textbf{Proof of \Cref{cor:brevelambdanullrhodist}}]
	\noindent
	\begin{enumerate}[leftmargin=*]
		\item The exposition in \textcite{HerwartzSiedenburg2010} shows $J_\alpha$ to converge to a positively-bounded r.v. $J_{\alpha,\,c}$ under \Cref{assum:lperrorsChangPark,assum:aroder}. The result then follows from the CMT.
		\item The result follows directly from part 2 of Proposition 2 in Arnold and Reinschlüssel (2023).\qedhere
	\end{enumerate}
\end{proof}

\newpage

\subsection{Details on the computation of the penalty weights}
\label{sec:ALURT_dcpw}

\begin{restatable}[$\Breve{w}_1$]{algo}{ALURT_wbrewedet}
    \label{algo:ALURT_wbrewedet}
    \noindent
    \begin{enumerate}
    \item After adjusting $y_t$ for the deterministic component $d_t$, estimate the LRV $\omega^2$ using model \eqref{eq:ALURT_adfreg} by
    \begin{align*}
        \widehat{\omega}^2_{\textup{AR}}(k) =&\, \frac{\widehat{\sigma}_k^2}{(1 - \sum_{j=1}^k \widehat{\delta}_j)^2}, \quad \widehat{\sigma}^2_k = (T-k)^{-1} \sum_{t=k+1}^T \widehat{e}_{t,k}^2
    \end{align*}
    for suitable $k$.
    \item Scale $y_t$ by $\widehat{\omega}_{\textup{AR}}(k)^{-1}$.
    \item For some fixed $\alpha\in(0,.5)$, obtain the range statistic $J_{\alpha} := \left\lvert\widehat{\zeta}^{(r)}_{1-\alpha/2}-\widehat{\zeta}^{(r)}_{\alpha/2}\right\rvert$ with $\widehat\zeta_\alpha^{(r)}$ the $\alpha$ quantile of simulated OLS estimates $\widehat\zeta^{(r)}$, in the model $$y_t = \mu^{(r)} + \varsigma^{(r)} t + \zeta^{(r)} x_t^{(r)} + \nu_t^{(r)}, \quad r=1,\dots,R,$$ where $x_t^{(r)}$ is a Gaussian random walk with zero initial condition.
    \item Compute $\Breve{w}_{1} := \left\lvert\widehat{\rho} / J_{\alpha}\right\rvert$ with $\widehat{\rho}$ the OLS estimator of $\rho^\star$ in model \eqref{eq:ALURT_adfreg}, adjusted for $d_t$ as in step 1.\qed
\end{enumerate}

\end{restatable}

\subsection{Additional Figures}

\begin{figure}[H]
\centering
\caption{Location of the groundwater measurement stations in NRW, Germany}
\label{fig:gwstationsmap}
\vspace{.25cm}
\includegraphics[width=.9\textwidth]{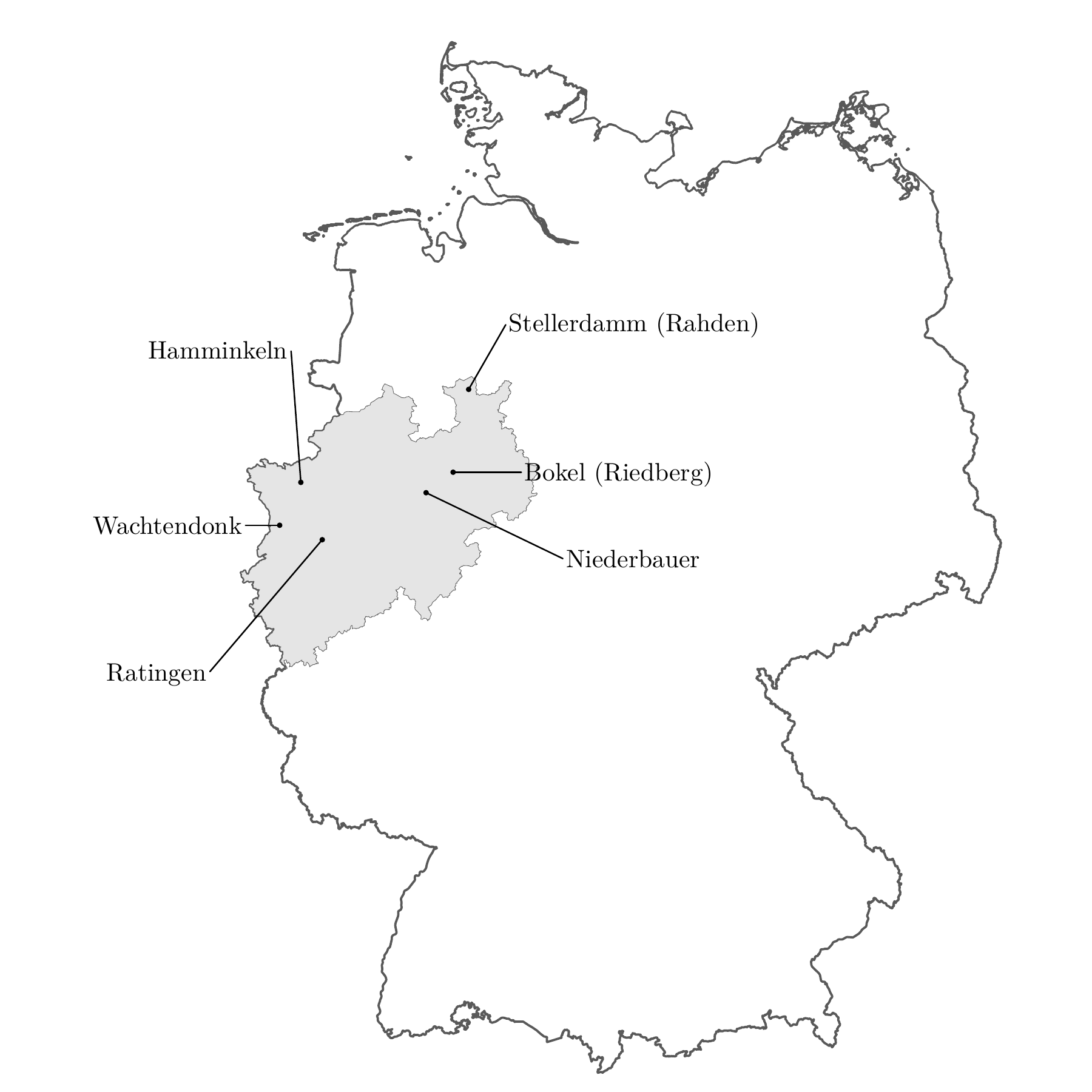}	
\begin{minipage}{\textwidth}
    \vspace{.25cm}
	\scriptsize\textit{Notes}: Black outlines show the borders of the Federal Republic of Germany. The federal state of North Rhine-Westphalia is shaded grey.
\end{minipage}
\end{figure}

\subsection{Additional simulation results}
\label{sec:ALURT_asr}

\begin{table}[H]
    \centering
    \caption{Critical values of activation knot tests under FD trend adjustment}
    \label{tab:CVsFDD}
    \vspace{.25cm}
    \setlength{\tabcolsep}{22pt}
    \renewcommand{\arraystretch}{1}
    \resizebox{\textwidth}{!}{
        \input{tabs/Tab_CVs_FDD.tex}
    }
    \begin{minipage}{\textwidth}
        \vspace{.25cm}
        \scriptsize\textit{Notes:} Critical values are computed based on the adaptive Lasso solution to \eqref{eq:ALURT_adfreg} with $d_t=0$ and $p=0$. $\Breve{\tau}$ is computed using $J_\alpha$ with $\alpha=.1$ and $\sigma^2$ estimated by $\widehat{\sigma}^2$. Data adjusted for a constant are obtained as $y^{\textup{FD}}_t = y_t - y_1$. Data adjusted for a linear time trend are obtained as $y^{\textup{FD}}_t = y_t - y_1 - t/T(y_T-y_1)$.  $5\cdot10^5$ replications.
    \end{minipage}
\end{table}


\renewcommand\floatpagefraction{0.1}
\begin{table}[H]
    \centering
    \caption{Empirical size: AR errors -- no adjustment}
    \label{tab:size_AR_nc}
    \vspace{.25cm}
    \setlength{\tabcolsep}{26pt}
    \renewcommand{\arraystretch}{.85}
    \resizebox{\textwidth}{!}{
        \input{tabs/Tab_AR_size_nc.tex}
    }
    \begin{minipage}{\textwidth}
        \vspace{.25cm}
        \scriptsize\textit{Notes:} DGP \eqref{eq:arerrorsdgp} with $c=0$ and $v_t=\varphi v_{t-1} + \epsilon_t$, $\epsilon_t\sim\,i.i.d.\,N(0,1)$. Model \eqref{eq:ALURT_adfreg} with $d_t = 0$ and $p=\lfloor12\cdot(T/100)^{.25}\rfloor$. $\tau$ and $\Breve{\tau}$ are activation knot tests based on FD-adjustment, if applicable. $\textup{ADF}^\textup{GLS}$ and $\textup{MZ}_t$ are the modified augmented Dickey-Fuller \parencite{Elliottetal1996} and Phillips-Perron $Z_t$ test \parencite{NgPerron2001} with GLS-adjusted data, if applicable. $J_\alpha$ is computed as stated in \textcite{HerwartzSiedenburg2010} using OLS-adjusted data, if applicable. The LRV is estimated as detailed in \Cref{sec:ALURT_ctutnif}. 5000 replications. 
    \end{minipage}
\end{table}

\begin{table}[H]
    \centering
    \caption{Empirical size: AR errors -- demeaning}
    \label{tab:size_AR_c}
    \vspace{.25cm}
    \setlength{\tabcolsep}{26pt}
    \renewcommand{\arraystretch}{.85}
    \resizebox{\textwidth}{!}{
        \input{tabs/Tab_AR_size_c.tex}
    }
    \begin{minipage}{\textwidth}
        \vspace{.25cm}
        \scriptsize\textit{Notes:} DGP \eqref{eq:arerrorsdgp} with $c=0$ and $v_t=\varphi v_{t-1} + \epsilon_t$, $\epsilon_t\sim\,i.i.d.\,N(0,1)$. Model \eqref{eq:ALURT_adfreg} with $d_t = 0$ and $p=\lfloor12\cdot(T/100)^{.25}\rfloor$. Data are adjusted for a constant. See \Cref{tab:size_AR_nc} for further details. 5000 replications.
    \end{minipage}
\end{table}

\begin{table}[H]
    \centering
    \caption{Empirical size: AR errors -- detrending}
    \label{tab:size_AR_ct}
    \vspace{.25cm}
    \setlength{\tabcolsep}{26pt}
    \renewcommand{\arraystretch}{.85}
    \resizebox{\textwidth}{!}{
        \input{tabs/Tab_AR_size_ct.tex}
    }
    \begin{minipage}{\textwidth}
        \vspace{.25cm}
        \scriptsize\textit{Notes:} DGP \eqref{eq:arerrorsdgp} with $c=0$ and $v_t=\varphi v_{t-1} + \epsilon_t$, $\epsilon_t\sim\,i.i.d.\,N(0,1)$. Model \eqref{eq:ALURT_adfreg} with $d_t = 0$ and $p=\lfloor12\cdot(T/100)^{.25}\rfloor$. Data are adjusted for a linear time trend. See \Cref{tab:size_AR_nc} for further details. 5000 replications.
    \end{minipage}
\end{table}


\begin{table}[H]
    \centering
    \caption{Size-adjusted local power: AR errors -- no adjustment}
    \label{tab:sapower_AR_nc}
    \vspace{.25cm}
    \setlength{\tabcolsep}{26pt}
    \renewcommand{\arraystretch}{.85}
    \resizebox{\textwidth}{!}{
        \input{tabs/Tab_AR_sapower_nc.tex}
    }
    \begin{minipage}{\textwidth}
        \vspace{.25cm}
        \scriptsize\textit{Notes:} DGP \eqref{eq:arerrorsdgp} with $c=-7$ and $v_t=\varphi v_{t-1} + \epsilon_t$, $\epsilon_t\sim\,i.i.d.\,N(0,1)$. Model \eqref{eq:ALURT_adfreg} with $d_t = 0$ and $p=\lfloor12\cdot(T/100)^{.25}\rfloor$. See \Cref{tab:size_AR_nc} for further details. 5000 replications.
    \end{minipage}
\end{table}

\begin{table}[H]
    \centering
    \caption{Size-adjusted local power: AR errors -- demeaning}
    \label{tab:sapower_AR_c}
    \vspace{.25cm}
    \setlength{\tabcolsep}{26pt}
    \renewcommand{\arraystretch}{.85}
    \resizebox{\textwidth}{!}{
        \input{tabs/Tab_AR_sapower_c.tex}
    }
    \begin{minipage}{\textwidth}
        \vspace{.25cm}
        \scriptsize\textit{Notes:} DGP \eqref{eq:arerrorsdgp} with $c=-7$ and $v_t=\varphi v_{t-1} + \epsilon_t$, $\epsilon_t\sim\,i.i.d.\,N(0,1)$. Model \eqref{eq:ALURT_adfreg} with $d_t = 0$ and $p=\lfloor12\cdot(T/100)^{.25}\rfloor$. Data are adjusted for a constant. See \Cref{tab:size_AR_nc} for further details. 5000 replications.
    \end{minipage}
\end{table}

\begin{table}[H]
    \centering
    \caption{Size-adjusted local power: AR errors -- detrending}
    \label{tab:sapower_AR_ct}
    \vspace{.25cm}
    \setlength{\tabcolsep}{26pt}
    \renewcommand{\arraystretch}{.85}
    \resizebox{\textwidth}{!}{
        \input{tabs/Tab_AR_sapower_ct.tex}
    }
    \begin{minipage}{\textwidth}
        \vspace{.25cm}
        \scriptsize\textit{Notes:} DGP \eqref{eq:arerrorsdgp} with $c=-13.5$ and $v_t=\varphi v_{t-1} + \epsilon_t$, $\epsilon_t\sim\,i.i.d.\,N(0,1)$. Model \eqref{eq:ALURT_adfreg} with $d_t = 0$ and $p=\lfloor12\cdot(T/100)^{.25}\rfloor$. Data are adjusted for a linear time trend. See \Cref{tab:size_AR_nc} for further details. 5000 replications.
    \end{minipage}
\end{table}


\begin{table}[H]
    \centering
    \caption{Empirical size: MA errors -- no adjustment}
    \label{tab:size_MA_nc}
    \vspace{.25cm}
    \setlength{\tabcolsep}{26pt}
    \resizebox{\textwidth}{!}{
        \input{tabs/Tab_MA_size_nc.tex}
    }
    \begin{minipage}{\textwidth}
        \vspace{.25cm}
        \scriptsize\textit{Notes:} DGP \eqref{eq:arerrorsdgp} with $c=0$ and $v_t=\vartheta \epsilon_{t-1} + \epsilon_t$, $\epsilon_t\sim\,i.i.d.\,N(0,1)$. Model \eqref{eq:ALURT_adfreg} with $d_t = 0$ and $p=\lfloor12\cdot(T/100)^{.25}\rfloor$. See \Cref{tab:size_AR_nc} for further details. 5000 replications.
    \end{minipage}
\end{table}

\begin{table}[H]
    \centering
    \caption{Empirical size: MA errors -- demeaning}
    \label{tab:size_MA_c}
    \vspace{.25cm}
    \setlength{\tabcolsep}{26pt}
    \resizebox{\textwidth}{!}{
        \input{tabs/Tab_MA_size_c.tex}
    }
    \begin{minipage}{\textwidth}
        \vspace{.25cm}
        \scriptsize\textit{Notes:} DGP \eqref{eq:arerrorsdgp} with $c=0$ and $v_t=\vartheta \epsilon_{t-1} + \epsilon_t$, $\epsilon_t\sim\,i.i.d.\,N(0,1)$. Model \eqref{eq:ALURT_adfreg} with $d_t = 0$ and $p=\lfloor12\cdot(T/100)^{.25}\rfloor$. Data are adjusted for a constant. See \Cref{tab:size_AR_nc} for further details. 5000 replications.
    \end{minipage}
\end{table}

\begin{table}[H]
    \centering
    \caption{Empirical size: MA errors -- detrending}
    \label{tab:size_MA_ct}
    \vspace{.25cm}
    \setlength{\tabcolsep}{26pt}
    \resizebox{\textwidth}{!}{
        \input{tabs/Tab_MA_size_ct.tex}
    }
    \begin{minipage}{\textwidth}
        \vspace{.25cm}
        \scriptsize\textit{Notes:} DGP \eqref{eq:arerrorsdgp} with $c=0$ and $v_t=\vartheta \epsilon_{t-1} + \epsilon_t$, $\epsilon_t\sim\,i.i.d.\,N(0,1)$. Model \eqref{eq:ALURT_adfreg} with $d_t = 0$ and $p=\lfloor12\cdot(T/100)^{.25}\rfloor$. Data are adjusted for a linear time trend. See \Cref{tab:size_AR_nc} for further details. 5000 replications.
    \end{minipage}
\end{table}


\begin{table}[H]
    \centering
    \caption{Size-adjusted local power: MA errors -- no adjustment}
    \label{tab:sapower_MA_nc}
    \vspace{.25cm}
    \setlength{\tabcolsep}{26pt}
    \resizebox{\textwidth}{!}{
        \input{tabs/Tab_MA_sapower_nc.tex}
    }
    \begin{minipage}{\textwidth}
        \vspace{.25cm}
        \scriptsize\textit{Notes:} DGP \eqref{eq:arerrorsdgp} with $c=-7$ and $v_t=\vartheta \epsilon_{t-1} + \epsilon_t$, $\epsilon_t\sim\,i.i.d.\,N(0,1)$. Model \eqref{eq:ALURT_adfreg} with $d_t = 0$ and $p=\lfloor12\cdot(T/100)^{.25}\rfloor$. See \Cref{tab:size_AR_nc} for further details. 5000 replications.
    \end{minipage}
\end{table}

\begin{table}[H]
    \centering
    \caption{Size-adjusted local power: MA errors -- demeaning}
    \label{tab:sapower_MA_c}
    \vspace{.25cm}
    \setlength{\tabcolsep}{26pt}
    \resizebox{\textwidth}{!}{
        \input{tabs/Tab_MA_sapower_c.tex}
    }
    \begin{minipage}{\textwidth}
        \vspace{.25cm}
        \scriptsize\textit{Notes:} DGP \eqref{eq:arerrorsdgp} with $c=-7$ and $v_t=\vartheta \epsilon_{t-1} + \epsilon_t$, $\epsilon_t\sim\,i.i.d.\,N(0,1)$. Model \eqref{eq:ALURT_adfreg} with $d_t = 0$ and $p=\lfloor12\cdot(T/100)^{.25}\rfloor$. Data are adjusted for a constant. See \Cref{tab:size_AR_nc} for further details. 5000 replications.
    \end{minipage}
\end{table}

\begin{table}[H]
    \centering
    \caption{Size-adjusted local power: MA errors -- detrending}
    \label{tab:sapower_MA_ct}
    \vspace{.25cm}
    \setlength{\tabcolsep}{26pt}
    \resizebox{\textwidth}{!}{
        \input{tabs/Tab_MA_sapower_ct.tex}
    }
    \begin{minipage}{\textwidth}
        \vspace{.25cm}
        \scriptsize\textit{Notes:} DGP \eqref{eq:arerrorsdgp} with $c=-13.5$ and $v_t=\vartheta \epsilon_{t-1} + \epsilon_t$, $\epsilon_t\sim\,i.i.d.\,N(0,1)$. Model \eqref{eq:ALURT_adfreg} with $d_t = 0$ and $p=\lfloor12\cdot(T/100)^{.25}\rfloor$. Data are adjusted for a linear time trend. See \Cref{tab:size_AR_nc} for further details. 5000 replications.
    \end{minipage}
\end{table}


%% file: tabs/Tab_CVs_FDD.tex
\begin{tabular}{lcccccc}
\toprule
\multicolumn{1}{c}{ } & \multicolumn{3}{c}{$\tau$} & \multicolumn{3}{c}{$\Breve{\tau}$} \\
\cmidrule(l{3pt}r{3pt}){2-4} \cmidrule(l{3pt}r{3pt}){5-7}
$T$ & 1\% & 5\% & 10\% & 1\% & 5\% & 10\%\\
\midrule
\addlinespace[0.3em]
\multicolumn{1}{l}{\textit{constant}}\\
50 & 7.40 & 4.28 & 3.06 & 13.21 & 5.32 & 3.08\\
75 & 7.25 & 4.23 & 3.02 & 13.55 & 5.40 & 3.12\\
100 & 7.18 & 4.21 & 3.02 & 13.72 & 5.49 & 3.17\\
150 & 7.12 & 4.19 & 3.01 & 14.02 & 5.59 & 3.22\\
250 & 7.08 & 4.16 & 2.99 & 13.98 & 5.58 & 3.21\\
500 & 6.97 & 4.15 & 2.98 & 13.96 & 5.62 & 3.23\\
1000 & 6.95 & 4.15 & 2.98 & 13.93 & 5.65 & 3.23\\
\addlinespace[0.3em]
\multicolumn{1}{l}{\textit{linear trend}}\\
50 & 10.97 & 7.22 & 5.65 & 20.15 & 10.49 & 7.24\\
75 & 10.65 & 7.07 & 5.56 & 20.85 & 10.85 & 7.51\\
100 & 10.51 & 7.03 & 5.55 & 21.41 & 11.08 & 7.65\\
150 & 10.37 & 6.94 & 5.49 & 21.68 & 11.21 & 7.72\\
250 & 10.21 & 6.90 & 5.47 & 21.84 & 11.31 & 7.83\\
500 & 10.15 & 6.85 & 5.46 & 22.06 & 11.41 & 7.88\\
1000 & 10.13 & 6.87 & 5.44 & 22.25 & 11.51 & 7.93\\
\bottomrule
\end{tabular}

%% file: tabs/Tab_AR_size_nc.tex
\begin{tabular}{lcccccc}
\toprule
$T$ & $\varphi$ & $\tau$ & $\Breve{\tau}$ & $\textup{ADF}^\textup{GLS}$ & $\textup{MZ}_t$ & $J_\alpha$\\
\midrule
\addlinespace[0.3em]
\multicolumn{1}{r}{\textbf{}}\\
50 &  & .056 & .031 & .039 & .004 & .019\\

75 &  & .054 & .035 & .040 & .003 & .024\\

100 &  & .052 & .033 & .040 & .005 & .024\\

150 &  & .051 & .042 & .043 & .010 & .032\\

250 &  & .049 & .041 & .044 & .017 & .031\\

500 &  & .055 & .049 & .048 & .029 & .041\\

1000 & \multirow{-7}{*}{\centering\arraybackslash $-.8$} & .051 & .049 & .050 & .038 & .045\\

\addlinespace[0.3em]
\multicolumn{1}{r}{\textbf{}}\\
50 &  & .064 & .045 & .046 & .014 & .024\\

75 &  & .056 & .039 & .037 & .010 & .021\\

100 &  & .061 & .044 & .046 & .019 & .030\\

150 &  & .054 & .044 & .044 & .019 & .029\\

250 &  & .050 & .044 & .043 & .024 & .035\\

500 &  & .051 & .049 & .050 & .034 & .041\\

1000 & \multirow{-7}{*}{\centering\arraybackslash $-.4$} & .048 & .050 & .050 & .041 & .043\\

\addlinespace[0.3em]
\multicolumn{1}{r}{\textbf{}}\\
50 &  & .051 & .028 & .045 & .014 & .023\\

75 &  & .045 & .026 & .042 & .014 & .020\\

100 &  & .042 & .028 & .044 & .016 & .020\\

150 &  & .044 & .035 & .046 & .020 & .027\\

250 &  & .044 & .036 & .042 & .022 & .030\\

500 &  & .052 & .045 & .050 & .036 & .037\\

1000 & \multirow{-7}{*}{\centering\arraybackslash $0$} & .045 & .043 & .046 & .037 & .042\\

\addlinespace[0.3em]
\multicolumn{1}{r}{\textbf{}}\\
50 &  & .066 & .018 & .042 & .012 & .014\\

75 &  & .062 & .025 & .044 & .014 & .020\\

100 &  & .053 & .028 & .041 & .021 & .025\\

150 &  & .050 & .033 & .044 & .022 & .028\\

250 &  & .046 & .037 & .041 & .028 & .031\\

500 &  & .045 & .037 & .044 & .032 & .035\\

1000 & \multirow{-7}{*}{\centering\arraybackslash $.4$} & .051 & .050 & .049 & .043 & .046\\

\addlinespace[0.3em]
\multicolumn{1}{r}{\textbf{}}\\
50 &  & .083 & .022 & .051 & .034 & .022\\

75 &  & .069 & .024 & .049 & .025 & .021\\

100 &  & .055 & .026 & .043 & .026 & .022\\

150 &  & .051 & .033 & .044 & .024 & .026\\

250 &  & .049 & .037 & .045 & .028 & .030\\

500 &  & .051 & .042 & .046 & .035 & .038\\

1000 & \multirow{-7}{*}{\centering\arraybackslash $.8$} & .049 & .048 & .052 & .045 & .043\\
\bottomrule
\end{tabular}

%% file: tabs/Tab_AR_size_c.tex
\begin{tabular}{lcccccc}
\toprule
$T$ & $\varphi$ & $\tau$ & $\Breve{\tau}$ & $\textup{ADF}^\textup{GLS}$ & $\textup{MZ}_t$ & $J_\alpha$\\
\midrule
\addlinespace[0.3em]
\multicolumn{1}{r}{\textbf{}}\\
50 &  & .063 & .057 & .027 & .006 & .048\\

75 &  & .054 & .056 & .031 & .007 & .045\\

100 &  & .051 & .049 & .030 & .007 & .043\\

150 &  & .046 & .044 & .031 & .013 & .043\\

250 &  & .052 & .054 & .045 & .031 & .045\\

500 &  & .046 & .046 & .041 & .034 & .045\\

1000 & \multirow{-7}{*}{\centering\arraybackslash $-.8$} & .048 & .049 & .049 & .044 & .045\\

\addlinespace[0.3em]
\multicolumn{1}{r}{\textbf{}}\\
50 &  & .078 & .077 & .039 & .040 & .059\\

75 &  & .063 & .061 & .031 & .033 & .044\\

100 &  & .056 & .063 & .037 & .036 & .047\\

150 &  & .058 & .057 & .039 & .039 & .047\\

250 &  & .051 & .051 & .041 & .041 & .044\\

500 &  & .052 & .053 & .046 & .045 & .048\\

1000 & \multirow{-7}{*}{\centering\arraybackslash $-.4$} & .048 & .049 & .046 & .046 & .047\\

\addlinespace[0.3em]
\multicolumn{1}{r}{\textbf{}}\\
50 &  & .047 & .047 & .037 & .040 & .042\\

75 &  & .040 & .038 & .030 & .032 & .035\\

100 &  & .045 & .047 & .037 & .038 & .047\\

150 &  & .052 & .049 & .046 & .049 & .044\\

250 &  & .046 & .046 & .043 & .043 & .049\\

500 &  & .051 & .046 & .049 & .049 & .049\\

1000 & \multirow{-7}{*}{\centering\arraybackslash $0$} & .055 & .055 & .054 & .055 & .053\\

\addlinespace[0.3em]
\multicolumn{1}{r}{\textbf{}}\\
50 &  & .037 & .021 & .012 & .038 & .027\\

75 &  & .038 & .029 & .026 & .041 & .036\\

100 &  & .041 & .031 & .028 & .043 & .040\\

150 &  & .046 & .040 & .038 & .050 & .042\\

250 &  & .049 & .046 & .045 & .052 & .044\\

500 &  & .050 & .043 & .045 & .047 & .045\\

1000 & \multirow{-7}{*}{\centering\arraybackslash $.4$} & .047 & .049 & .044 & .046 & .046\\

\addlinespace[0.3em]
\multicolumn{1}{r}{\textbf{}}\\
50 &  & .053 & .042 & .030 & .119 & .059\\

75 &  & .050 & .037 & .032 & .083 & .042\\

100 &  & .050 & .037 & .035 & .064 & .037\\

150 &  & .048 & .041 & .040 & .057 & .037\\

250 &  & .047 & .045 & .043 & .053 & .041\\

500 &  & .046 & .045 & .046 & .048 & .039\\

1000 & \multirow{-7}{*}{\centering\arraybackslash $.8$} & .051 & .050 & .050 & .052 & .046\\
\bottomrule
\end{tabular}

%% file: tabs/Tab_AR_size_ct.tex
\begin{tabular}{lcccccc}
\toprule
$T$ & $\varphi$ & $\tau$ & $\Breve{\tau}$ & $\textup{ADF}^\textup{GLS}$ & $\textup{MZ}_t$ & $J_\alpha$\\
\midrule
\addlinespace[0.3em]
\multicolumn{1}{r}{\textbf{}}\\
50 &  & .073 & .078 & .025 & .003 & .070\\

75 &  & .068 & .070 & .028 & .002 & .062\\

100 &  & .056 & .061 & .025 & .002 & .048\\

150 &  & .050 & .052 & .026 & .004 & .050\\

250 &  & .047 & .047 & .032 & .011 & .050\\

500 &  & .047 & .048 & .037 & .021 & .050\\

1000 & \multirow{-7}{*}{\centering\arraybackslash $-.8$} & .050 & .050 & .044 & .032 & .048\\

\addlinespace[0.3em]
\multicolumn{1}{r}{\textbf{}}\\
50 &  & .117 & .104 & .040 & .044 & .072\\

75 &  & .099 & .098 & .032 & .035 & .070\\

100 &  & .080 & .080 & .030 & .031 & .054\\

150 &  & .073 & .070 & .032 & .031 & .045\\

250 &  & .064 & .066 & .035 & .035 & .046\\

500 &  & .047 & .049 & .037 & .037 & .047\\

1000 & \multirow{-7}{*}{\centering\arraybackslash $-.4$} & .051 & .052 & .044 & .043 & .048\\

\addlinespace[0.3em]
\multicolumn{1}{r}{\textbf{}}\\
50 &  & .040 & .037 & .029 & .031 & .045\\

75 &  & .040 & .041 & .033 & .034 & .039\\

100 &  & .044 & .042 & .037 & .040 & .044\\

150 &  & .044 & .043 & .034 & .036 & .045\\

250 &  & .043 & .043 & .036 & .036 & .041\\

500 &  & .044 & .039 & .035 & .035 & .042\\

1000 & \multirow{-7}{*}{\centering\arraybackslash $0$} & .049 & .050 & .048 & .049 & .051\\

\addlinespace[0.3em]
\multicolumn{1}{r}{\textbf{}}\\
50 &  & .008 & .007 & .003 & .035 & .014\\

75 &  & .015 & .014 & .012 & .045 & .023\\

100 &  & .025 & .024 & .022 & .051 & .036\\

150 &  & .034 & .032 & .030 & .049 & .043\\

250 &  & .038 & .036 & .032 & .044 & .045\\

500 &  & .047 & .045 & .044 & .051 & .047\\

1000 & \multirow{-7}{*}{\centering\arraybackslash $.4$} & .049 & .045 & .042 & .046 & .044\\

\addlinespace[0.3em]
\multicolumn{1}{r}{\textbf{}}\\
50 &  & .038 & .038 & .028 & .238 & .068\\

75 &  & .042 & .033 & .032 & .143 & .041\\

100 &  & .040 & .030 & .031 & .103 & .032\\

150 &  & .045 & .036 & .037 & .077 & .038\\

250 &  & .045 & .038 & .040 & .059 & .031\\

500 &  & .045 & .043 & .042 & .052 & .036\\

1000 & \multirow{-7}{*}{\centering\arraybackslash $.8$} & .043 & .044 & .043 & .049 & .043\\
\bottomrule
\end{tabular}

%% file: tabs/Tab_AR_sapower_nc.tex
\begin{tabular}{lcccccc}
\toprule
$T$ & $\varphi$ & $\tau$ & $\Breve{\tau}$ & $\textup{ADF}^\textup{GLS}$ & $\textup{MZ}_t$ & $J_\alpha$\\
\midrule
\addlinespace[0.3em]
\multicolumn{1}{r}{\textbf{}}\\
50 &  & .474 & .608 & .374 & .518 & .603\\

75 &  & .466 & .605 & .408 & .569 & .607\\

100 &  & .468 & .617 & .439 & .585 & .592\\

150 &  & .460 & .554 & .478 & .568 & .519\\

250 &  & .489 & .573 & .503 & .559 & .532\\

500 &  & .438 & .505 & .503 & .518 & .481\\

1000 & \multirow{-7}{*}{\centering\arraybackslash $-.8$} & .458 & .506 & .504 & .511 & .459\\

\addlinespace[0.3em]
\multicolumn{1}{r}{\textbf{}}\\
50 &  & .485 & .578 & .351 & .556 & .573\\

75 &  & .501 & .596 & .431 & .630 & .591\\

100 &  & .456 & .577 & .429 & .608 & .553\\

150 &  & .474 & .564 & .483 & .601 & .551\\

250 &  & .487 & .560 & .510 & .597 & .500\\

500 &  & .469 & .510 & .503 & .524 & .463\\

1000 & \multirow{-7}{*}{\centering\arraybackslash $-.4$} & .462 & .496 & .506 & .513 & .461\\

\addlinespace[0.3em]
\multicolumn{1}{r}{\textbf{}}\\
50 &  & .404 & .605 & .414 & .661 & .601\\

75 &  & .475 & .610 & .462 & .682 & .612\\

100 &  & .450 & .616 & .474 & .650 & .600\\

150 &  & .471 & .568 & .474 & .606 & .550\\

250 &  & .470 & .573 & .509 & .603 & .535\\

500 &  & .433 & .522 & .499 & .521 & .493\\

1000 & \multirow{-7}{*}{\centering\arraybackslash $0$} & .474 & .518 & .529 & .531 & .477\\

\addlinespace[0.3em]
\multicolumn{1}{r}{\textbf{}}\\
50 &  & .138 & .479 & .271 & .537 & .536\\

75 &  & .237 & .505 & .355 & .564 & .532\\

100 &  & .319 & .509 & .420 & .544 & .507\\

150 &  & .383 & .548 & .462 & .602 & .541\\

250 &  & .451 & .553 & .542 & .604 & .499\\

500 &  & .478 & .548 & .533 & .548 & .502\\

1000 & \multirow{-7}{*}{\centering\arraybackslash $.4$} & .431 & .478 & .503 & .506 & .438\\

\addlinespace[0.3em]
\multicolumn{1}{r}{\textbf{}}\\
50 &  & .110 & .360 & .203 & .310 & .309\\

75 &  & .192 & .400 & .273 & .439 & .401\\

100 &  & .269 & .457 & .335 & .453 & .416\\

150 &  & .331 & .465 & .406 & .476 & .452\\

250 &  & .396 & .473 & .453 & .473 & .446\\

500 &  & .408 & .500 & .479 & .492 & .447\\

1000 & \multirow{-7}{*}{\centering\arraybackslash $.8$} & .418 & .468 & .457 & .464 & .426\\
\bottomrule
\end{tabular}

%% file: tabs/Tab_AR_sapower_c.tex
\begin{tabular}{lcccccc}
\toprule
$T$ & $\varphi$ & $\tau$ & $\Breve{\tau}$ & $\textup{ADF}^\textup{GLS}$ & $\textup{MZ}_t$ & $J_\alpha$\\
\midrule
\addlinespace[0.3em]
\multicolumn{1}{r}{\textbf{}}\\
50 &  & .185 & .228 & .225 & .191 & .230\\

75 &  & .203 & .239 & .228 & .206 & .234\\

100 &  & .213 & .257 & .238 & .217 & .252\\

150 &  & .234 & .295 & .279 & .268 & .272\\

250 &  & .218 & .256 & .244 & .231 & .255\\

500 &  & .263 & .308 & .321 & .312 & .268\\

1000 & \multirow{-7}{*}{\centering\arraybackslash $-.8$} & .299 & .347 & .328 & .327 & .301\\

\addlinespace[0.3em]
\multicolumn{1}{r}{\textbf{}}\\
50 &  & .217 & .246 & .223 & .216 & .214\\

75 &  & .239 & .277 & .264 & .249 & .251\\

100 &  & .234 & .248 & .241 & .235 & .231\\

150 &  & .234 & .279 & .265 & .264 & .255\\

250 &  & .238 & .282 & .269 & .277 & .248\\

500 &  & .244 & .294 & .296 & .298 & .253\\

1000 & \multirow{-7}{*}{\centering\arraybackslash $-.4$} & .309 & .363 & .357 & .363 & .278\\

\addlinespace[0.3em]
\multicolumn{1}{r}{\textbf{}}\\
50 &  & .217 & .263 & .272 & .271 & .236\\

75 &  & .251 & .312 & .316 & .311 & .273\\

100 &  & .209 & .247 & .259 & .253 & .230\\

150 &  & .199 & .252 & .243 & .233 & .236\\

250 &  & .245 & .298 & .291 & .286 & .253\\

500 &  & .251 & .320 & .293 & .297 & .266\\

1000 & \multirow{-7}{*}{\centering\arraybackslash $0$} & .268 & .317 & .306 & .307 & .259\\

\addlinespace[0.3em]
\multicolumn{1}{r}{\textbf{}}\\
50 &  & .091 & .170 & .176 & .130 & .159\\

75 &  & .144 & .229 & .227 & .205 & .202\\

100 &  & .183 & .247 & .263 & .248 & .218\\

150 &  & .188 & .260 & .261 & .243 & .243\\

250 &  & .211 & .270 & .261 & .252 & .252\\

500 &  & .236 & .310 & .296 & .291 & .266\\

1000 & \multirow{-7}{*}{\centering\arraybackslash $.4$} & .304 & .340 & .354 & .354 & .292\\

\addlinespace[0.3em]
\multicolumn{1}{r}{\textbf{}}\\
50 &  & .099 & .145 & .163 & .069 & .082\\

75 &  & .125 & .178 & .176 & .137 & .148\\

100 &  & .133 & .196 & .193 & .164 & .176\\

150 &  & .162 & .215 & .207 & .201 & .183\\

250 &  & .205 & .242 & .246 & .234 & .207\\

500 &  & .250 & .299 & .286 & .295 & .265\\

1000 & \multirow{-7}{*}{\centering\arraybackslash $.8$} & .274 & .326 & .318 & .314 & .275\\
\bottomrule
\end{tabular}

%% file: tabs/Tab_AR_sapower_ct.tex
\begin{tabular}{lcccccc}
\toprule
$T$ & $\varphi$ & $\tau$ & $\Breve{\tau}$ & $\textup{ADF}^\textup{GLS}$ & $\textup{MZ}_t$ & $J_\alpha$\\
\midrule
\addlinespace[0.3em]
\multicolumn{1}{r}{\textbf{}}\\
50 &  & .216 & .263 & .269 & .176 & .307\\

75 &  & .241 & .277 & .309 & .212 & .305\\

100 &  & .268 & .312 & .310 & .252 & .346\\

150 &  & .300 & .335 & .331 & .288 & .315\\

250 &  & .318 & .360 & .329 & .332 & .313\\

500 &  & .363 & .389 & .397 & .392 & .334\\

1000 & \multirow{-7}{*}{\centering\arraybackslash $-.8$} & .372 & .409 & .408 & .406 & .355\\

\addlinespace[0.3em]
\multicolumn{1}{r}{\textbf{}}\\
50 &  & .315 & .343 & .262 & .245 & .302\\

75 &  & .289 & .310 & .276 & .281 & .271\\

100 &  & .324 & .363 & .333 & .322 & .323\\

150 &  & .332 & .365 & .356 & .333 & .342\\

250 &  & .339 & .371 & .367 & .369 & .351\\

500 &  & .384 & .404 & .374 & .374 & .336\\

1000 & \multirow{-7}{*}{\centering\arraybackslash $-.4$} & .369 & .388 & .394 & .396 & .348\\

\addlinespace[0.3em]
\multicolumn{1}{r}{\textbf{}}\\
50 &  & .376 & .400 & .360 & .376 & .333\\

75 &  & .364 & .385 & .382 & .381 & .349\\

100 &  & .351 & .374 & .345 & .352 & .319\\

150 &  & .358 & .371 & .367 & .366 & .316\\

250 &  & .367 & .401 & .385 & .394 & .352\\

500 &  & .376 & .423 & .409 & .408 & .348\\

1000 & \multirow{-7}{*}{\centering\arraybackslash $0$} & .371 & .388 & .390 & .385 & .333\\

\addlinespace[0.3em]
\multicolumn{1}{r}{\textbf{}}\\
50 &  & .190 & .201 & .188 & .056 & .176\\

75 &  & .178 & .194 & .151 & .108 & .178\\

100 &  & .226 & .249 & .216 & .181 & .215\\

150 &  & .281 & .296 & .294 & .293 & .278\\

250 &  & .342 & .352 & .349 & .348 & .291\\

500 &  & .323 & .341 & .328 & .331 & .306\\

1000 & \multirow{-7}{*}{\centering\arraybackslash $.4$} & .362 & .399 & .396 & .396 & .358\\

\addlinespace[0.3em]
\multicolumn{1}{r}{\textbf{}}\\
50 &  & .083 & .070 & .072 & .018 & .044\\

75 &  & .136 & .150 & .146 & .080 & .129\\

100 &  & .180 & .188 & .181 & .135 & .163\\

150 &  & .208 & .219 & .214 & .184 & .175\\

250 &  & .251 & .279 & .262 & .251 & .260\\

500 &  & .307 & .318 & .308 & .299 & .291\\

1000 & \multirow{-7}{*}{\centering\arraybackslash $.8$} & .370 & .376 & .370 & .361 & .322\\
\bottomrule
\end{tabular}

%% file: tabs/Tab_MA_size_nc.tex
\begin{tabular}{lcccccc}
\toprule
$T$ & $\vartheta$ & $\tau$ & $\Breve{\tau}$ & $\textup{ADF}^\textup{GLS}$ & $\textup{MZ}_t$ & $J_\alpha$\\
\midrule
\addlinespace[0.3em]
\multicolumn{1}{r}{\textbf{}}\\
50 &  & .240 & .187 & .071 & .041 & .083\\

75 &  & .242 & .181 & .062 & .023 & .077\\

100 &  & .240 & .179 & .064 & .022 & .073\\

150 &  & .226 & .176 & .069 & .017 & .072\\

250 &  & .205 & .163 & .075 & .015 & .066\\

500 &  & .161 & .136 & .067 & .023 & .060\\

1000 & \multirow{-7}{*}{\centering\arraybackslash $-.8$} & .132 & .120 & .075 & .045 & .064\\

\addlinespace[0.3em]
\multicolumn{1}{r}{\textbf{}}\\
50 &  & .082 & .061 & .054 & .022 & .031\\

75 &  & .073 & .055 & .047 & .017 & .029\\

100 &  & .069 & .055 & .044 & .022 & .033\\

150 &  & .076 & .065 & .054 & .025 & .038\\

250 &  & .065 & .058 & .047 & .031 & .036\\

500 &  & .072 & .068 & .057 & .043 & .050\\

1000 & \multirow{-7}{*}{\centering\arraybackslash $-.4$} & .056 & .056 & .046 & .038 & .046\\

\addlinespace[0.3em]
\multicolumn{1}{r}{\textbf{}}\\
50 &  & .060 & .025 & .043 & .015 & .019\\

75 &  & .047 & .025 & .038 & .014 & .017\\

100 &  & .057 & .037 & .047 & .022 & .029\\

150 &  & .047 & .036 & .038 & .021 & .033\\

250 &  & .053 & .044 & .046 & .027 & .035\\

500 &  & .051 & .046 & .045 & .031 & .041\\

1000 & \multirow{-7}{*}{\centering\arraybackslash $.4$} & .048 & .045 & .046 & .038 & .040\\

\addlinespace[0.3em]
\multicolumn{1}{r}{\textbf{}}\\
50 &  & .069 & .028 & .040 & .019 & .022\\

75 &  & .065 & .034 & .042 & .020 & .023\\

100 &  & .066 & .038 & .042 & .021 & .028\\

150 &  & .057 & .036 & .039 & .025 & .029\\

250 &  & .060 & .048 & .043 & .032 & .039\\

500 &  & .051 & .048 & .042 & .038 & .043\\

1000 & \multirow{-7}{*}{\centering\arraybackslash $.8$} & .048 & .047 & .045 & .043 & .043\\
\bottomrule
\end{tabular}

%% file: tabs/Tab_MA_size_c.tex
\begin{tabular}{lcccccc}
\toprule
$T$ & $\vartheta$ & $\tau$ & $\Breve{\tau}$ & $\textup{ADF}^\textup{GLS}$ & $\textup{MZ}_t$ & $J_\alpha$\\
\midrule
\addlinespace[0.3em]
\multicolumn{1}{r}{\textbf{}}\\
50 &  & .370 & .372 & .207 & .177 & .291\\

75 &  & .298 & .291 & .131 & .093 & .202\\

100 &  & .275 & .263 & .103 & .059 & .162\\

150 &  & .252 & .237 & .081 & .033 & .124\\

250 &  & .217 & .203 & .068 & .021 & .085\\

500 &  & .170 & .163 & .071 & .027 & .085\\

1000 & \multirow{-7}{*}{\centering\arraybackslash $-.8$} & .139 & .134 & .070 & .040 & .072\\

\addlinespace[0.3em]
\multicolumn{1}{r}{\textbf{}}\\
50 &  & .117 & .119 & .067 & .066 & .082\\

75 &  & .109 & .109 & .057 & .056 & .073\\

100 &  & .092 & .093 & .045 & .045 & .060\\

150 &  & .090 & .094 & .053 & .052 & .062\\

250 &  & .073 & .076 & .050 & .049 & .053\\

500 &  & .063 & .066 & .049 & .050 & .052\\

1000 & \multirow{-7}{*}{\centering\arraybackslash $-.4$} & .061 & .063 & .049 & .049 & .049\\

\addlinespace[0.3em]
\multicolumn{1}{r}{\textbf{}}\\
50 &  & .046 & .028 & .015 & .035 & .033\\

75 &  & .039 & .036 & .023 & .042 & .035\\

100 &  & .048 & .041 & .030 & .042 & .046\\

150 &  & .051 & .047 & .035 & .044 & .047\\

250 &  & .052 & .052 & .041 & .048 & .045\\

500 &  & .053 & .051 & .047 & .049 & .048\\

1000 & \multirow{-7}{*}{\centering\arraybackslash $.4$} & .049 & .047 & .049 & .051 & .047\\

\addlinespace[0.3em]
\multicolumn{1}{r}{\textbf{}}\\
50 &  & .059 & .045 & .017 & .059 & .042\\

75 &  & .055 & .043 & .018 & .046 & .035\\

100 &  & .047 & .044 & .018 & .045 & .044\\

150 &  & .047 & .045 & .024 & .044 & .043\\

250 &  & .053 & .054 & .034 & .051 & .046\\

500 &  & .051 & .047 & .041 & .050 & .046\\

1000 & \multirow{-7}{*}{\centering\arraybackslash $.8$} & .055 & .056 & .049 & .055 & .052\\
\bottomrule
\end{tabular}

%% file: tabs/Tab_MA_size_ct.tex
\begin{tabular}{lcccccc}
\toprule
$T$ & $\vartheta$ & $\tau$ & $\Breve{\tau}$ & $\textup{ADF}^\textup{GLS}$ & $\textup{MZ}_t$ & $J_\alpha$\\
\midrule
\addlinespace[0.3em]
\multicolumn{1}{r}{\textbf{}}\\
50 &  & .558 & .582 & .359 & .348 & .498\\

75 &  & .502 & .503 & .244 & .216 & .387\\

100 &  & .442 & .432 & .176 & .132 & .300\\

150 &  & .377 & .353 & .102 & .054 & .218\\

250 &  & .348 & .311 & .074 & .021 & .152\\

500 &  & .289 & .250 & .059 & .010 & .108\\

1000 & \multirow{-7}{*}{\centering\arraybackslash $-.8$} & .232 & .194 & .072 & .014 & .085\\

\addlinespace[0.3em]
\multicolumn{1}{r}{\textbf{}}\\
50 &  & .204 & .184 & .097 & .108 & .132\\

75 &  & .176 & .152 & .061 & .072 & .099\\

100 &  & .168 & .149 & .057 & .061 & .086\\

150 &  & .135 & .119 & .046 & .050 & .071\\

250 &  & .117 & .104 & .043 & .045 & .059\\

500 &  & .101 & .089 & .041 & .044 & .053\\

1000 & \multirow{-7}{*}{\centering\arraybackslash $-.4$} & .081 & .073 & .048 & .049 & .050\\

\addlinespace[0.3em]
\multicolumn{1}{r}{\textbf{}}\\
50 &  & .014 & .011 & .004 & .030 & .018\\

75 &  & .028 & .027 & .013 & .042 & .037\\

100 &  & .034 & .030 & .020 & .046 & .037\\

150 &  & .047 & .046 & .034 & .057 & .055\\

250 &  & .051 & .048 & .032 & .046 & .045\\

500 &  & .048 & .046 & .033 & .042 & .042\\

1000 & \multirow{-7}{*}{\centering\arraybackslash $.4$} & .046 & .049 & .039 & .045 & .048\\

\addlinespace[0.3em]
\multicolumn{1}{r}{\textbf{}}\\
50 &  & .033 & .027 & .005 & .069 & .033\\

75 &  & .047 & .042 & .007 & .060 & .039\\

100 &  & .044 & .043 & .009 & .056 & .040\\

150 &  & .043 & .043 & .011 & .055 & .042\\

250 &  & .044 & .046 & .021 & .051 & .050\\

500 &  & .051 & .056 & .033 & .063 & .056\\

1000 & \multirow{-7}{*}{\centering\arraybackslash $.8$} & .051 & .053 & .037 & .054 & .053\\
\bottomrule
\end{tabular}

%% file: tabs/Tab_MA_sapower_nc.tex
\begin{tabular}{lcccccc}
\toprule
$T$ & $\vartheta$ & $\tau$ & $\Breve{\tau}$ & $\textup{ADF}^\textup{GLS}$ & $\textup{MZ}_t$ & $J_\alpha$\\
\midrule
\addlinespace[0.3em]
\multicolumn{1}{r}{\textbf{}}\\
50 &  & .643 & .618 & .453 & .471 & .564\\

75 &  & .646 & .639 & .413 & .400 & .570\\

100 &  & .623 & .585 & .373 & .355 & .535\\

150 &  & .593 & .584 & .415 & .377 & .522\\

250 &  & .561 & .565 & .430 & .373 & .509\\

500 &  & .562 & .552 & .497 & .486 & .503\\

1000 & \multirow{-7}{*}{\centering\arraybackslash $-.8$} & .462 & .464 & .468 & .431 & .418\\

\addlinespace[0.3em]
\multicolumn{1}{r}{\textbf{}}\\
50 &  & .558 & .605 & .354 & .499 & .564\\

75 &  & .586 & .615 & .415 & .559 & .570\\

100 &  & .602 & .620 & .459 & .578 & .565\\

150 &  & .522 & .557 & .438 & .565 & .520\\

250 &  & .563 & .583 & .526 & .560 & .538\\

500 &  & .484 & .482 & .472 & .496 & .450\\

1000 & \multirow{-7}{*}{\centering\arraybackslash $-.4$} & .520 & .529 & .544 & .559 & .470\\

\addlinespace[0.3em]
\multicolumn{1}{r}{\textbf{}}\\
50 &  & .210 & .482 & .269 & .519 & .514\\

75 &  & .357 & .560 & .377 & .596 & .572\\

100 &  & .352 & .513 & .393 & .576 & .516\\

150 &  & .439 & .536 & .454 & .593 & .524\\

250 &  & .439 & .521 & .472 & .558 & .498\\

500 &  & .454 & .534 & .507 & .532 & .474\\

1000 & \multirow{-7}{*}{\centering\arraybackslash $.4$} & .450 & .528 & .520 & .540 & .483\\

\addlinespace[0.3em]
\multicolumn{1}{r}{\textbf{}}\\
50 &  & .231 & .466 & .218 & .427 & .485\\

75 &  & .274 & .469 & .280 & .458 & .479\\

100 &  & .283 & .462 & .296 & .446 & .473\\

150 &  & .348 & .516 & .385 & .518 & .489\\

250 &  & .365 & .457 & .408 & .475 & .437\\

500 &  & .432 & .485 & .479 & .477 & .452\\

1000 & \multirow{-7}{*}{\centering\arraybackslash $.8$} & .441 & .493 & .491 & .487 & .455\\
\bottomrule
\end{tabular}

%% file: tabs/Tab_MA_sapower_c.tex
\begin{tabular}{lcccccc}
\toprule
$T$ & $\vartheta$ & $\tau$ & $\Breve{\tau}$ & $\textup{ADF}^\textup{GLS}$ & $\textup{MZ}_t$ & $J_\alpha$\\
\midrule
\addlinespace[0.3em]
\multicolumn{1}{r}{\textbf{}}\\
50 &  & .163 & .204 & .197 & .195 & .222\\

75 &  & .202 & .238 & .208 & .212 & .229\\

100 &  & .185 & .206 & .170 & .171 & .207\\

150 &  & .184 & .213 & .173 & .148 & .211\\

250 &  & .210 & .214 & .177 & .160 & .245\\

500 &  & .220 & .235 & .224 & .197 & .242\\

1000 & \multirow{-7}{*}{\centering\arraybackslash $-.8$} & .286 & .302 & .287 & .274 & .262\\

\addlinespace[0.3em]
\multicolumn{1}{r}{\textbf{}}\\
50 &  & .231 & .254 & .214 & .209 & .235\\

75 &  & .231 & .236 & .214 & .208 & .208\\

100 &  & .246 & .266 & .239 & .234 & .239\\

150 &  & .234 & .246 & .231 & .227 & .221\\

250 &  & .272 & .280 & .264 & .261 & .248\\

500 &  & .295 & .318 & .293 & .293 & .278\\

1000 & \multirow{-7}{*}{\centering\arraybackslash $-.4$} & .334 & .369 & .354 & .353 & .303\\

\addlinespace[0.3em]
\multicolumn{1}{r}{\textbf{}}\\
50 &  & .101 & .193 & .196 & .152 & .185\\

75 &  & .183 & .235 & .216 & .193 & .203\\

100 &  & .180 & .245 & .243 & .235 & .213\\

150 &  & .193 & .250 & .251 & .250 & .209\\

250 &  & .222 & .258 & .256 & .249 & .243\\

500 &  & .227 & .291 & .284 & .284 & .241\\

1000 & \multirow{-7}{*}{\centering\arraybackslash $.4$} & .283 & .351 & .323 & .323 & .281\\

\addlinespace[0.3em]
\multicolumn{1}{r}{\textbf{}}\\
50 &  & .112 & .170 & .172 & .100 & .152\\

75 &  & .150 & .212 & .210 & .162 & .193\\

100 &  & .181 & .221 & .206 & .187 & .180\\

150 &  & .185 & .233 & .225 & .212 & .217\\

250 &  & .188 & .237 & .232 & .220 & .231\\

500 &  & .224 & .298 & .273 & .267 & .253\\

1000 & \multirow{-7}{*}{\centering\arraybackslash $.8$} & .260 & .300 & .297 & .293 & .260\\
\bottomrule
\end{tabular}

%% file: tabs/Tab_MA_sapower_ct.tex
\begin{tabular}{lcccccc}
\toprule
$T$ & $\vartheta$ & $\tau$ & $\Breve{\tau}$ & $\textup{ADF}^\textup{GLS}$ & $\textup{MZ}_t$ & $J_\alpha$\\
\midrule
\addlinespace[0.3em]
\multicolumn{1}{r}{\textbf{}}\\
50 &  & .192 & .255 & .271 & .137 & .302\\

75 &  & .222 & .283 & .288 & .279 & .329\\

100 &  & .237 & .294 & .274 & .268 & .317\\

150 &  & .224 & .271 & .246 & .237 & .292\\

250 &  & .228 & .263 & .215 & .183 & .274\\

500 &  & .260 & .304 & .268 & .219 & .326\\

1000 & \multirow{-7}{*}{\centering\arraybackslash $-.8$} & .282 & .326 & .303 & .251 & .328\\

\addlinespace[0.3em]
\multicolumn{1}{r}{\textbf{}}\\
50 &  & .294 & .334 & .247 & .269 & .319\\

75 &  & .287 & .322 & .264 & .258 & .301\\

100 &  & .281 & .306 & .263 & .251 & .303\\

150 &  & .316 & .352 & .303 & .288 & .313\\

250 &  & .336 & .372 & .317 & .318 & .327\\

500 &  & .340 & .390 & .373 & .364 & .355\\

1000 & \multirow{-7}{*}{\centering\arraybackslash $-.4$} & .391 & .417 & .398 & .400 & .365\\

\addlinespace[0.3em]
\multicolumn{1}{r}{\textbf{}}\\
50 &  & .254 & .288 & .316 & .140 & .261\\

75 &  & .202 & .224 & .204 & .121 & .199\\

100 &  & .242 & .262 & .214 & .179 & .244\\

150 &  & .283 & .284 & .252 & .252 & .246\\

250 &  & .322 & .338 & .313 & .317 & .300\\

500 &  & .361 & .383 & .368 & .355 & .325\\

1000 & \multirow{-7}{*}{\centering\arraybackslash $.4$} & .382 & .401 & .383 & .381 & .333\\

\addlinespace[0.3em]
\multicolumn{1}{r}{\textbf{}}\\
50 &  & .158 & .182 & .183 & .042 & .147\\

75 &  & .192 & .225 & .215 & .121 & .206\\

100 &  & .236 & .261 & .248 & .161 & .227\\

150 &  & .261 & .291 & .261 & .201 & .248\\

250 &  & .292 & .316 & .301 & .258 & .260\\

500 &  & .289 & .311 & .281 & .270 & .282\\

1000 & \multirow{-7}{*}{\centering\arraybackslash $.8$} & .334 & .359 & .352 & .342 & .314\\
\bottomrule
\end{tabular}